/
\documentclass[12pt]{iopart}

\newcommand{\lambdabar}{\mbox{\makebox[-0.5ex][l]{$\lambda$} \raisebox{0.7ex}[0pt][0pt]{--}}}
\usepackage{graphicx}
\usepackage{citesort}
\usepackage{listings}
\usepackage{amssymb}
\usepackage{url}
\usepackage{enumerate}
\usepackage{fancyhdr}
\begin{document}
\title[Nuclear Physics of the Outer Layers of Accreting Neutron Stars]{Nuclear Physics of the Outer Layers of Accreting Neutron Stars}

\author{Zach Meisel}
\address{Department of Physics \& Astronomy, Ohio University, Athens, OH, 45701 USA}
\ead{meisel@ohio.edu}

\author{Alex Deibel}
\address{Department of Astronomy, Indiana University, Bloomington, IN 47405, USA}
\ead{adeibel@iu.edu}

\author{Laurens Keek}
\address{University of Maryland, College Park, MD 20742, USA}
\ead{lkeek@umd.edu}

\author{Peter Shternin}
\address{Ioffe Institute, Politekhnicheskaya 26, St. Petersburg 194021, Russia}
\ead{pshternin@gmail.com}

\author{Justin Elfritz}
\address{Anton Pannekoek Institute, University of Amsterdam, Postbus 94249, NL-1090GE Amsterdam, the Netherlands}
\ead{j.g.elfritz@uva.nl}

\begin{abstract}
Now 50 years since the existence of the neutron star crust was
proposed, we review the current understanding of the nuclear physics of the outer layers of accreting neutron stars. Nuclei produced during nuclear burning replace the nascent composition of the neutron star ocean and crust. Non-equilibrium nuclear reactions driven by compression alter the  outer thermal structure and chemical composition, leaving observable imprints on astronomical phenomena. As observations of 
bursting neutron stars and cooling neutron stars have increased, the recent volume of astronomical data allows new insights into the microphysics of the neutron star interior and the possibility to test nuclear physics input in model calculations. Despite numerous advances in our understanding of neutron star interiors and observed neutron star phenomena, many challenges remain in the astrophysics theory of accreting neutron stars, the nuclear theory of neutron-rich nuclei, and  
the reach 
and precision of terrestrial nuclear physics experiments.
\end{abstract}


\tableofcontents{\thispagestyle{empty}}
\maketitle

\newpage
\setcounter{page}{1}

\section{Introduction}

It has been nearly a century since the first nuclear reaction was intentionally measured in the laboratory ($^{14}\rm{N}(\alpha,p)$ in 1919 by Rutherford~\cite{Ruth35}) and a half century since the existence of the neutron star crust was proposed by Ruderman~\cite{Rude68}. These milestones, and the later discovery of the first X-ray burst in the source 3U~1820-30~\cite{grindlay75}, laid the foundation for a collaborative effort between nuclear physics and astrophysics to advance our understanding of dense matter. 
It was soon postulated that X-ray bursts on accreting neutron stars arise from the unstable ignition of accreted hydrogen and helium material \cite{Woos76,joss77}, an idea later supported by numerical models~\cite{fujimoto81,ayasli1982,Woos04}. Burst models now serve as an important probe of nuclear reactions~\cite{Cybu16} because nuclear burning in X-ray bursts proceeds through a combination of the triple-$\alpha$ reaction, rapid proton-capture (rp)-process, and $\alpha$-capture proton-emission ($\alpha \rm{p}$)-process \cite{Wall81}. Moreover, burst models reveal the thermal structure of the neutron star's outer layers where the bursts originate. 

As accreted material accumulates on the neutron star 
surface it compresses underlying material to greater depths and higher mass densities. Before the ashes of hydrogen and helium burning join the crust, further nuclear burning takes place. For example, any $^4$He still present will be captured on heavier isotopes. Runaway $^{12}$C+$^{12}$C fusion powering superbursts is the most dramatic instance of burning in the ashes layer \cite{cumming2001,Stro02}. These superbursts are a thousand times more energetic, a thousand times longer, and occur much deeper than the standard X-ray bursts. These rare events allow one to probe deeper layers of the stellar envelope.

Cooling neutron stars provide another avenue to explore the thermal properties of dense matter and the nuclear reactions therein. In transient systems, the neutron star accretes material in episodes lasting days to years (see Reference~\cite{Dege15} for a summary) and the accretion-induced compression
of the neutron star crust triggers non-equilibrium nuclear reactions~\cite{bisnovatyi1979,Sato79} that heat the neutron star crust~\cite{Haen90,Haen01}. When an accretion episode ends, the neutron star cools on observable timescales~\cite{ushomirsky2001,rutledge2002} and the cooling trend probes the microphysics of the interior~\cite{Shte07,Brow09}. In addition to constraints on the thermal properties of the outer layers, the observed cooling reveals the strength and location of nuclear reaction heating, and by extension the details of the nuclear reactions themselves. For example, observations of cooling neutron stars can be explained by models including 
$e^{-}$-capture heating~\cite{Gupt07} and pycnonuclear fusion heating~\cite{haensel2008} in the neutron star crust. Furthermore, crust cooling has the potential to reveal strong Urca cooling layers~\cite{Scha14,Deib16} that are suspected to be present in crusts enriched by X-ray burst ashes~\cite{Meis17}.

Experimental nuclear data are critical physics input for neutron star models. Dense matter constraints derived from neutron star observations are therefore limited by the precision of nuclear physics input in addition to other systematics. 
The reduction of nuclear physics uncertainties, 
as well as the development of more sophisticated neutron star models, will improve future observational constraints on dense matter. For example, current investigations of the critical reaction rates in X-ray bursts~\cite{Cybu16,Scha17}, studies of key properties of nuclei in the accreted crust~\cite{Deib16,Meis17}, and nuclear reaction network calculations of accreted crust compositions~\cite{Gupt07,Scha14,Lau18}, all promise to improve the observational constraints on dense matter derived from accreting neutron stars. 

We begin in section~\ref{section:structure} by briefly outlining the neutron star structure. Sections~\ref{subsection:pristine} and \ref{section:accretion} describe the original composition of the neutron star crust and the accretion process which drives the system from equilibrium. In section~\ref{section:production}, we discuss the nuclear burning that can occur on the surface of accreting neutron stars and the nuclei produced during the different possible burning regimes. In actively accreting neutron stars, the ashes of prior surface nuclear burning are compressed to greater depths by newly accreted material. We discuss in section~\ref{section:interaction} the nuclear interactions involving the ashes that take place as the ambient mass density increases. In section~\ref{section:impact}, we investigate the impact of interior nuclear interactions on observable neutron star phenomena. We summarize and discuss prospects for future work in section~\ref{section:summary}. 

\section{Neutron Star Structure} \label{section:structure}
A neutron star is born in the death of a massive star (from 8 and up to 50\ $\rm{M}_{\odot}$ \cite{Heger2003,Pejcha2015ApJ}, where $M_\odot=1.99\times 10^{33}\,\mathrm{g}$ is the mass of the Sun), when the lack of sufficient outward pressure from the core to balance the inward compression from gravity leads to a dramatic collapse. The iron-group nuclei in the progenitor core are rapidly transmuted by electron-captures and photodisintegration, transforming the core into a hot and extremely dense neutron-rich sphere~\cite{Woos05,Burr86}. The collapse halts when the central density approaches the mass density of atomic nuclei, $\rho_0=2.8\times 10^{14}$~g~cm$^{-3}$ (baryon density $n_{0}=0.16~\rm{fm}^{-3}$), and the neutron degeneracy pressure and the repulsion from the strong nuclear force bounce back the infalling matter. This can finally result in a core collapse supernova event, leaving a dense compact remnant -- a neutron star  
-- behind \cite{Heger2003,Pejcha2015ApJ}. 

As a consequence, neutron stars 
contain $\sim1\textrm{--}2~\rm{M}_{\odot}$ in a sphere of radius $R\sim10\textrm{--}15 \, \mathrm{km}$~\cite{steiner2010,steiner13} and have average mass densities of several $\rho_0$. Almost all (99 per cent) of the mass is concentrated in the bulky core composed of uniform nuclear (or possibly more exotic) matter. The equation of state and even the composition of the neutron star core are unknown and their elucidation are some of the fundamental problems of neutron star astrophysics \cite{HPY2007}.  

At mass densities $\rho\lesssim 0.5\, \rho_0$ the uniform nuclear matter is unstable and arranges into 
nuclear clusters, which at the low densities are familiar albeit neutron-rich nuclei \cite{Peth95}. These outer layers of the neutron star, though comprising only roughly one percent of the 
mass, provide the settings for all of the astronomical observables used to characterize accreting neutron stars,
and are the primary subjects of the present review. 

\begin{figure} [ht]
\centering
\includegraphics[scale=0.7]{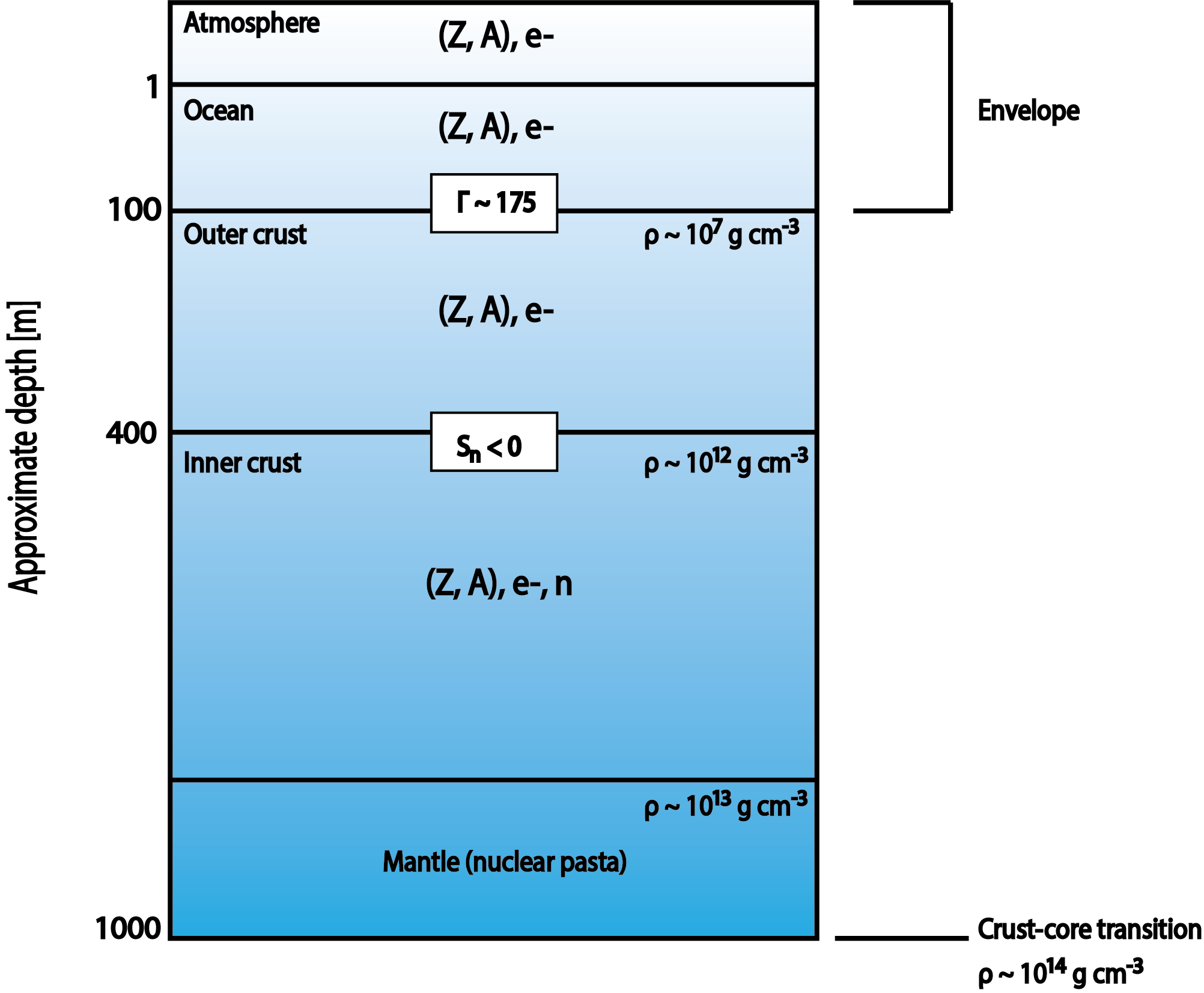}
\caption{Schematic of the outer layers of a neutron star.}
\label{figure:schematic}
\end{figure}

Much like a terrestrial planet, the neutron star outer layers consist, from the outside-in, of an atmosphere and liquid ocean (collectively referred to as the envelope), solid crust (outer and inner), and the mantle, roughly 1~km thick in total as schematically shown in figure~\ref{figure:schematic}. 
In contrast to planetary envelopes, matter in these layers is 
under extreme conditions, for example, the enormous gravity $g=(GM/R^2)(1+z)$, where $G$ is the gravitational constant, $M$ is the neutron star mass, $(1+z)\equiv \left(1-2GM/(Rc^2)\right)^{-1/2}$ is the gravitational redshift at the neutron star surface, and $c$ is the speed of light. For a canonical neutron star of $M=1.4 M_\odot$ and $R=10$~km, $g=2.44\times 10^{14}$~cm~s$^{-2}$ and $1+z= 1.31$. Large gravity ensures that General Relativity effects are important when neutron star phenomena are studied. Since the outer layers are thin, the gravity and the redshift can be set constant and equal to the surface values. 

Throughout this paper we will use various physical quantities to specify the current position in the global structure of the envelope depending on the aspects of the problem discussed. The equation of state and composition of dense matter mainly depend on the mass density $\rho$ or baryon number density $n_B$. However, these quantities may be discontinuous at composition changes (see below) and thus may not be the appropriate variables to follow the crustal structure. Instead, the pressure $P$ or the column density $y\equiv P/g$ are continuous and increase monotonically, and are convenient measures of the depth. The column density is especially useful when accretion phenomena are studied, since the total baryon mass above a layer with $y=\mathrm{constant}$ is $\Delta M_B(y)=4\pi R^2 y$. Note that the gravitational mass of the same layer is smaller, $\Delta M = \Delta M_B/(1+z)$, because of the gravitational binding energy.
One also distinguishes between the radial coordinate $r$ (so that the surface area is $4\pi r^2$) and the proper depth measured by a local observer in the envelope $\zeta=\int_r^R (1+z(r))\mathrm{d}r\approx (1+z)(R-r)$.

The outer neutron star envelope -- atmosphere and ocean -- contains a non-ideal plasma of electrons and ions, and except in the outermost layers, the plasma is fully ionized. There is usually no phase transition between the gaseous and liquid state, except for some special cases \cite{Potekhin2014PhyU}. The atmosphere is distinguished as the region where the neutron star surface emission is formed 
and extends up to a density of 
$\sim 10^{-4}-10^{6}~\rm{g}~\rm{cm}^{-3}$ \cite{Potekhin2010PhyU}, depending on the physical conditions. Both the atmosphere and ocean may host convective processes, driven by gradients of temperature and chemical potentials~\cite{Malone2011,Medin2011}. 
In most of the envelope, electrons are strongly degenerate, relativistic (at $\rho\gg 10^6$~g~cm$^{-3}$), and provide the main contribution to the pressure. For ultrarelativistic electrons, their contribution to the pressure is $P_e=\mu_e^4/(12\pi^2 \hbar^3 c^3)$, where $\mu_e$ is the electron chemical potential and  $\hbar$ is the reduced Planck constant. Thus $\mu_e$ is directly related to the column depth $y$ (such that $\mu_e\propto y^{1/4}$) and provides another useful measure of the depth. $\mu_e$ is a particularly convenient coordinate to use when the crust composition 
and nuclear reactions are studied (sections~\ref{subsection:pristine}, \ref {section:interaction}). 
For convenience, we give the relation between various measures of depth in the neutron star outer envelope and crust, where the dominant pressure is from degenerate electrons:
\begin{eqnarray}
y&\approx& 7.2\times 10^9\, \mathrm{g}\,\mathrm{cm}^{-2} \left(\frac{\mu_e} {1~\mathrm{MeV}}\right)^4\frac{2.44\times 10^{14}~\mathrm{cm}~\mathrm{s}^{-2}}{g},\label{eq:y-mu}\\
\rho&\approx&7.2\times 10^6\, \mathrm{g}\,\mathrm{cm}^{-3} \left(\frac{\mu_e} {1~\mathrm{MeV}}\right)^3\frac{1}{Y_e},\label{eq:rho-mu}\\
\zeta&\approx&40~\mathrm{m}\, \frac{\mu_e} {1~\mathrm{MeV}}\frac{2.44\times 10^{14}~\mathrm{cm}~\mathrm{s}^{-2}}{g} Y_e \label{eq:z-mu},
\end{eqnarray}
where $Y_e$ is the electron fraction, 
and in equation~(\ref{eq:z-mu}) it is a depth-averaged quantity.

The properties of ions are determined by the ratio of their Coulomb energy to the thermal energy. For a plasma containing ions of one species with charge number $Z$, the plasma coupling parameter $\Gamma=Z^2 e^2/ (a k_\mathrm{B} T)$, where $a=[3/(4\pi n_i)]^{1/3}$ is the ion-sphere (Wigner-Seitz) radius, $n_i$ is the ion number density, $e$ is the elementary electric charge, and $k_\mathrm{B}$ is the Boltzmann constant. When $\Gamma\gtrsim 1$, ion correlations are important so they form a non-ideal Coulomb liquid.
In the region where electrons are ultrarelativistic,
\begin{equation}
\Gamma\approx 105\, \left(\frac{Z}{26}\right)^{5/3} \left(\frac{T}{10^8~{\rm K}}\right)^{-1} \frac{\mu_e}{1~\mathrm{MeV}}\approx 361 \left(\frac{Z}{26}\right)^{5/3} \left(\frac{T}{10^8~{\rm K}}\right)^{-1} y_{12}^{1/4},
\label{eq:Gamma}
\end{equation}
where $y_{12}$ is the column depth measured in units of $10^{12}$~g~cm$^{-2}$ and the canonical value of $g$ is used. When there are multiple ion species $Z_j$, which is the case for the accreted crust, partial $\Gamma_j$ are introduced, and an average $\langle\Gamma_j\rangle$ can be used to describe the state of the whole mixture. An important 
quantity is the so-called impurity parameter which describes variance
of the charge mixture around the mean charge $\langle Z\rangle$. It is defined as
\begin{equation}
Q_{\rm imp} \equiv \frac{1}{n_{\rm ion}} \sum_j n_j (Z_j - \langle Z \rangle)^2 \ ,
\label{equation:Qimp}
\end{equation}
where $n_{\rm ion}$ is the total local number density of ions, and $n_j$ is the local number density of nuclei species $j$.
The impurity parameter is important when the transport properties such as thermal conductivity are discussed (see section~\ref{section:impact}).

At a depth of several tens of meters, the ocean solidifies~\cite{Rude68,Medi10}. The pure Coulomb plasma solidifies at about $\Gamma\approx175$ \cite{Potekhin2000} 
and forms a perfect crystal with a body-centered cubic lattice.  As follows from equation~(\ref{eq:Gamma}), for a typical temperature $T=10^8$~K solidification occurs at $\mu_e\approx1.7$~MeV ($y\approx6\times 10^{10}$~g~cm$^{-2}$, $\zeta\approx 140$~m) for $Z=26$ (iron). This point corresponds to mass density $\rho\approx 7\times 10^7$~g~cm$^{-3}$ and shifts deeper inside the neutron star for decreasing $Z$.

The solidification of the multi-component mixture is thought to occur at a similar order of magnitude of the average $\langle \Gamma \rangle$ \cite{Mcki16}. However, the exact structure of the  multicomponent solid crust and solidification layer at the ocean-crust boundary are not well understood. There is strong theoretical and observational evidence that the solid mixture also arranges in a regular but impure crystalline structure (see sections \ref{subsection:impact_inner} and \ref{sec:cooling_transients}). 

A few-hundred meters deeper, the rising electron chemical potential results in $\beta$-equilibrium nuclei closer to the neutron dripline, until ultimately the neutron separation energy for nuclei in the plasma becomes negative.
At this point it is energetically favorable for neutrons to drip out of nuclei and form a neutron gas~\cite{Chamel2008LRR,HPY2007} 
marking the upper boundary of the inner crust. This neutron drip occurs at densities $\rho_{\rm{nd}}\sim 4-10 \times 10^{11}$~g~cm$^{-3}$ 
depending on the composition and the theoretical model used. For the accreted crust, generally, $\rho_{\rm{nd}}$ is on the high end of the aforementioned range.
Thus in the inner crust an ion lattice coexists with a gas of free neutrons which soon becomes degenerate and starts giving the dominant contribution to the pressure. At this point, equations~(\ref{eq:y-mu})--(\ref{eq:z-mu}) are now inapplicable.

Due to the attractive component of the strong interaction, the degenerate neutron gas is subject to a pairing instability. It is believed that in the crust paring occurs in the singlet $^1S_0$ channel and the critical temperature $T_\mathrm{cn}$ is density-dependent with the maximal value of order of $\sim 5\times 10^9$~K \cite{Lombardo2001LNP}.
Thus a large portion of the crustal neutrons is thought to be superfluid, although the precise form of the $T_\mathrm{cn}(\rho)$ profile is very model-dependent. At large densities the nuclear interaction in the $^1S_0$ channel becomes repulsive and superfluidity ceases at the lower bound of the inner crust. However, in some models, $^1S_0$ superfluidity penetrates the core. We 
discuss the observable impact of neutron superfluidity in section~\ref{sec:quasipers}.

When the density increases further in the inner crust, groupings
of protons and neutrons are no longer accurately considered as nuclei 
and can rather be called nuclear clusters. Eventually, the clusters become so large and closely spaced that the competition between the nuclear surface energy and Coulomb energy distorts the nuclei in into complex shapes called nuclear pasta \cite{ravenhall1983}.
Molecular dynamics simulations demonstrate that nuclear pasta can begin to appear near mass densities of $\rho \gtrsim 8\times 10^{13} \ \mathrm{g\,cm^{-3}}$ \cite{hashimoto1984}.
As a consequence,  this `mantle' layer can contain half the mass of the whole crust. 
Note that the pasta layer does not exist for all equations of state \cite{Douchin2000PhLB,Oyamatsu2007PhRvC}. Finally, when proton clustering ceases at $\rho\gtrsim 0.5\rho_0$, the transition to a liquid core occurs.

The structure and properties of the crust as described above can be modified by the presence of fast rotation or high magnetic fields. For instance, rotation deforms the crust, making it non-spherical. The presence of a strong magnetic field in the outer layers can result in the absence of the ocean and a first order phase transition between the thin gaseous atmosphere and the condensed neutron star surface. These effects are outside the scope of this paper  (see, e.g., References~\cite{Chamel2008LRR,Potekhin2014PhyU}).

For the upper reaches of the inner crust and above, these regions are composed of nuclei which can be made in current and near-future nuclear physics laboratories. As such, accreting neutron stars
provide a medium to explore many interesting multi-physics questions through astronomical observations and theory, as well as nuclear physics theory and experiment. The following sections will explore the dominant nuclear physics processes occurring in accreting neutron stars'  
outer layers and the astronomical observables these processes impact. But first, a brief discussion of the crust composition in the absence of accretion is beneficial in order to provide context for the dramatic transformation accretion induces.

 \section{Pristine Crust Composition}
 \label{subsection:pristine}
  
The extreme temperatures ($\gtrsim 10^{11}$~K) achieved during the supernova collapse drive 
matter in the unaccreted crust to its ground state in terms of nuclear and $\beta$ equilibrium, as determined by the local environment conditions and nuclear masses~\cite{Beth70,Buch71,Baym71a,Baym71b}. 
This means that the unaccreted outer crust is stratified into several layers, each of which is comprised of a single species of nuclide. 
Here we discuss the cold-catalyzed crust, where matter is in its absolute ground state~\cite{Baym71a}. 
In principle, the composition may freeze-in at a higher temperature, modifying the equilibrium abundance distribution due to thermal fluctuations (see, e.g., Reference~\cite{Gori11}).

In the outer crust, it is energetically favorable for nuclei to arrange themselves in a Coulomb lattice within an electron Fermi gas~\cite{Rude68}. As such, beyond minimizing the usual liquid drop terms describing the nuclear mass in vacuum, the additional energy contributions from the lattice and the electron gas must be considered~\cite{Buch71,Baym71a}. The dominant liquid drop terms include the volume energy, which can be modified due to compression by the dense environment; the surface energy, which generally favors large-$A$ nuclides;
the Coulomb energy, which favors low-$A$ nuclides; and the symmetry energy, which favors nuclides with neutron number $N=Z$. The lattice energy favors the existence of large-$A$ nuclei, providing a Coulomb energy of opposite sign to its liquid-drop partner. The resulting total energy per nucleon is
$E_{\mathrm{total}}/A = E_{\mathrm{nuclear}}/A + E_{e^{-}\mathrm{-gas}}/A + E_{\mathrm{lattice}}/A$,
whose minimum defines the equilibrium nucleus. In practice, the equilibrium composition is solved for at a given pressure, which is a proper continuous variable. Therefore, it is generally the Gibbs free energy per nucleon that is minimized~\cite{Baym71b,Haen01,Roca08}.
  
  \begin{figure} [h]
\centering
\includegraphics[scale=0.5]{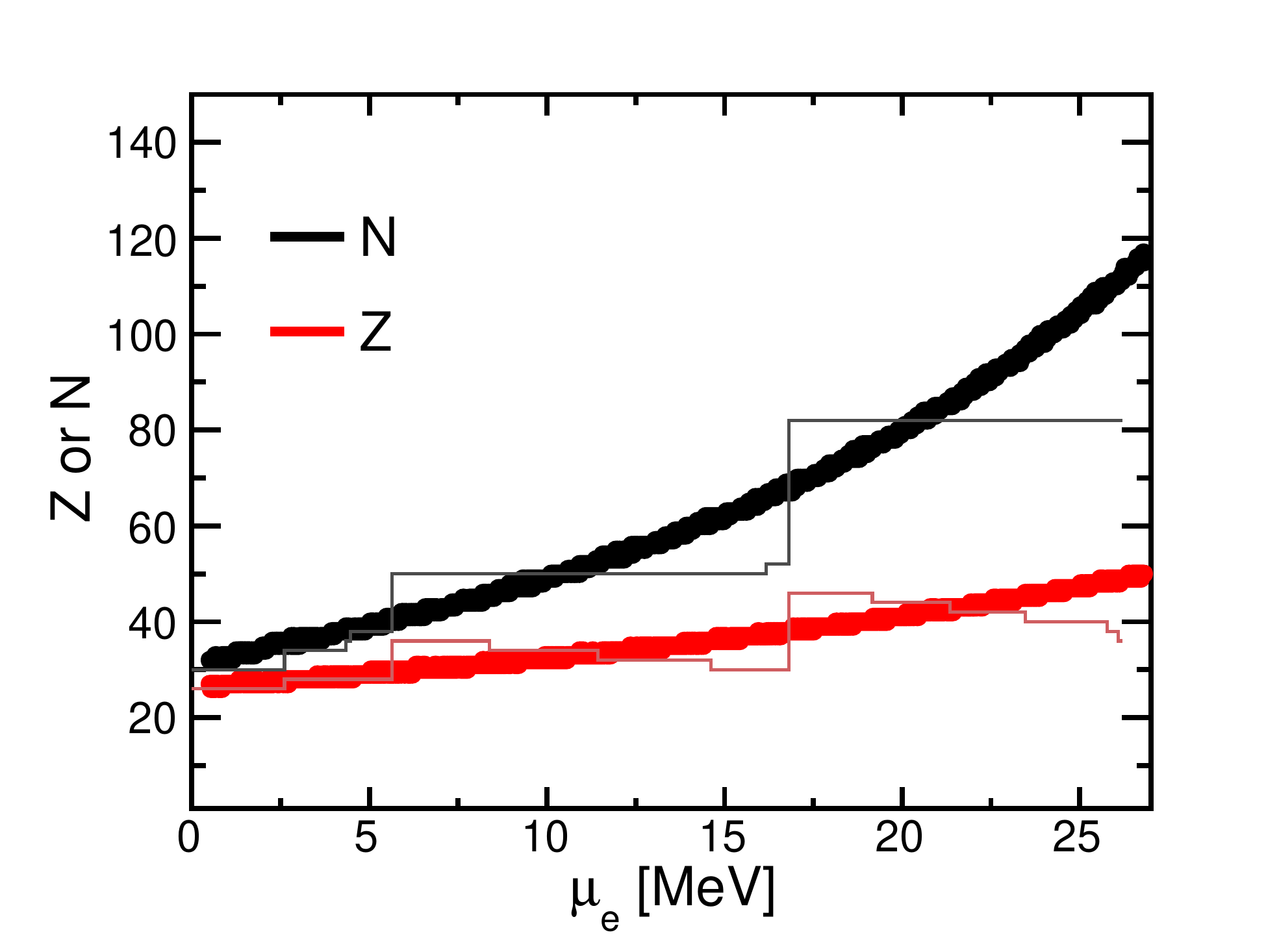}
\caption{Equilibrium composition for a cold-catalyzed neutron star outer crust calculated using a liquid drop model for nuclear masses (thick lines). To demonstrate the impact of shell effects, the equilibrium composition obtained using experimental masses when available and the BSk8 Skyrme model otherwise, as reported by Reference~\cite{Rust06}, is also shown (thin lines). See figure~\ref{figure:ECcrustcomposition} to compare to the accreted crust composition.}
\label{figure:pristinecomp}
\end{figure}
  
At relatively low density near the neutron star 
surface, the nuclear energy term will dominate and the nucleus with the minimum mass per nucleon, $^{56}\rm{Fe}$, is optimal. At deeper depths, the increasing electron chemical potential $\mu_e$ creates an energetic incentive to lower the electron fraction, $Y_{e}=Z/A$, favoring more neutron-rich nuclides, in competition with the nuclear symmetry energy which favors $Y_{e}=1/2$. A decent estimate of the equilibrium nuclide at a given depth can be obtained by employing the semi-empirical mass formula for the nuclear binding energy in conjunction with the $e^{-}$-gas chemical potential and lattice energy, resulting in the total energy per nucleon
\begin{eqnarray}
\label{eqn:CrustE}
E_{\rm{total}}\left(A,Z,\mu_{e}\right)/A= &m_\mathrm{p}c^{2}Y_{e}+m_\mathrm{n}c^{2}\left(1-Y_{e}\right)-a_{v}+\frac{a_{s}}{A^{1/3}}+a_{c}A^{2/3}Y_{e}^{2}\\
&+a_{a}\left(1-2Y_{e}\right)^{2}+\frac{3}{4}
Y_{e}
\mu_{e}-C_{\ell}A^{2/3}
Y_{e}^{5/3}\mu_{e},\nonumber
\end{eqnarray}
where $m_\mathrm{p}$ and $m_\mathrm{n}$ are the proton and neutron masses, respectively; $a_{v}$, $a_{s}$, $a_{c}$, and $a_{a}$ are the volume, surface, Coulomb, and asymmetry coefficients of the semi-empirical mass formula;  
and $C_{\ell}$ is a (relatively small) constant scale factor for the Coulomb lattice energy~\cite{Roca08}. This assumes an ultrarelativistic form for the electron Fermi gas contribution.  
The equilibrium nuclide is found by minimizing equation~(\ref{eqn:CrustE}) with respect to $Y_{e}$
and separately with respect to $A^{1/3}$, which gives $Z$ and $A$ at the minimum.
 
To demonstrate how liquid-drop nuclear binding, electron gas, and Coulomb lattice energy contributions impact the equilibrium nucleus, an approach similar to Reference~\cite{Roca08} is followed here. We fit the nuclear binding contribution from equation~(\ref{eqn:CrustE}) to the experimental masses of the 2012 Atomic Mass Evaluation~\cite{Audi12}, though including an additional term for the nuclear pairing energy that goes as $i{a_{p}}{A^{-3/2}}$~\cite{Mart09}, where $i=+1,-1,$ or $0$ to enhance, penalize, or leave alone nuclear binding for even-even, odd-odd, and even-odd nuclides, respectively. For such a fit, $a_{v}$=15.302, $a_{s}$=16.518, $a_{c}$=0.687, $a_{a}$=22.243, and $a_{p}$=5.898, each with units of MeV. Using these parameters and $C_{\ell}$=3.40665$\times10^{-3}$ (for a body-centered cubic lattice, e.g.  Reference~\cite{Roca08}), simultaneous minimization of $A^{1/3}$ and $Y_{e}$ results in the equilibrium nucleus trend shown with thick lines in figure~\ref{figure:pristinecomp}.

The trend in composition for a liquid drop model is modified by the presence of nuclear shell closures present in more realistic nuclear mass models, as shown with thin lines in figure~\ref{figure:pristinecomp}. 
Though models disagree on the detailed composition depending on which nuclear mass model is employed, generally, the equilibrium nuclide is near the $Z=28$ shell closure at shallow depths 
($n_B\sim$$10^{-7}$~fm$^{-3}$, i.e. $\sim$50 m), transitioning to $N=50$ nuclides ($n_B\sim$$10^{-5}$~fm$^{-3}$, i.e. $\sim$150 m), and finally resulting in $N=82$ nuclides prior to neutron-drip ($n_B\sim$$10^{-4}$~fm$^{-3}$, i.e. $\sim$300 m)~\cite{Baym71b,Roca08,Pear11,Rust06}.
Nuclear mass measurements at radioactive ion beam facilities are working their way toward more exotic nuclides to resolve existing discrepancies in theoretical predictions, where experimental constraints are presently available down to a depth of $\sim$200~m~\cite{Wolf13}. 
Beyond this, theoretical estimates are necessary~\cite{Utam17}.
\begin{figure}[ht]
\begin{center}
\begin{minipage}{0.4\textwidth}
\includegraphics[width=\textwidth]{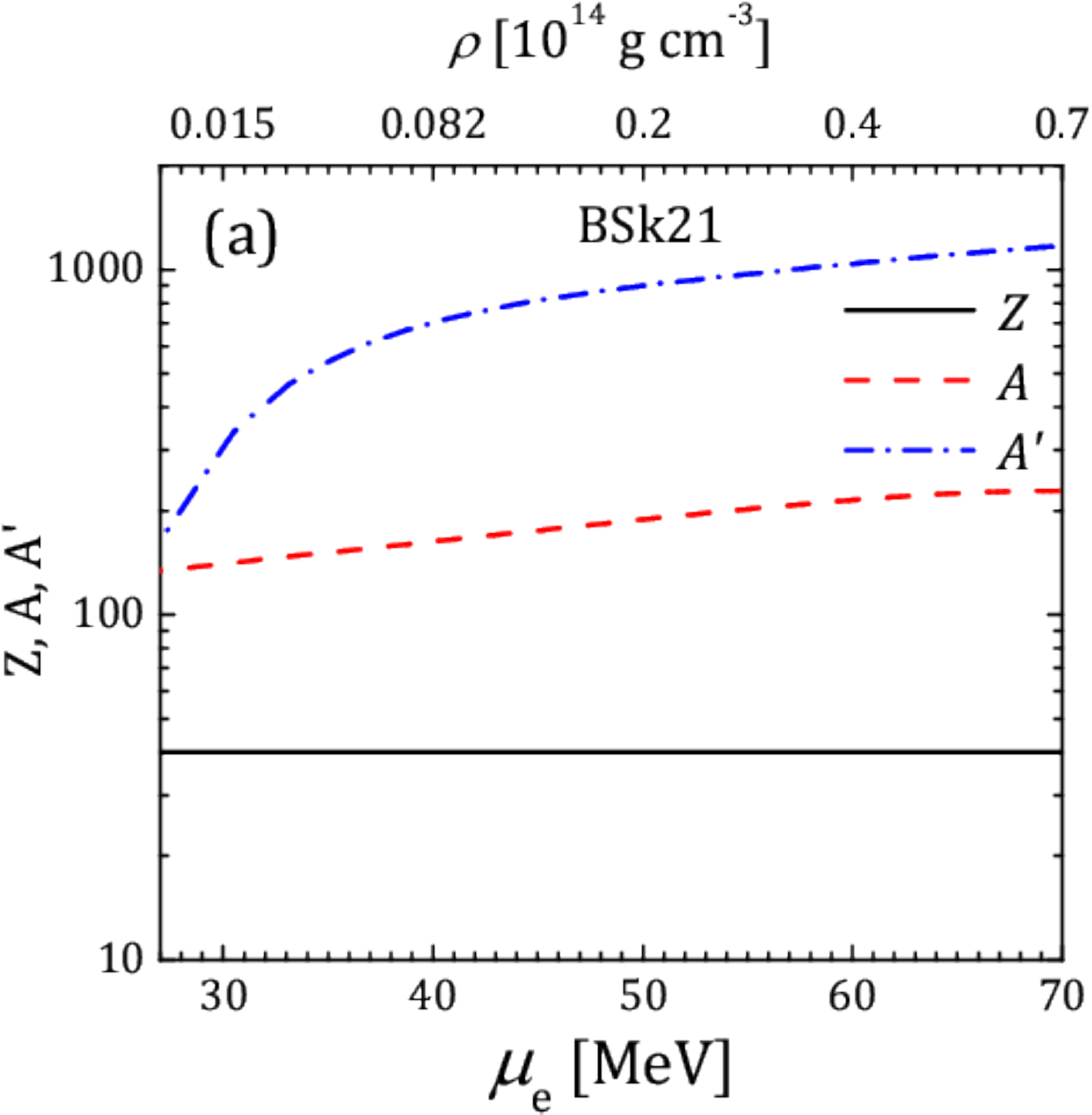}
\end{minipage}
\hspace{0.05\textwidth}
\begin{minipage}{0.5\textwidth}
\includegraphics[width=\textwidth]{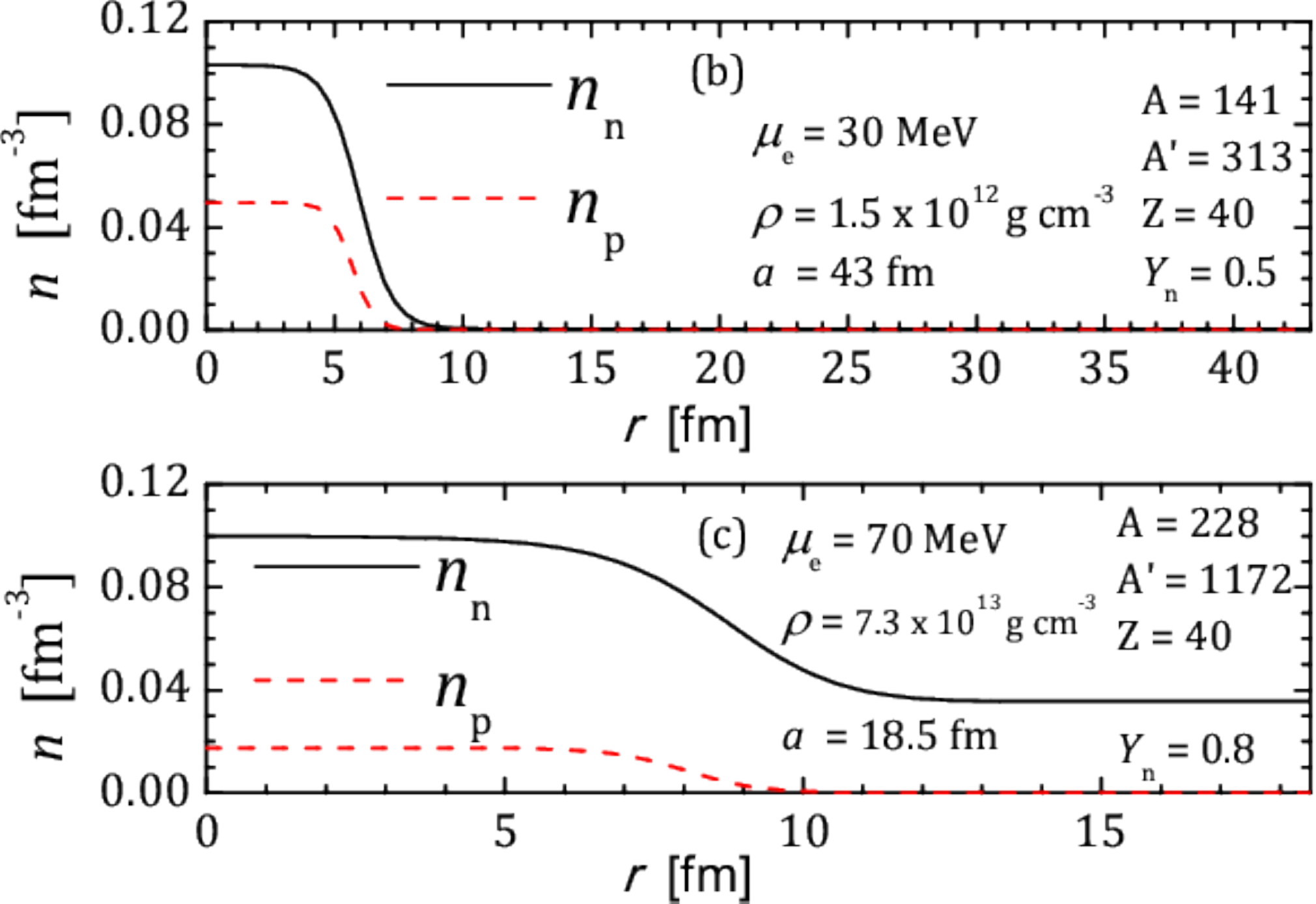}
\end{minipage}
\end{center}
\caption{{\it (a)}: Composition in the inner crust in the BSk21 model after Reference \cite{Pote13}. Solid, dashed, and dash-dotted lines show charge number $Z$, mass number in the cluster $A$, and total mass number in the Wigner-Seitz cell $A'$, respectively. {\it Right:} Neutron and proton number density distributions  in the Wigner-Seitz cell. {\it (b)}: nucleon density profiles near the top of the inner crust, at $\mu_\mathrm{e}=30$~MeV. Relevant nuclear quantities are given in the plot. {\it (c)}: the same for  $\mu_\mathrm{e}=70$~MeV, close to the nuclear pasta onset. }\label{fig:inner}
\end{figure}

By contrast, the pristine inner crust composition is far more uncertain. The presence of the neutron gas makes describing the composition in terms of nuclei less adequate  and, instead, it is more precise to use the concept of proton clusters~\cite{Chamel2008LRR}. Example equilibrium ``nuclei"  are $^{1269}\rm{Zr}$ and $^{633}\rm{Ca}$~\cite{Bald06}, where the total mass number per cluster is used to mark the isotopes. In this respect, one usually wants to distinguish between the total number of nucleons $A'$ per cluster and the number of nucleons inside the cluster $A$. Such a distinction is not a well-defined procedure and one usually takes the nucleon density far from the cluster center as the `gas' phase density.
One approach to describe this region is the classical approach employing a compressible liquid drop model~\cite{Baym71a}. Conceptually, this technique is similar to the method described above for the outer crust, however, extra terms are included in the total energy to account for contributions from the neutron gas and modifications to the liquid-drop semi-empirical mass formula to account for nuclear compressibility. Instead, semi-classical approaches describe the inner crust using energy density functionals, e.g. via the Thomas-Fermi approximation~\cite{Onsi08}. The most sophisticated and computationally-intensive approach is a quantum mechanical solution of the wave function for the cluster existing at each depth~\cite{Bald06}. Each of the three approaches feature compositions dominated by proton shells $Z=20, 40,$ and $50$, though disagreement remains as to which shell the composition converges (see tables 3 and 4 of Reference~\cite{Chamel2008LRR}). The example of the composition resulting from calculations in the BSk21 model \cite{Pote13} is shown in the left panel of figure~\ref{fig:inner} where we plot it until the density where the possible pasta phase would appear. In this model, the proton shell $Z=40$ is found everywhere in the inner crust. 
The concept of a ``cluster" as opposed to a nucleus is illustrated in the right panels in figure~\ref{fig:inner} where we plot the nucleon density profiles in the same model  for $\mu_e=30$~MeV (panel (b)) and near the bottom of the crust $\mu_e=70$~MeV (panel (c)). In the second case there is a significant free neutron fraction 
$Y_\mathrm{n}\approx 0.8$ and a broad neutron distribution is clearly seen.  We refer the reader to References~\cite{Chamel2008LRR,HPY2007} and more recent works \cite{Sharma2015A&A,Pote13,Lim2017PhRvC,Pastore2017JPhG} for a detailed discussion of the inner crust composition in the absence of accretion.

\section{Accretion onto the Neutron Star Surface}\label{section:accretion}

The neutron star 
crust undergoes a significant transformation from the pristine state due to accretion.
Neutron stars 
in binary systems with a lower-mass companion (low-mass X-ray binaries) can accrete material onto their surfaces through Roche-lobe overflow. Matter from the outer layers of the companion is transferred to an accretion disk surrounding the neutron star.
There it loses angular momentum and subsequently falls onto the neutron star. 
Accretion can proceed continuously and we refer to the system as a ``persistent X-ray source'', or can proceed episodically in ``transient'' sources. A transient outburst may have any duration from weeks or months to years (section~\ref{sec:cooling_transients}). 
The composition of the accreted material is that of the outer layers of the companion star, which is often assumed to be similar to that of the Sun with mass fractions of $0.739$\,$^1$H, $0.247$\,$^4$He, and the rest heavier isotopes \cite{Lodders2010}. In Ultra-Compact X-ray Binaries~\cite{intZand2007} with a binary period shorter than $80$ minutes, the companion star lacks a hydrogen-rich envelope, and the accretion composition is mostly $^4$He. The signatures of hydrogen and helium have been detected in the optical spectra of low-mass X-ray binaries, but it is challenging to determine the mass fractions precisely~\cite{Nele06}.

When the companion mass is higher than the mass of the compact object, the accretion via Roche-lobe overflow is not possible. However, the neutron star in the binary system with massive ($>10\,M_\odot$) early-type companion (high-mass X-ray binaries) can accrete matter from the stellar wind or Be-star decretion disk. Neutron stars in high-mass X-ray binaries usually have relatively large magnetic fields ($\sim 10^{12}$~G) and show regular X-ray pulsations, while the magnetic fields of neutron stars in low-mass X-ray binaries are low ($10^{8}-10^{9}$~G). We do not consider high-mass X-ray binaries below and focus on low-mass X-ray binary sources, since the latter provide the main observational manifestations of nuclear processes during accretion.

When an accreted nucleon finally settles at the neutron star surface, it releases $z/(1+z)\times m_u c^2\approx 220$~MeV of the gravitational energy (as measured in a distant observer's frame), where  $m_u$ is the nucleon mass unit \cite{ShaprioTeukolsky1986}. Part of this energy is radiated from the accretion disk and the remaining part is assumed to be radiated away from the surface. Therefore the estimate for the total accretion luminosity (mostly in X-rays), as seen by the distant observer, is
\begin{equation}
L_A^\infty=\frac{z}{1+z}\dot{M} c^2,
\end{equation}
where $\dot{M}$ is the mass accretion rate as seen at infinity (the mass accretion rate as seen by the observer on the neutron star surface is $\dot{M}(1+z)$). Thus a measurement of the persistent accretion flux in principle allows one to estimate $\dot{M}$. This is usually done by assuming spherically-symmetric accretion and introducing the appropriate bolometric correction factor to convert from the observed X-ray luminosity $L_X^\infty$ to bolometric luminosity $L_A^\infty$. 
It is customary to quantify $\dot{M}$ in fractions of the Eddington mass accretion rate $\dot{M}_\mathrm{Edd}$ at which the radiation pressure of the emitted X-ray photons balances the gravitational pull on the infalling material. The critical Eddington luminosity is
\begin{equation}\label{eq:Ledd}
L^\infty_\mathrm{Edd}=\frac{4\pi G M c}{\kappa_\mathrm{es}}/(1+z_\mathrm{ph}),
\end{equation}
where $\kappa_{\rm es}=0.2\,(1+X)$~g$^{-1}$~cm$^2$, with $X$ being the hydrogen mass fraction in the infalling material, is the electron scattering opacity and  $(1+z_\mathrm{ph})$ is the gravitational redshift of the emission region (photosphere); this region 
can be high above the neutron star surface. 
Usually the Eddington luminosity without the latter factor is used, which is the truly maximal limiting luminosity a distant observer can see \cite{ShaprioTeukolsky1986}. Alternatively, one sometimes sets $z_\mathrm{ph}=z$ in equation~(\ref{eq:Ledd}). For a 10~km Newtonian neutron star (neglecting all General Relativity corrections) and a solar composition,
one obtains $\dot{M}_\mathrm{Edd}=1.72\times 10^{-8}\,M_\odot$~yr$^{-1}$. 

During the time interval $\Delta t$, a neutron star 
accretes $\Delta M_B = \dot{M}\Delta t$  baryon mass. According to section~\ref{section:structure}, the base of this slab will be at the column density $y=\dot{M}\Delta t/(4\pi R^2)$. When accretion continues, this layer is compressed to higher $y$. In this sense, by following the crust structure in section~\ref{section:structure} 
with increasing column depth we at the same time are following the journey
of a given accreted fluid element deep through the crust, where it undergoes various nuclear transformations, as discussed in the following sections, to finally merge with the neutron star core. It is clear {\it a fortiori} that the composition of the  accreted layers shall be very different from the pristine crust composition discussed in section~\ref{subsection:pristine}. A neutron star accreting steadily at $0.01\dot{M}_\mathrm{Edd}$, will replace its entire crust with the accreted crust in $10^8$~yr.

\section{Production of Nuclei on Neutron Star Surfaces} \label{section:production}

A variety of processes are responsible for nucleosynthesis on and near the surface of accreting neutron stars.  
Though the products of these burning processes, the ashes, are likely unable to escape the neutron star
gravitational potential and contribute to the cosmic abundances~\cite{Wall81}, they 
have a significant impact on the accreted crust thermal and compositional structure. Here we discuss the primary mechanisms through which nuclei are produced in and on accreting neutron stars.

\begin{figure} [ht]
\centering
\includegraphics[scale=0.8]{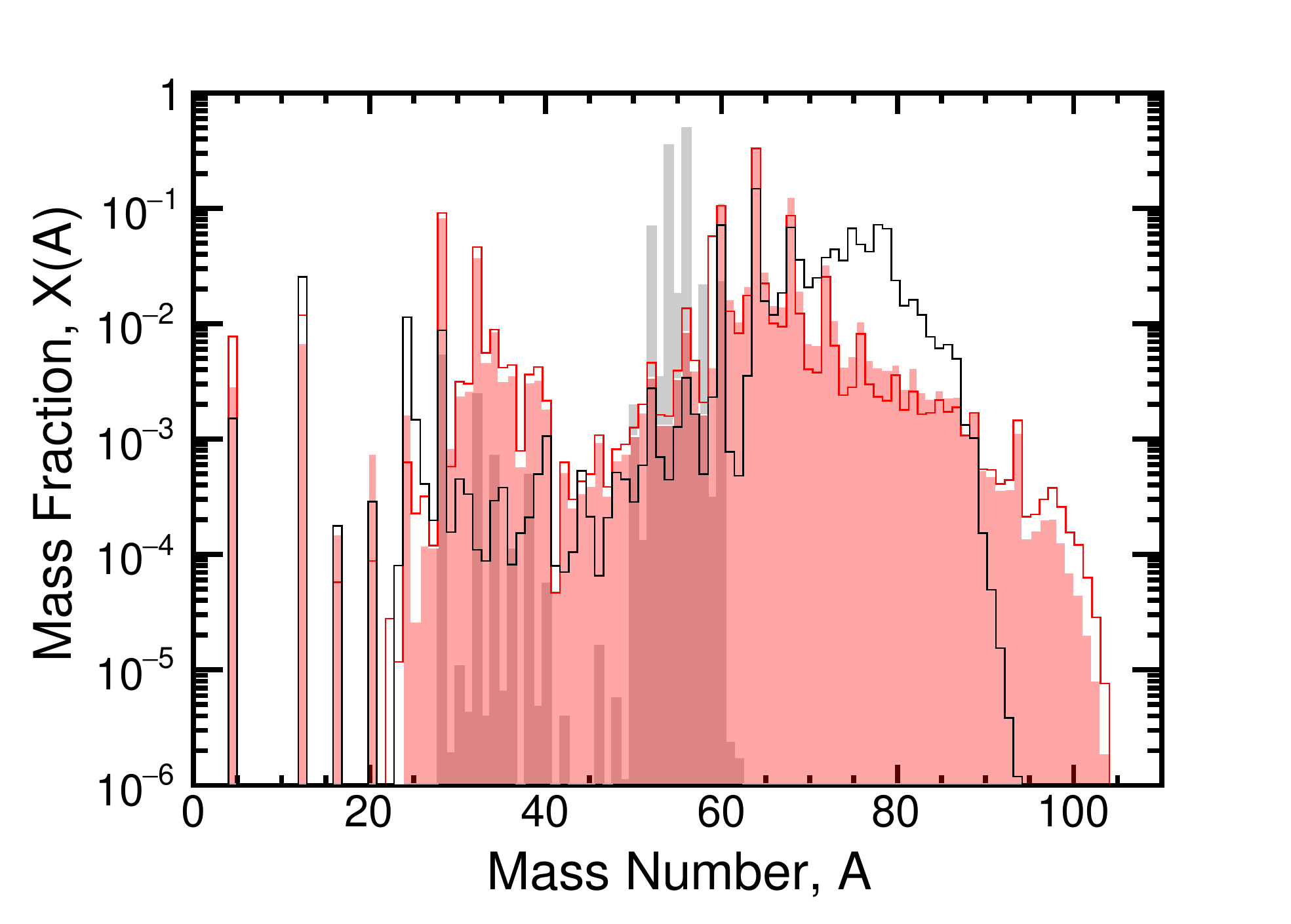}
\caption{Ash abundances predicted for stable surface burning at near-Eddington acretion rate 
(black-lined unfilled histogram), superbursts (gray-filled histogram), and Type-I X-ray bursts for a nominal reaction rate library (red-filled translucent histogram) and, to highlight the impact of nuclear physics uncertainties, for the $^{59}\rm{Cu}(p,\gamma)$ rate reduced by a factor of 100 (red-lined unfilled histogram). Calculation details are described in Reference~\cite{Meis17}. General ash properties are listed in Table~\ref{table:AvgZAQTable}.\label{figure:ashes}}
\end{figure}
  
  \subsection{Ash Production from Type I X-ray Bursts and Other Hydrogen/Helium Burning Regimes}
   The most prominent departure from the single-species-per-depth composition of pristine crusts is realized for accreting neutron stars 
exhibiting Type I X-ray bursts (see figure~\ref{figure:ashes})~\cite{Meis17}. Here we describe the key features of X-ray burst nucleosynthesis, focusing on helium-ignited hydrogen and helium burning bursts, as this case requires the most extensive nuclear physics input as compared to other burning regimes. For completeness, we first summarize the key observational and theoretical aspects of X-ray bursts before discussing the nuclear physics sensitivities in detail (see also the recent review \cite{GallowayKeek2017}).

\subsubsection{Observation} 
\label{sec:burst_observation}
 \hfill\\
     
\begin{figure} [h]
\centering
\includegraphics[scale=0.8]{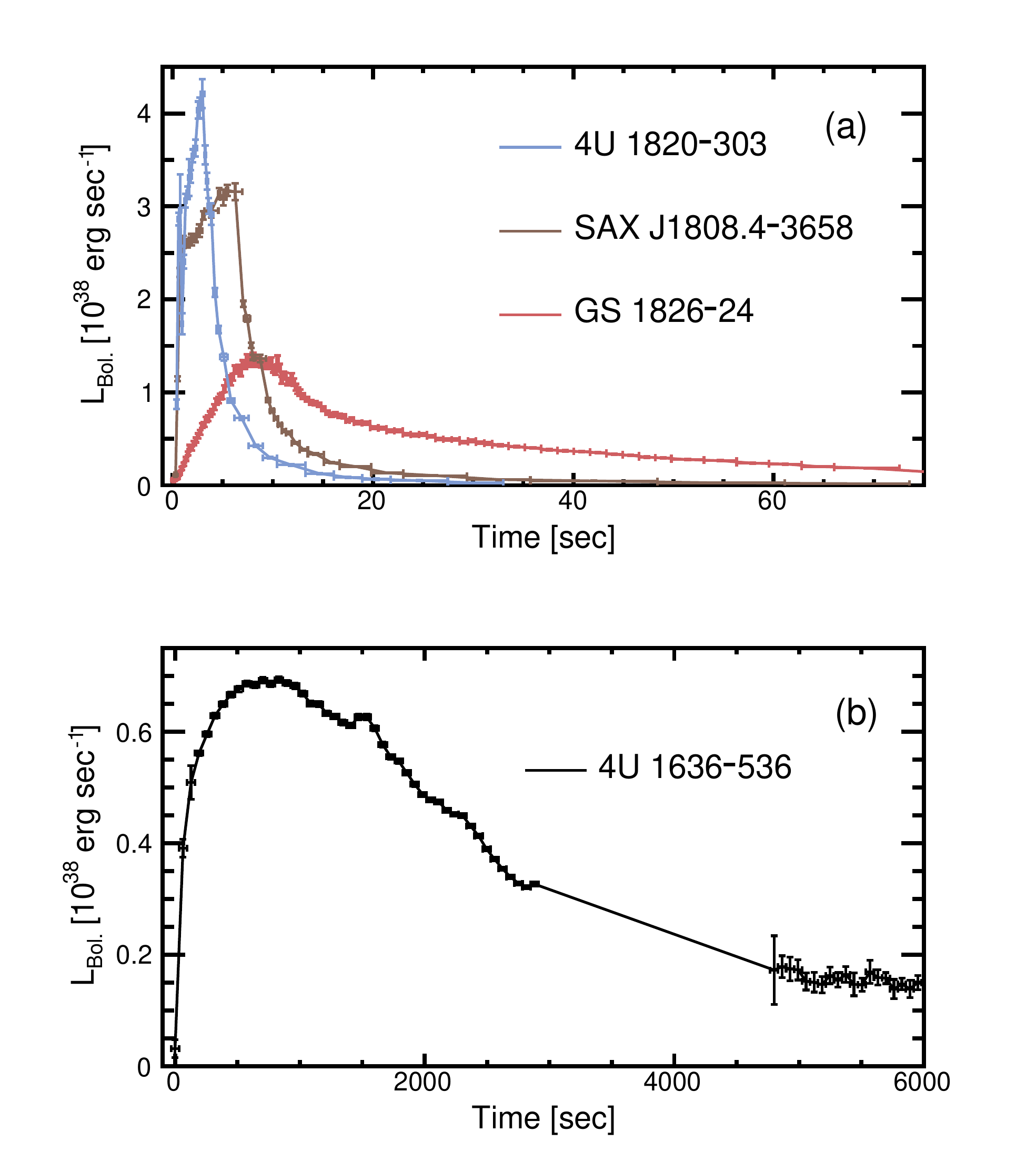}
\caption{Bolometric luminosity light curves observed with the RXTE/PCA
for sources 4U~1820$-$303, SAX~J1808.4$-$3658, GS~1826$-$24 (panel (a)), and 4U~1636$-$536 (panel (b)), reproduced from Reference~\cite{Gall17}. These bursts are thought to be pure-He Type-I, He-rich Type-I, H-rich Type-I, and a superburst, respectively. 
Note that the precursor burst for 4U~1636$-$536~\cite{Stro02} is not visible in panel (b) due to the coarse time binning.
The gap in the data for panel (b) is due to occultation by the Earth. Data are courtesy of the Multi-Instrument Burst Archive (MINBAR): \url{https://burst.sci.monash.edu/minbar/}.
\label{figure:LCobs}}
\end{figure}

Over $100$ X-ray bursting systems are currently known in our Galaxy, including both persistent and transient sources\footnote{\url{https://personal.sron.nl/~jeanz/bursterlist.html}}; figure~\ref{figure:LCobs} shows examples of bursting systems. For many systems the mass accretion rate is observed to be time variable, and the bursting behavior changes accordingly, such that for a specific source different kinds of bursts may be observed over time. X-ray bursts were first reported in 1974 from observations with the Astronomical Netherlands Satellite (ANS; \cite{grindlay75}). At present several thousands of bursts have been observed with a wide variety of durations, recurrence times, and energetics \cite{Galloway2008catalog}. Notable instruments include the {\it European X-ray Observatory Satellite} (EXOSAT; active from 1983 to 1986) and the {\it BeppoSAX Wide-Field Cameras} (WFCs; active from 1996 to 2002),
which observed large burst samples from a range of sources, and demonstrated how the bursting behavior changes as a function of the accretion rate \cite{Paradijs1988,Cornelisse2003}. The majority of bursts are observed at accretion rates $\dot{M}\simeq 0.01-1.00\ \dot{M}_\mathrm{Edd}$, with bursts lasting tens-to-hundreds of seconds and recurring every few hours-to-days. Typically, the burst rate increases with $\dot{M}$: bursts repeat faster when their fuel is replenished more rapidly. Around $0.1-0.3\ \dot{M}_\mathrm{Edd}$, however, the burst rate reaches a maximum, and decreases at higher $\dot{M}$. For most systems, no bursts are observed near $\dot{M}_\mathrm{Edd}$. In that case, it is thought that the accreted material burns 
in a stable way, 
unlike a thermonuclear runaway that would produce a burst.
Note that we do not consider Type-II X-ray bursts, which generally have much shorter recurrence times (tens of seconds to minutes)
and $\sim$100 times greater time-averaged luminosities than Type-I bursts~\cite{Lewi93}. While Type-I bursts are attributed to nuclear burning (see section~\ref{sec:burst_theory}), Type-II bursts are attributed to impulses of accretion onto the neutron star~\cite{Hoffman1978}.

The most detailed burst observations have been performed with the Proportional Counter Array (PCA) on-board the {\it Rossi X-ray Timing Explorer} (RXTE; active from 1995 to 2012) \cite{Strohmayer2006,Galloway2008catalog}. With its large collecting area, RXTE detected detailed burst light curves that can be compared to nucleosynthesis models \cite{Hege07}. For example, the contribution of the rp-process to the burst tail has been quantified for a large sample \cite{Zand17}, and the burst rise is significantly shaped by flame spreading across the neutron star surface \cite{Maurer2008}. Recently, a 
reference sample of RXTE bursts has been created for different fuel compositions (mixed hydrogen and helium, pure helium, and carbon) as a benchmark for nucleosynthesis calculations \cite{Gall17}.

In many astrophysical sites, for example novae and supernovae, the ash composition can be inferred from spectral lines or edges from the debris of the explosion. Unfortunately this is not the case for most X-ray bursts. The strong surface gravity binds the ashes to the neutron star, 
and they are quickly covered by newly accreted material. An exception is a small group of bursts with strong photospheric radius expansion.
At the peak of these powerful bursts, the luminosity reaches the Eddington limit. Potentially, the upper layers of the star are blown off, and the burst ashes are exposed for a few seconds. Observations of such bursts hint at the presence of spectral edges, but the data quality has been insufficient to identify the ions involved \cite{Zand2013,Barr15,Kaja17}. Another possibility is that the ashes could be ejected in the photospheric radius expansion wind~\cite{Wein06}.

RXTE has ceased operations, but at present, similar high quality burst observations are performed with the {\it Nuclear Spectroscopic Telescope Array} (NuSTAR \cite{Harrison2013}), ASTROSAT \cite{Verdhan2017} and the {\it Neutron Star Interior Composition Explorer} (NICER \cite{Gendreau2012NICER}).
The latter is sensitive in a lower energy band, which will allow for a more accurate separation of the X-ray flux from nuclear burning and from the surrounding environment.

    \subsubsection{Theory} 
    \label{sec:burst_theory}
     \hfill\\
The first X-ray burst observations occurred just after the publication of a study that predicted shell flashes from neutron stars
\cite{Hansen1975}, such that the thermonuclear nature was quickly established \cite{Woos76,Maraschi1977,Lamb1978}. Simple one-zone ignition models were used to map out the different burning regimes as a function of $\dot{M}$ \cite{fujimoto81} (see also \cite{Fushiki1987ApJ,Nara03}), and one-dimensional multi-zone models with increasingly sophisticated nuclear networks have been used to simulate the nuclear burning during bursts in great detail (e.g., \cite{joss77,Wall81,Taam1996,Woos04,Fisker2008,Jose10}). Many of the observed features of bursts and other burning modes are reproduced, including a burst rate that increases with $\dot{M}$ and stable burning at high $\dot{M}$. For those systems that accrete a mixture containing both hydrogen and helium, the burning of both is investigated separately. Unstable hydrogen burning via the CNO cycle produces bursts at ignition temperatures of $T\lesssim0.7\times 10^8\ \mathrm{K}$. These conditions are reached at the lowest accretion rates, $\dot{M}\lesssim 0.004\dot{M}_\mathrm{Edd}$. If hydrogen burning ignites helium as well, a mixed H/He burst results. However, for $\dot{M}\gtrsim0.001\dot{M}_\mathrm{Edd}$ this is not the case: helium continues to pile up during several (relatively weak) hydrogen bursts until a powerful helium burst ignites. The weak hydrogen bursts have not been identified in observations. 
At these low accretion rates, there is sufficient time for gravitational settling to separate the ions in the accreted composition, which may have an effect on the boundaries of these regimes \cite{Peng2007}.

For $\dot{M}\gtrsim0.004\dot{M}_\mathrm{Edd}$, hydrogen burns to helium via the ``hot'' or ``$\beta$-limited'' CNO cycle, where the burning time scale is set by the combined half lives of $^{14}$O and $^{15}$O \cite{Wall81}. This burning rate is independent of temperature, hence hydrogen burning cannot run away, and instead proceeds stably. Bursts in this regime are, therefore, produced by unstable helium burning, starting with runaway $3\alpha$ burning. For $\dot{M}\lesssim0.1\dot{M}_\mathrm{Edd}$ there is sufficient time to burn all hydrogen prior to the helium runaway. The burst then ignites in a layer where all hydrogen has burned to helium: a so-called pure-helium burst. Runaway $3\alpha$ burning of helium to carbon raises the temperature and enables a series of $\alpha$-captures on carbon, creating a chain of $\alpha$ elements up to calcium. Interestingly, when $T\gtrsim 1$~GK,  $(\alpha,\rm{p})$ reactions take place which create a small amount of protons \cite{Wein06}. This expands the accessible reactions with proton-captures, even though no hydrogen was present at ignition. The by-pass of  $^{12}\mathrm{C}(\alpha,\gamma)$ by the faster $^{12}\mathrm{C}(\rm{p},\gamma)$, has been suggested as the explanation for observed bursts with exceptionally short rise times of $\sim 1$\,ms \cite{Wein06,Zand2014a}.

At mass accretion rates in excess of $0.1\dot{M}_\mathrm{Edd}$, some hydrogen remains when helium burning ignites, and again
a mixed hydrogen/helium burst results. In the presence of hydrogen, the $3\alpha$ runaway is more complicated. Helium burns to carbon, which boosts the burning of hydrogen in the $\beta$CNO cycle. In turn, the CNO-cycle burning increases the helium abundance as well as the temperature, which further boosts $3\alpha$. This interplay continues until at a temperature of $5\times 10^8$~K breakout from the CNO cycle via $^{15} \mathrm{O}(\alpha,\gamma)$ and via $^{18} \mathrm{Ne}(\alpha,\rm{p})$ at $6\times 10^8$~K becomes efficient~\cite{Wies86}. The break-out reactions open the door for two long reaction chains. First the $\alpha \rm{p}$-process: a series of $(\alpha,\rm{p})(\rm{p},\gamma)$ reactions. These are fast reactions that mostly take place at the burst onset. This process feeds into the rp-process: a series of $(\rm{p},\gamma)$ reactions and $\beta^+$-decays. The proton captures are typically fast, and the timescale of the rp-process is set by the half lives of the decays. Therefore, the initial part of the reaction chain goes fast, until the first slow decay is reached at $^{30}$S \cite{Woos04,Fisker2008} within roughly a second. 
A small amount of the material in high-temperature regions enables this waiting point to be bypassed, but the reaction sequence nonetheless stalls at $^{56}\rm{Ni}$~\cite{Fisker2008}. The rest of the rp-process burning takes place in the tail of the X-ray burst, where it powers the light curve. The length of the reaction chain depends on the amount of hydrogen that is still present at this time and on the maximum temperature that is reached. The end of the rp-process lies at the closed SnSbTe cycle \cite{Scha01,Jose10}, but for typical burst conditions the  composition of ashes peaks at lighter elements near Ge and Ga \cite{Woos04}.

The rp-process can follow multiple branches in the nuclear reaction network. 
Which branch is dominant depends on the reaction rates, which are often only weakly constrained by theoretical calculations (e.g., \cite{Raus00}). Experimental constraints on the rates are crucial for accurate predictions of the ash composition. Changes to the reaction path that substantially change the ashes often also produce observable changes in the X-ray light curve (e.g., \cite{Heger2005}). Large numerical studies identify which reactions are most important in this respect and require new nuclear physics experiments to improve their accuracy \cite{Pari13,Cybu16,Scha17}.

Both the pure-helium and mixed hydrogen/helium bursts as described by theory have been observed. Furthermore, theory predicts that helium burning becomes stable at high mass accretion rates $\dot{M}\geq\dot{M}_\mathrm{Edd}$.
Observations see stable burning already at a three times lower rate. Moreover, the observed reduced burst rate is not predicted by theory. Possibly a hot crust \cite{Keek2009,Zamfir2014} or mixing processes \cite{Piro2007,Keek2009,Cave17} in the envelope stabilize nuclear burning already at lower accretion rates. In addition, the reaction rates of the CNO breakout reactions influence the stability transition \cite{Fisker2006,Keek14}. A further observational constraint is that hydrogen and helium burning must produce sufficient carbon to power the rare superbursts (Section \ref{sec:superbursts}). Whereas burst models do not create enough carbon, stable burning and shallow heating may be of importance here \cite{Stev14,Reic17}. 

Here we discussed one-dimensional models, which resolve the neutron star 
envelope in the radial direction. They employ an approximation for turbulent mixing, which hints at the fact that convective mixing of ashes and fresh fuel has an important effect 
on the burst ignition and the resulting ash composition (``compositional inertia" \cite{Woos04}). In recent years 2D and 3D hydrodynamics models have been created to study convection at the burst onset \cite{Fryx1982,Zingale2001,Malone2011,Malone2014,Zingale2015} and flame spreading across the stellar surface \cite{Spit2002,Cavecchi2013,Cavecchi2015,Cavecchi2016}. The computational demands on such simulations are substantial, and only small approximate nuclear networks are employed that lack the $\alpha \rm{p}$- and rp-processes.

    \subsubsection{Nuclear Sensitivities and Recent Uncertainty Reduction Efforts}\label{sec:rpprocess}
     \hfill\\
 The nuclear reaction sequence powering Type-I X-ray bursts involves more than a thousand reactions on over three-hundred nuclei, posing a daunting experimental challenge at first glance. Fortunately, not all nuclei and their associated reaction rates are of equal importance. Model calculations of X-ray burst light curves and nucleosynthesis have played and continue to play a critical role in identifying nuclear physics uncertainties that are of the highest priority. The key nuclear physics uncertainties associated with Type-I X-ray bursts are summarized here. Since X-ray burst nucleosynthesis is covered extensively elsewhere~\cite{Lewi93,Stro03,Scha06,Pari13,Jose16}, we will focus on past highlights and important results from roughly the past decade.
 
\begin{figure} [h]
\centering
\includegraphics[scale=0.6]{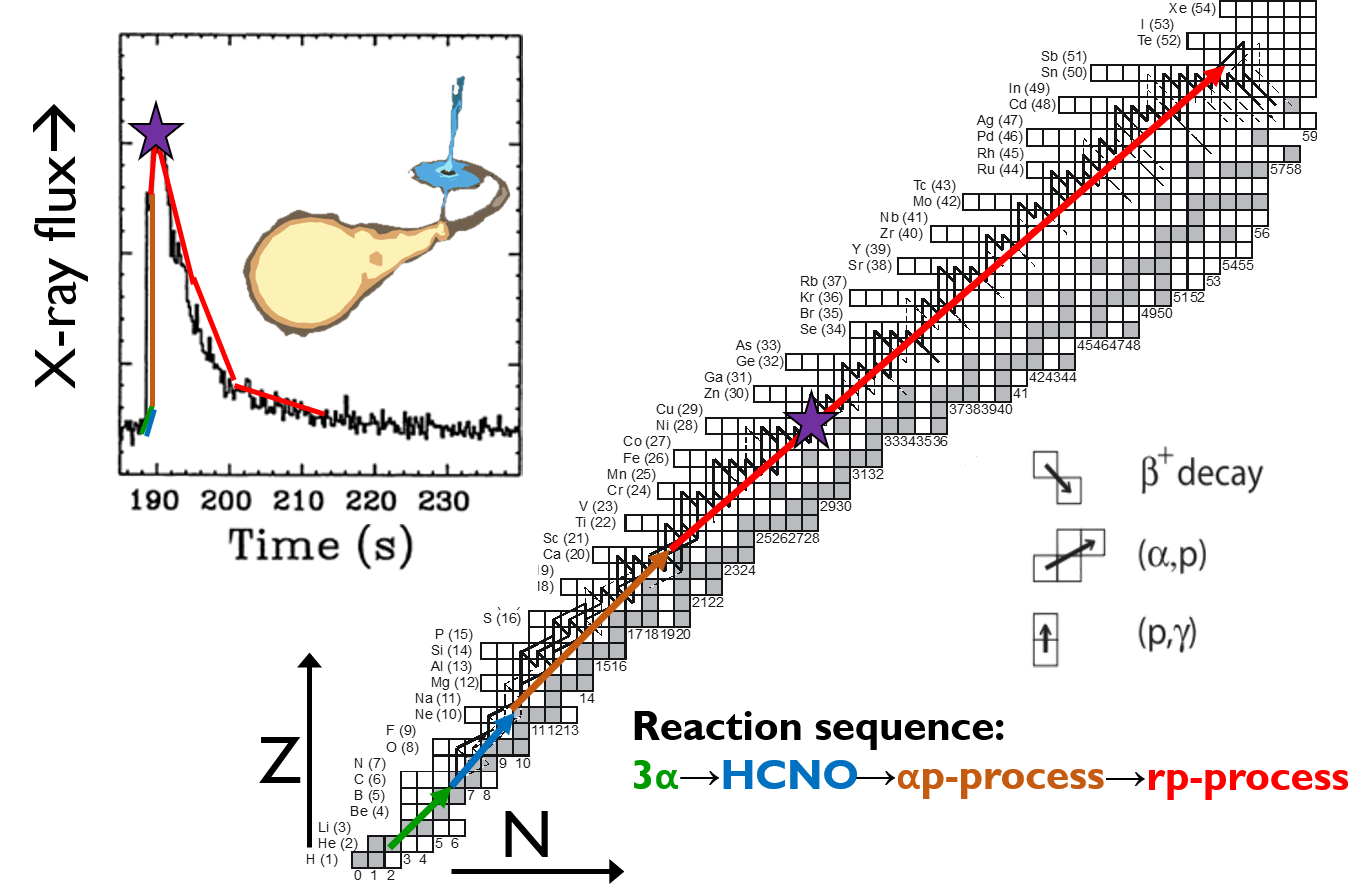}
\caption{Nuclear reaction sequence powering Type I X-ray bursts~\cite{Scha01} with colored lines indicating rates driving particular parts of the X-ray burst light-curve~\cite{Lewi93}.
\label{figure:RPprocess}}
\end{figure}

 Detailed calculations of the rp-process demonstrate that, of the large number of reactions present, only a handful play a significant role in X-ray bursts~\cite{Scha98,Wies00,Woos04}. Broadly speaking, these reactions can be categorized into groups (1) ignition/breakout, (2) branch points, (3) waiting points, and (4) cycles. For a rough orientation on the nuclear chart, (1) and (2) are generally located within the HCNO and $\alpha \rm{p}$-process portions of figure~\ref{figure:RPprocess}, respectively, whereas (3) and (4) are generally located within the rp-process reaction sequence.

 Group (1) reactions are responsible for triggering the thermonuclear runaway that initiates bursts. Namely, these are reaction rates that connect and break-out of the CNO cycles via $\alpha$-capture~\cite{Wies99}. $\alpha$-capture enables material to flow along the valley of $\beta$-stability in the nuclear landscape and avoid unbound nuclides that easily photodisintegrate. The important $\alpha$-captures occur on nuclides with large equilibrium abundances in the $\beta$-limited CNO cycles, i.e. those whose destruction while in the cycle happens via the relatively slow $\beta^{+}$-decay 
 process~\cite{Wies99}. 
 
 Group (2) reactions occur at nuclides whose competing destruction rates are of a similar order of magnitude, meaning the details of the competition determine the subsequent reaction network flow. Branch points are especially prevalent during the burst rise, when $(\alpha,\rm{p})$ and $(\rm{p},\gamma)$ reactions on relatively low-$Z$ nuclides compete~\cite{Fisk04}. These reactions are all located below $^{40}\rm{Ca}$, as the Coulomb barrier is too large at and above this $Z$ for $\alpha$-capture to be competitive.
 
 Group (3) reactions are nuclides without a swift destruction process, which temporarily brings most energy generation to a stand-still until a sufficient quantity of material has been transmuted beyond that point. Generally, a waiting-point nucleus has a low or negative proton-capture $Q$-value $Q_{\rm{p},\gamma}$, meaning the photodisintegration reaction 
 is competitive with the corresponding radiative captures. 
 Therefore, the rp-process flow is funneled into much slower $\beta^{+}$-decay 
 or $e^{-}$-capture decay, while a small percentage of the flow bypasses the waiting-point via proton-capture on the equilibrium abundance of the first proton-capture daughter~\cite{Gorr95,Scha98}. The waiting-point nuclide location and equilibrium abundance are defined by the local proton and photon densities, $n_{\rm{p}}$ and $n_{\gamma}$, respectively, and the environment temperature $T$. Waiting points occur when $Q_{\rm{p},\gamma}$
 is such that the photodisintegration rate $\lambda_{\gamma,\rm{p}}$ is related to the radiative proton-capture rate $\lambda_{\rm{p},\gamma}$ via
 \begin{equation}
 \label{eqn:equilibrium}
 \frac{\lambda_{\gamma,\rm{p}}}{\lambda_{\rm{p},\gamma}}=\frac{n_{\rm{p}}}{n_{\gamma}}\left(\frac{\mu_\mathrm{red} c^{2}}{k_\mathrm{B}T}\right)^{3/2}\mathrm{exp}{\left(-\frac{Q_{\rm{p},\gamma}}{k_\mathrm{B}T}\right)}\gtrsim1,\\
 \end{equation}
 where 
 $\mu_\mathrm{red}$ is the reduced mass of the $(\rm{p},\gamma)$ reaction, which, for most cases of relevance, $\mu_\mathrm{red}\approx1$.
 From the Planck distribution, $n_{\gamma}=\pi\left(k_{B}T\right)^{3}/(13c^{3}\hbar^{3})$. %
 Using typical rp-process conditions, $\rho\approx10^{6}$~g~cm$^{-3}$,  $X(\mathrm{H})\approx0.7$, and $T\approx1$~GK, relation~(\ref{eqn:equilibrium}) is satisfied for $Q_{\mathrm{p},\gamma}\lesssim$1~MeV.
 
Group (4) refers to reaction sequences that operate collectively like a single-nucleus waiting point~\cite{VanW94}. The most well known of which is probably the SnSbTe cycle which marks the rp-process end-point~\cite{Scha01}. In these cases, $(\rm{p},\alpha)$ and $(\rm{p},\gamma)$ reactions compete, where the former continues the cycle and the latter enables the matter flow to break free.
 
Early X-ray burst calculations explored the impact of reaction rates on features of the burst light-curve and ashes. For instance, $(\rm{p},\gamma)$ reactions beyond $^{56}\rm{Ni}$ strongly impact the late time light curve in single-zone calculations~\cite{Hana83}, as 
confirmed later through calculations with higher-fidelity nuclear physics input~\cite{Koik99} and multiple zones~\cite{Woos04}. These results prompted efforts to improve nuclear physics inputs, such 
as implementing Hauser-Feshbach calculations and empirically-based direct capture and resonant rates computed for 
astrophysical calculations~\cite{VanW94}.
 
The impact of nuclear physics uncertainties on X-ray burst ashes has been explored by several groups. 
Early 
studies 
tested different reaction rate~\cite{Koik99} and mass~\cite{Clem03} libraries. Due to limited computing power, early multi-zone work was restricted to modifying half-lives to approximate the behavior of waiting-points~\cite{Woos04}. Until recently, detailed sensitivity studies investigating individual reaction rates were necessarily limited to post-processing studies, which lack feedback between nuclear energy generation and environmental conditions, but enabled exploring the impact of rates for a variety of astrophysical conditions~\cite{Pari08,Pari09}. To date only one X-ray burst sensitivity study has been performed for the whole rp-process with feedback between energy generation and the environment~\cite{Cybu16}. Even in this case, multi-zone reaction rate variations were limited to cases highlighted as important by a single-zone study, as a train of a dozen or so bursts used to assess rate sensitivity required 
a week or more to calculate (see Reference~\cite{GallowayKeek2017} for a discussion of burst models with different dimensionality). The 
first single-zone mass sensitivity study with coupled energy generation and hydrodynamics was only 
recently  performed~\cite{Scha17}. 
Note that progress in high-precision mass measurements (e.g. References within~\cite{Kank17}) means that 
only a handful of masses with insufficient precision remain. Half-life sensitivities need not be investigated, as almost all relevant half-lives are experimentally well-constrained~\cite{Scha06,Scha17}. That said, uncertainties remain in the modifications to terrestrial half-lives that are required to account for high-temperature and high-density effects, such as thermal population of excited states, inhibition of bound state electron capture, and continuum electron capture~\cite{Full1982,Oda1994,Prue2003,Lau18b}.
 
 A cautionary point regarding what the ``most important'' nuclear sensitivities are is that this will depend on the astrophysical conditions. For instance, it is known that the rp-process reaction network path depends on metallicity~\cite{Jose10} and very different bursts result from models with varied accretion rates~\cite{Woos04,Lamp16}. 
 Additionally, at least one model~\cite{Davi11} has found discrepant sensitivity with respect to other models~\cite{Fisker2006,Cybu16} for $^{15}\rm{O}(\alpha,\gamma)$, indicating more cross-code comparisons (e.g., \cite{Meis17,Meis18}) are needed.
 
 Efforts to remove or reduce the most critical nuclear physics uncertainties have primarily consisted of indirect measurements to constrain nuclear reaction rates. These include measurements of nuclear masses, spectroscopy of low-lying nuclear excited states, and determinations of statistical properties for more highly excited compound nuclei. While direct measurements are preferable, the necessary radioactive beams at astrophysically relevant energies and sufficient intensities are not yet available.

We continue with a 
brief summary of recent uncertainty reduction efforts for relevant rp-process reaction rates grouped by reaction categories defined above.

\begin{figure} [h]
\centering
\includegraphics[scale=0.7]{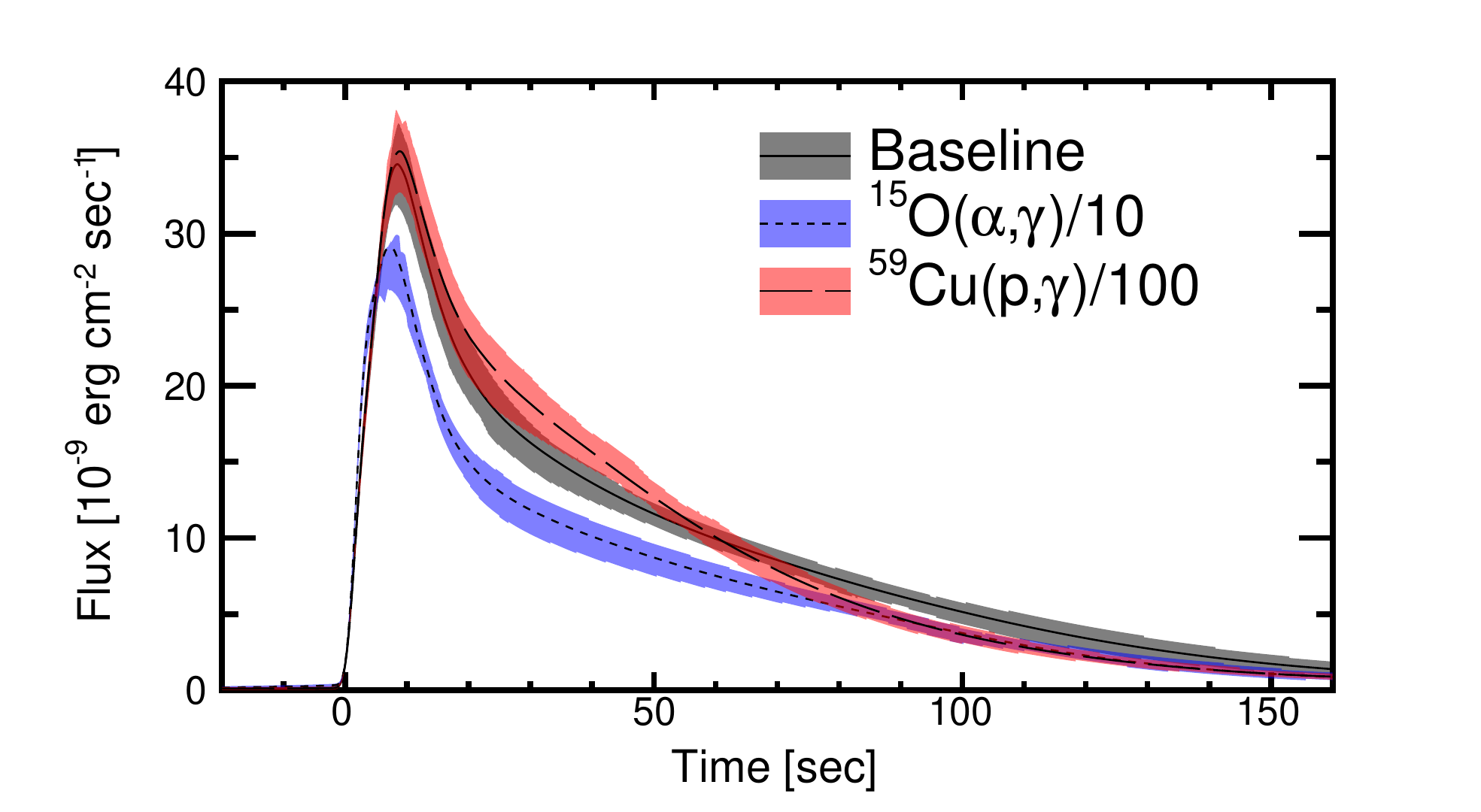}
\caption{Averaged X-ray burst light curves and asymmetric 1-$\sigma$ uncertainties corresponding to sequences of $>$10 bursts, for multi-zone X-ray burst calculations performed with MESA~\cite{Paxt11,Paxt13,Paxt15}. Calculations employ a solar accreted composition at $\dot{M}$=0.17$\dot{M}_\mathrm{Edd}$ onto an 11.2~km neutron star with $0.1$~MeV per accreted nucleon of base heating, where the distance and redshift of GS~1826$-$24 reported in Reference~\cite{Gall17} are used to convert luminosity (from the calculation results) to flux. Results are shown using the ReacLibV2.2 
rate library~\cite{Cybu10} (gray band), with a reduced $^{15}\rm{O}(\alpha,\gamma)$ rate (blue band), and a reduced $^{59}\rm{Cu}(p,\gamma)$ rate (red band). All other calculation details are as described in Reference~\cite{Meis17}.
\label{figure:modelLC}}
\end{figure}

\paragraph{Ignition/Breakout: } Experimental efforts have primarily focused on the two key break-out rates, $^{18}\rm{Ne}(\alpha,p)$ and $^{15}\rm{O}(\alpha,\gamma)$, and the $^{14}\rm{O}(\alpha,p)$ rate which links the CNO cycles on the way to breakout. $^{14}\rm{O}(\alpha,p)$ has most recently been summarized in References~\cite{Alma12b} and \cite{Hu14}. Stated briefly, this rate may be known to sufficient precision. However, further experimental investigations are merited in the $\sim6-7$~MeV excited state energy region of the $^{18}\rm{Ne}$ compound nucleus to confirm important resonance information. Furthermore, theoretical confirmation is required to assess the impact of $^{14}\rm{O}(\alpha,p)$ in more sophisticated X-ray burst calculations. $^{15}\rm{O}(\alpha,\gamma)$ has most recently been evaluated in Reference~\cite{Davi11}, though the most recent experimental results come from References~\cite{Tan07,Tan09,Lund16,Wred17}. For this case the uncertainty stems from the unknown resonance strength for capture into the $\sim 4$~MeV excited state of $^{19}\rm{Ne}$. The small $\alpha$-branching ratio dominates the rate uncertainty and has thus far evaded direct measurement. An example of this rate's impact on the X-ray burst light curve is shown in figure~\ref{figure:modelLC}. The state of $^{18}\rm{Ne}(\alpha,p)$ is summarized in Reference~\cite{Mohr14}, relying primarily on complementary measurements from References~\cite{Mati09,Salt12,He13,Zhan14}. Further progress for this rate would require experimental resonance strength determinations in the $\sim9-11$~MeV excitation energy region in $^{22}\rm{Mg}$.  A less significant but nonetheless notable development is the recent set of experimental constraints placed on $^{19}\rm{Ne}(p,\gamma)$, which connects $^{15}\rm{O}(\alpha,\gamma)$ to the rp-process, including the first direct measurement of capture onto a radioactive ion excited state~\cite{Wilk17,Bela16}.

 \paragraph{Branch Points: } Rates in the $\alpha \rm{p}$-process of particular importance are those involving nuclides where destruction via $(\alpha,\rm{p})$ is competitive with the $(\rm{p},\gamma)$ or $\beta^{+}$-decay alternatives. Chief among these are $^{26}\rm{Si}(\alpha,\rm{p})$, $^{30}\rm{S}(\alpha,\rm{p})$ and its competitor $^{30}\rm{S}(\rm{p},\gamma)$, and $^{34}\rm{Ar}(\alpha,\rm{p})$. A single measurement is the source of experimental constraints for $^{26}\rm{Si}(\alpha,\rm{p})$, providing 
 level energies but largely lacking certain spin assignments and $\alpha$-widths~\cite{Alma12a}. However, the relevant level densities are large enough that a statistical reaction rate approach may be adequate. $^{30}\rm{S}(\alpha,\rm{p})$ has been the subject of more investigations, including direct measurements of the time-reversed reaction~\cite{Deib11} and spectroscopic studies of the compound nucleus $^{34}\rm{Ar}$~\cite{Obri09,Kahl18}. As with the previous case, absent a direct measurement of the forward reaction,
 $\alpha$-widths are the source of the main experimental uncertainty.  The competing reaction $^{30}\rm{S}(\rm{p},\gamma)$ involves lower-lying excitation energies in the compound nucleus $^{31}\rm{Cl}$ and has thus been more amenable to spectroscopic constraints, resulting in a relatively well-constrained reaction rate~\cite{Wred09,Lang14b}. In the past few years $^{34}\rm{Ar}(\alpha,\rm{p})$ has been the subject of a series of complementary direct and indirect measurements. These include a spectroscopic measurement using the $(\rm{p},t)$ mechanism~\cite{Long17}, a direct measurement above the Gamow-window~\cite{Schm17}, and another announced work on spectroscopy of the compound nucleus using $\rm{p},\rm{p}'$~\cite{Laue16}.
A comprehensive analysis of these works will likely vastly improve the situation regarding this branch-point nucleus.
 
\paragraph{Waiting Points: }
The nuclides which satisfy
relation~(\ref{eqn:equilibrium})
and have several-second half-lives
are $^{56}$Ni, $^{64}$Ge, $^{68}$Se, $^{72}$Kr, and $^{100}$Sn. For these cases, $Q_{\rm{p},\gamma}$ for the waiting-point nucleus is essential to determine the equilibrium abundance of the proton-capture daughter. $Q_{\rm{p},\gamma}$ for the first proton-capture daughter and the structure of the second proton-capture daughter are equally important in order to determine the flow of material through the waiting-point. For waiting-points away from the proton drip-line, the possibility of material flowing around the waiting-point also needs to be investigated. $^{56}$Ni falls into the latter category, where experimental constraints suggest a strong waiting-point~\cite{Lang14}, but 
it is possible this waiting-point could be bypassed 
at high temperatures and densities~\cite{Ong17,Valv18}. $^{64}$Ge has been declared to not be a waiting-point~\cite{Wing93,Tu11}, but this claim was likely premature. The large remaining uncertainty in the nuclear masses of $^{65}$As and $^{66}$Se leave open the possibility for a strong waiting-point, as do uncertain properties of relevant resonances for the $^{64}\rm{Ge}(\rm{p},\gamma)^{65}\rm{As}(\rm{p},\gamma)^{66}\rm{Se}$ sequence~\cite{Lam16}.  Independent measurements have determined a rather negative proton-separation energy of $^{69}\rm{Br}$, resulting in $<$13\% of material flowing through proton-capture on $^{68}$Se, solidifying $^{68}$Se as a strong waiting-point~\cite{Blan95,Pfaf96,Roge11,DelS14}. Similar recently completed measurements will soon resolve the waiting-point status of $^{72}$Kr, but results are not yet published~\cite{Roge17}.
Studies of $^{100}\rm{Sn}$ are limited to a confirmation of its long half-life~\cite{Bazi08}, but the proximity of this reaction to the rp-process end-point limits its importance for all but the most energetic X-ray bursts~\cite{Scha01}.

\begin{figure} [h]
\centering
\includegraphics[scale=0.8]{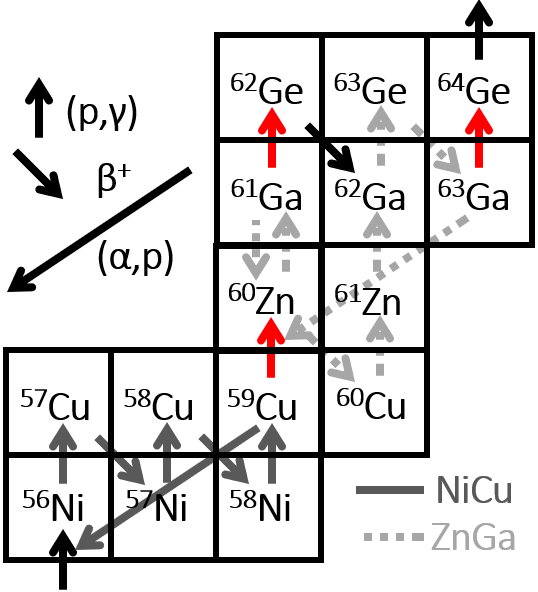}
\caption{Portion of the rp-process reaction sequence featuring the NiCu and ZnGa reaction cycles. $^{59}\rm{Cu}(p,\gamma)$, $^{61}\rm{Ga}(p,\gamma)$, and $^{63}\rm{Ga}(p,\gamma)$ (red arrows) significantly affect the reaction flow in this region.
\label{figure:cycles}}
\end{figure}

\paragraph{Cycles: }
For cases where the $\alpha$ emission threshold is lower than the proton threshold,  $(\rm{p},\gamma)$ and $(\rm{p},\alpha)$ reactions can compete~\cite{Remb97}. In the vicinity of $Z=28-31$, just beyond $^{56}\rm{Ni}$ in the rp-process path, this leads to the NiCu and ZnGa cycles~\cite{VanW94}. For these cycles, radiative proton-capture onto $^{59}\rm{Cu}$, $^{61}\rm{Ga}$, and $^{63}\rm{Ga}$ are of particular importance, as indicated in figure~\ref{figure:cycles}. Proton-capture onto $^{59}\rm{Cu}$ either returns the cycle to $^{56}\rm{Ni}$ or breaks-out via $^{59}\rm{Cu}(\rm{p},\gamma)$, depending on the $(\rm{p},\gamma)/(\rm{p},\alpha)$ rate ratio. Once out of the NiCu cycle, the several minute half-life of $^{60}\rm{Zn}$ stalls the rp-process unless it is bypassed by proton-capture. The $\sim250$~keV $Q$-value for $^{60}\rm{Zn}(p,\gamma)^{61}\rm{Ga}$ enables an equilibrium abundance of $^{61}\rm{Ga}$ to be built-up, which can then bypass $^{60}\rm{Zn}(\beta^{+})$ via $^{61}\rm{Ga}(\rm{p},\gamma)$. $^{63}\rm{Ga}$ plays the analogous role in the ZnGa cycle to the role of $^{59}\rm{Cu}$ in the NiCu cycle. At present these rates are primarily determined by theoretical estimates~\cite{Fisk01,Raus00} since the distance from the valley of $\beta$-stability makes even indirect measurements challenging.

\paragraph{}
The majority of rp-process reaction rates lack any experimental constraints beyond nuclear masses and rough structure details. In their stead, the most common approach is to employ shell model rates when available (and applicable), particularly in the mid-mass region~\cite{Fisk01}, and statistical model rates otherwise, e.g.~\cite{Raus00}. A commonly used reaction library which follows this strategy is the JINA-CEE ReacLib database~\cite{Cybu10}.

However, rapid experimental progress is anticipated in the near future. Reaccelerated beams at the Facility for Rare Isotope Beams studied with devices such as the JENSA gas-jet target~\cite{Bard16} and SECAR recoil separator~\cite{Berg17} will enable direct measurements of several interesting reaction rates at astrophysically relevant energies. Nonetheless, indirect measurements will continue to guide direct measurement studies and to cover the large number of cases for which direct measurements will still not be possible. Newer techniques, such as the $\beta$-Oslo method~\cite{Spyr14}, are poised to significantly grow the ranks of experimentally-constrained reaction rates for nuclides far off stability. 

Though only discussed tangentially here, much of the discussion above also pertains to nuclosynthesis during stable burning on the neutron star 
surface. For near Eddington and super-Eddington accretion rates, nuclear reaction sequences often resemble the rp-process~\cite{Scha99,Stev14,Keek16,Meis17}, resulting in burst-like ash distributions (see figure~\ref{figure:ashes}). However, the detailed abundance distribution depends on the astrophysical conditions, and so a wide variety of ash abundances can be produced in stable burning.
For instance, recently a regime of stable burning has been proposed around $0.1\dot{M}_\mathrm{Edd}$, with carbon-rich ashes \cite{Keek16}.
The exact reactions of interest for these processes will depend on the hydrogen/helium composition
of the accreted material and the accretion rate
~\cite{Nara03,Keek14}. 
 
 \subsection{X-ray Superbursts:}
\label{sec:superbursts}

  \subsubsection{Observation}
   \label{sssec:SuperburstObs}
     \hfill\\
Superbursts were first observed in 1996 with the BeppoSAX/WFCs 
and RXTE~\cite{Cornelisse2000,Stro02}. To date, 26 superbursts have been detected from 15 neutron stars \cite{Zand17}. Superbursts reach a peak flux that is similar to normal bursts, but their decay lasts several hours, and the total emitted energy is $\sim 10^3$ times larger than for hydrogen/helium flashes 
which justifies the ``super'' designation \cite{wijnands2001}.
The X-ray observatories that detected superbursts are in low Earth orbits of $\sim 90$ minutes, which is shorter than the typical superburst duration. Therefore, superburst observations are interrupted by data gaps of up to $30$ minutes, when the view of the X-ray source is blocked by the Earth (see  figure~\ref{figure:LCobs}, bottom panel).
Often the start of the superburst falls in a data gap, and it is challenging to accurately determine the burst's properties and confirm its superburst nature, so it is instead referred to as a ``superburst candidate"~\cite{Altamirano2012}. 
Most superbursts have been observed with wide-field or all-sky instruments, which produce data of modest quality. Only in two cases have detailed observations been performed with RXTE/PCA \cite{Stro02,Strohmayer2002a}.

The observed superburst light curves are fit with numerical models of cooling envelopes to determine the ignition conditions. From the tail of the light curve the ignition column depth is measured, 
and the fluence constrains the energetics of the fuel \cite{2004CummingMacBeth,Cumming2006}. The inferred parameters  suggest that unstable carbon burning in the neutron star ocean ignites superbursts \cite{cumming2001,Stro02}. Ignition column
depths are found to be in the range of $y_\mathrm{ign}\simeq(0.2-3)\times 10^{12}\ \mathrm{g\ cm^{-2}}$, and the fuel energetics are equivalent to a carbon mass fraction of $\sim 20\%$ \cite{Cumming2006,Zand17}. The rise of the light curve is shaped by the temperature profile left behind by the carbon flame (measured only once \cite{Keek16}).

All superbursting sources also exhibit normal (short) bursts. Most have mass accretion rates in the range of $0.1-0.3\ \dot{M}_\mathrm{Edd}$, where the burst rate drops and a substantial fraction of the accreted hydrogen/helium burns in a stable manner \cite{Zand2004}. Both explosive and stable hydrogen/helium burning may be necessary for superbursts. As the normal bursts do not produce sufficient carbon, stable burning is likely needed to create the carbon fuel (e.g., \cite{Stev14,Keek16}; Section~\ref{sec:burst_theory}). However, the bursts produce heavy isotopes (iron group or heavier), which increase the opacity, allowing the base of the ocean to reach the temperature required for runaway carbon fusion \cite{cumming2001}.

Because of their long recurrence times (typically a year) \cite{Zand17}, superbursts are rare, and each new observation brings new insight. 

   \subsubsection{Theory}
      \hfill\\
The nuclear processes that power superbursts show a resemblance to those in models of Type Ia supernovae. The runaway is initiated by $^{12}$C+$^{12}$C burning. One of the dominant channels is $\mathrm{^{12}C(^{12}C,\alpha)^{20}Ne}$. A fraction of the $\alpha$ particles capture on carbon to form oxygen, enabling the follow-up reactions $^{12}$C+$^{16}$O and $^{16}$O+$^{16}$O. The ashes of these reactions are rich in $^{28}$Si. At sufficiently high temperatures ($T\gtrsim 10^9$~K) photodisintegration of silicon occurs (``silicon melting''). A large number of $\alpha$ particles is 
created, the majority of which are quickly captured again to form a composition that is dominated by $^{56}$Ni. The precise ash composition is set by the nuclear statistical equilibrium of photodisintegration and capture reactions. Any isotopes in the hydrogen/helium ashes substantially heavier than nickel will also photodisintegrate, and this can contribute up to 50\% of the superburst energy~\cite{Scha03}. From the carbon runaway, all these burning processes take less than a second to complete. Afterwards, as the layer cools, electron captures transform $^{56}$Ni into $^{56}$Fe, making $A=56$ the primary component of the ashes as seen in figure~\ref{figure:ashes}. 

Similar to the hydrogen/helium bursts (section~\ref{sec:burst_theory}), one-dimensional multi-zone models are employed to study superburst ignition and the various burning processes. Self-consistent simulations should model hundreds of hydrogen/helium flashes to build up a carbon layer and produce a superburst, but this approach suffers from two problems. First, it is computationally expensive, and studies with reduced rp-process networks are likely not accurate (compare Reference~\cite{Taam1996} to Reference~\cite{Woos04}). Second, as noted in section~\ref{sssec:SuperburstObs}, current models of hydrogen/helium burning do not produce enough carbon. 
Therefore, superburst models typically directly accrete a carbon-rich fuel  
to study 
ignition conditions \cite{Keek11}. Shortly before ignition, the accretion composition can be switched to hydrogen/helium-rich material to study the effect of a superburst on the atmosphere \cite{Keek2012}. Importantly, the heated ocean after a superburst quenches short bursts in the atmosphere: for several days hydrogen and helium burning becomes stable.

No multi-dimensional models have yet been created for superbursts. One-dimensional simulations show, however, that the hydrodynamics of the carbon flame is important for shaping the observable X-ray light curve. Initially, a convective region forms around the ignition depth, until 
heating by thermonuclear burning in one  zone becomes faster than cooling by convection.
A local runaway ensues, burning all carbon in that zone. The burning time scale becomes shorter than the sound-crossing time scale, raising the question whether the flame will spread as a detonation \cite{Weinberg2006sb,Weinberg2007}, or whether convection can spread the flame to lower depths as a deflagration. Either of these processes can generate a shock that produces a short precursor burst at the start of the superburst light curve \cite{Keek2012}. 
The carbon flame leaves behind a temperature profile in the envelope, which shapes the observable light curve. It encodes information about the ignition depth, fuel energetics, and the flame spreading (e.g., whether the flame reached the surface or stalled) \cite{2004CummingMacBeth,Cumming2006,Keek16}.

The ignition depths inferred from observations are smaller than those predicted by theory \cite{Keek11}, leading to questions about carbon fusion as the process that powers superbursts.
The problem of carbon ignition became more pressing with the discovery of superbursts from transient sources \cite{Keek08,Altamirano2012,Chenevez2011ATel,Serino2016}. The neutron star 
envelope heats up during accretion, but in 
most of these cases accretion was active for mere days or weeks, which was thought to be insufficient to reach the temperature needed for runaway carbon fusion \cite{Keek08}. Therefore, stronger heating of the base of the crust is required on the relatively short time scales of the transient accretion events. This may be related to the shallow heat source inferred for 
quasi-persistent transients
when they cool after accretion ceases (section~\ref{sec:cooling_transients}). Simulations confirm that the ignition depth is reduced with the addition of extra heating \cite{Keek11}. Interestingly, additional shallow heating has also been proposed for enhancing carbon production during stable burning~\cite{Reic17}.

   \subsubsection{Nuclear Sensitivities and Recent Uncertainty Reduction Efforts}
   \label{sssec:superburstnuc}
     \hfill\\
     The several gigakelvin temperature achieved in superbursts is more than sufficient to drive material to nuclear statistical equilibrium~\cite{Scha03}.
     As such, the final abundance distribution is largely determined by the nuclear masses, environment temperature, and environment 
     $Y_{e}$. In principle the abundance distribution will be modified by light charged-particle capture after nuclear statistical equilibrium freeze-out, but low proton and $\alpha$ abundance predictions result in a negligible modification~\cite{Scha03}. 
Since the majority of the relevant nuclear masses are well known, uncertainties which may affect $Y_{e}$ are of interest. Namely, these are the reactions modifying the ashes of the stable burning and X-ray bursts 
already discussed above.

Though inconsequential in terms of the resulting crust composition, the  $^{12}$C$+$$^{12}$C fusion rate is nonetheless of interest due to its role in triggering superburst ignition~\cite{Coop09}. The relatively low energy of interest for the large Coulomb barrier involved (see section~\ref{sec:pycno})
have made advances in this area extremely challenging~\cite{Cost09}. Only very recently have direct 
laboratory measurements pushed into the astrophysical energy region
~\cite{Buch15,Jian18}. Of particular interest is the possible existence of a resonance in $^{12}$C$+$$^{12}$C at a center of mass energy of $\sim$1.5~MeV~\cite{Coop09}. This resonance has been posited to explain the discrepancy between the observed and predicted superburst ignition depth, much in the way the Hoyle state was predicted to explain cosmic carbon production~\cite{Hoyl54}.
However, present experimental constraints are contradictory and rely on theoretical extrapolations~\cite{Buch15b}.

\section{Interaction of Nuclei Within the neutron star Ocean and Crust} \label{section:interaction}

The ashes of surface burning experience compression under the pressure of the continuously infalling material (section~\ref{section:accretion}). As they move to higher mass densities (and greater electron chemical potentials), they undergo a variety of non-equilibrium reactions.
These reactions, primarily $e^{-}$-captures, cause the compressed nuclei to become increasingly neutron-rich. Near the neutron drip point, the  reaction mechanisms operating in the crust transitions from  $e^{-}$-capture reactions (with only $\nu$ and $\gamma$ emission) 
to neutron emission reactions \cite{Gupt07,Gupt08,Lau18}. The presence of pycnonuclear (density-driven) fusion reactions~\cite{Came59,Harr64,Salp69} was also found in the inner crust~\cite{Sato79}. These reactions
directly modify the crust composition, but more importantly, they can drastically alter the thermal structure of the crust, impacting thermonuclear processes on and near the surface 
and related observables~(\cite{Mira90,Keek11,steiner2012,Scha14,Deib16,Meis17}; section~\ref{section:impact}).

Nuclear reactions deposit heat in the neutron star  
crust during an accretion episode~\cite{Haen90,Haen03,haensel2008} and an accurate accounting of crustal heating during active accretion is required to match the neutron star's 
surface temperature when accretion ends~\cite{Brow09}. The impact of $e^{-}$-captures on heating was substantially increased when it was realized that nuclear de-excitations following electron captures increase heat deposition~\cite{Gupt07,Gupt08,Zhan15}. Experimental investigations along these lines include References~\cite{Estr11,Meis15}. 

Non-equilibrium reactions may also cool the neutron star  
crust. Urca cooling\textrm{---} cycles of $e^{-}$-capture and $\beta^{-}$-decay that
generate neutrino emission\textrm{---} was introduced for white dwarf stars~\cite{Gamo40,Gamo41}. 
Only recently was it realized that the finite temperature of the neutron star ocean and crust allows sufficient phase-space for Urca cooling to exist~\cite{Gupt07,Scha14}. 
Shortly after, this was limited to odd-A nuclides~\cite{Meis15} and efforts have been made to employ more experimentally-based nuclear data and identify the presence of Urca cooling in the neutron star's 
outer layers 
based on observations of neutron star crustal cooling~\cite{Deib16,Meis17}.

    \subsection{$e^{-}$-Capture Reactions}
    \label{ssec:ECrxns}
\begin{figure} [h]
\centering
\includegraphics[scale=0.65]{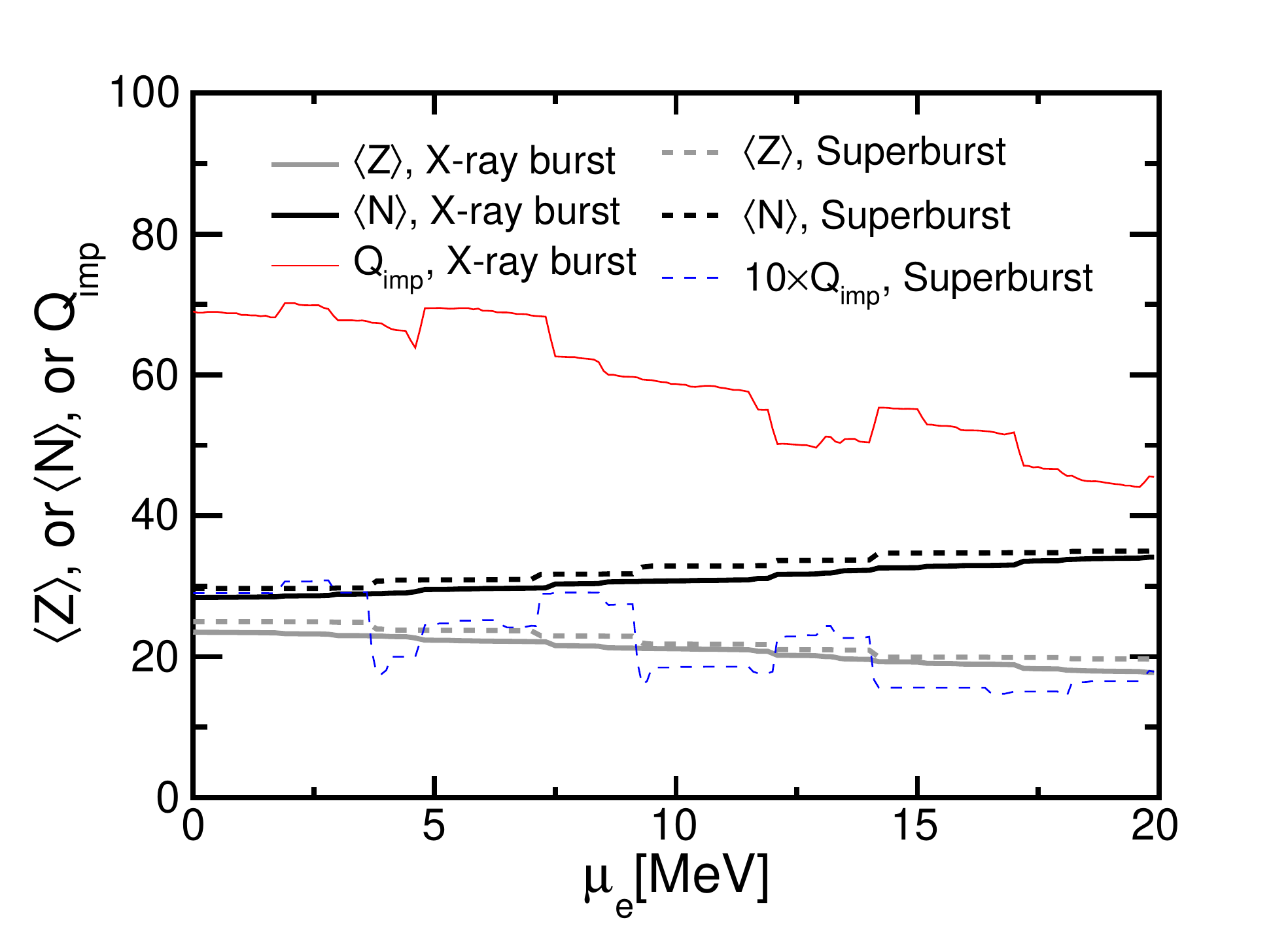}
\caption{Evolution of the average proton number $\langle Z\rangle$, average neutron number $\langle N\rangle$, and impurity $Q_\mathrm{imp}$ (see equation~\ref{equation:Qimp}) 
due to $e^{-}$-capture for X-ray burst and superburst ashes (see figure~\ref{figure:ashes}), using experimental masses when available~\cite{Audi12} and the WS3 global mass model otherwise~\cite{Wang10}. The superburst impurity is multiplied by 10 for visibility. 
This plot does not account for neutron emission and fusion reactions (sections \ref{sec:nuc_emission} and \ref{sec:pycno}).
See figure~\ref{figure:pristinecomp} to compare to the pristine crust composition.
\label{figure:ECcrustcomposition}}
\end{figure}

The first transmutation ashes experience as they are buried by accretion of additional material is a sequence of $e^{-}$-captures.
The rising electron chemical potential 
$\mu_e$ with increased depth makes it energetically possible for $e^{-}$-capture to proceed for nuclei with steadily more negative $e^{-}$-capture $Q$-values, $Q_{\rm{EC}}$. Once $\mu_e>|Q_{\rm{EC}}|$, $e^{-}$-capture immediately proceeds.
The end result is that the crust becomes steadily more neutron-rich with increasing depth
as shown in figure~\ref{figure:ECcrustcomposition}. Surface values of $\langle Z\rangle$, $\langle A\rangle$, and $Q_{\rm{imp}}$ for the ashes shown in figure~\ref{figure:ashes} are listed in table~\ref{table:AvgZAQTable}.

   \begin{table}
\caption{Properties on the neutron star surface of the ashes shown in figure~\ref{figure:ashes}.}
\begin{center}
\begin{tabular}{c c c c  }
\hline
Quantity & X-ray Burst & Superburst & Stable Burning \\
\hline
$\langle Z\rangle$ & 24 & 25 & 26 \\
$\langle A\rangle$ & 52 & 55 & 60 \\
$Q_{\mathrm{imp}}$ & 69 & 3 & 101 \\
\hline
 \label{table:AvgZAQTable}
\end{tabular}
\end{center}
\end{table}

The interesting pattern in the impurity shown for X-ray burst and superburst ashes in figure~\ref{figure:ECcrustcomposition} deserves some discussion. The overall decline in impurity with increasing depth is due to the fact that higher-$Z$ elements near stability undergo $e^{-}$-capture at lower $\mu_e$ relative to their lower-$Z$ counterparts because the nuclear mass surface is more shallow near the valley of $\beta$-stability at high $A$.
The jumps in impurity are due to $e^{-}$-captures on the few isotopes which are the most abundant (see figure~\ref{figure:ashes}). For superbursts, the dominant species are $A=52,54$, and $56$, while for X-ray bursts $A=60,64$, and $68$ dominate the composition with significant but less influential contributions from $A=28$ and $32$. This can be confirmed 
by a comparison of the location for jumps in the impurity in figure~\ref{figure:ECcrustcomposition} and $|Q_{\rm{EC}}|$ for neutron-rich isotopes of the aforementioned isobars.

   \begin{figure} [h]
\centering
\includegraphics[scale=0.6]{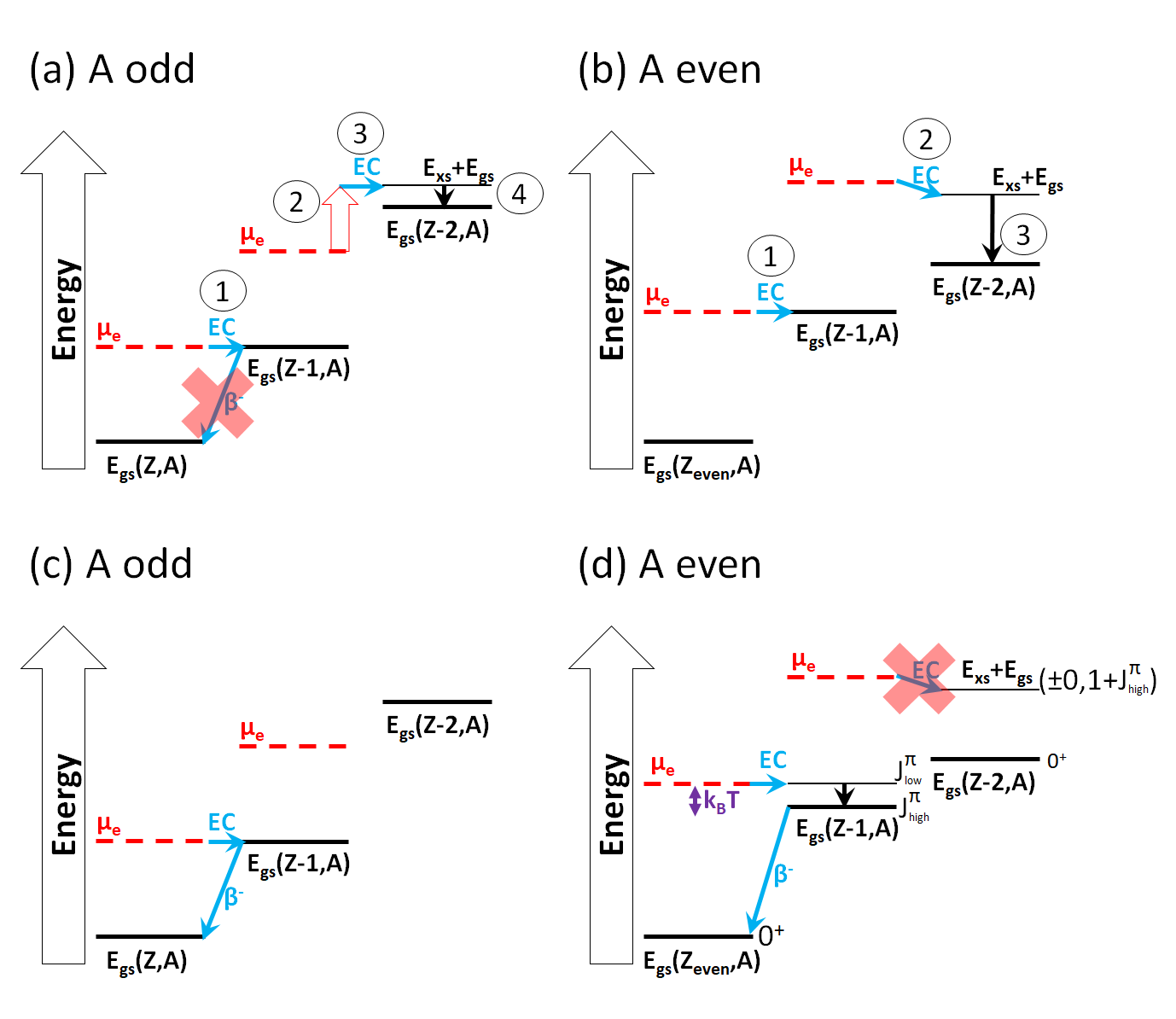}
\caption{Scenarios for $e^{-}$-capture in the accreted crust at a depth indicated by the electron chemical potential $\mu_{e}=|Q_{\rm{EC}}|$, resulting in heating (a and b), which happen in steps indicated by the encircled numbers, or cooling (c and d), which are cyclic.
For $e^{-}$-capture on an odd-$A$ nucleus, a subsequent $e^{-}$-capture cannot occur until the nucleus has been buried to a larger $\mu_{e}$. Therefore whether heating (a) or cooling (c) occurs depends on whether $\beta^{-}$-decay from the first $e^{-}$-capture daughter is favorable. For $e^{-}$-capture on an even-$A$ nucleus, odd-even mass staggering generally enables an immediate subsequent $e^{-}$-capture, resulting in heating (b). However, special circumstances, namely a low-lying (relative to $k_\mathrm{B}T$) low-$J^{\pi}$ isomer in the odd-odd daughter of the first $e^{-}$-capture, can enable cooling via $e^{-}$-capture-$\beta^{-}$-cycling to proceed instead (d). 
\label{figure:ECschematic}}
\end{figure}

$e^{-}$-capture reactions are the richest in terms of potential for experimental nuclear physics constraints with present and near-future facilities. Depending on the nuclear masses and structure of the nuclei involved in an $e^{-}$-capture reaction sequence,
one of four scenarios can take place once $\mu_e\approx|Q_{\rm{EC}}|$, as depicted in figure~\ref{figure:ECschematic}. For these cases, the two-step $e^{-}$-captures will generally create local heat sources~\cite{Gupt07}, whereas the $e^{-}$-captures that can be reversed via $\beta^{-}$-decays can create local cooling sources~\cite{Scha14}.  The heating predominantly comes from radiative de-excitation of states populated in $e^{-}$-capture, whereas cooling is due to neutrinos produced in the $e^{-}$-capture and $\beta^{-}$-decay escaping the neutron star  
crust. 

The majority of $e^{-}$-capture heating comes from the scenario depicted in panel (b) of figure~\ref{figure:ECschematic}, where odd-even mass staggering makes it energetically favorable for an $e^{-}$-capture on an even-even nucleus to be immediately followed by $e^{-}$-capture on the odd-odd daughter of the first reaction. For this case, $\mu_e\approx|Q_{\rm{EC}}(Z,A)|>|Q_{\rm{EC}}(Z-1,A)|$. The first $e^{-}$-capture happens near threshold (unless there is a large change in $J^{\pi}$ required for a ground-state to ground-state transition), 
so little heating or cooling is achieved. In the second $e^{-}$-capture, roughly one-quarter of the surplus energy will be deposited into the crust~\cite{Haen03}. 
Usually the second $e^{-}$-capture proceeds into an excited state of energy $E_\mathrm{xs}$, and all of the de-excitation energy 
is deposited into the crust, resulting in the lion's share of the heating.
The energy deposited for a two-step $e^{-}$-capture starting on isotope $Z,A$ at a depth $\mu_e=|Q_{\rm{EC}}(Z,A)|$ is 
\begin{equation}
E_{\rm{heat}}=\eta\left(|Q_{\rm{EC}}(Z,A)|-|Q_{\rm{EC}}(Z-1,A)|-E_\mathrm{xs}\right)+E_\mathrm{xs},
\label{equation:NucHeating}
\end{equation}
with $1/6\lesssim\eta\lesssim1/4$, where the exact pre-factor $\eta$ requires an explicit calculation of $e^{-}$-capture to excited states~\cite{Haen03,Gupt07}.
In the case depicted in figure~\ref{figure:ECschematic}a, $\mu_e$ is not large enough for the second e$^-$-capture to occur immediately. However, the $\beta^{-}$-decay from the first $e^{-}$-capture daughter is much slower than the accretion timescale. Therefore the second $e^{-}$-capture happens only after the  chemical potential has increased enough for the next $e^{-}$-capture to proceed. Little heating is produced in this case.
Evidently the nuclear masses and low-lying excited states of neutron-rich isobars of nuclides predicted to be abundant in the ashes of surface burning processes are of significant interest for $e^{-}$-capture heating in the accreted neutron star outer crust. 
The amount of heating produced by $e^{-}$-capture reactions in the outer crust depends on the ashes composition.  For X-ray burst ashes, the total heating is on the order of $0.1-0.15$~MeV/u, or about $10^{35}$~erg~s$^{-1}$ at Eddington accretion rate \cite{Gupt07}. Notice that this is more than five times larger than the earlier estimates for a single-component composition \cite{Haen90} which missed the role of excited states shown in figure~\ref{figure:ECschematic}b.  Significant heating mostly results from predominant nuclides with $X(A)\gtrsim 10\%$. It was found that when the heating through excited states is included, the simple single-component model give similar $e^{-}$-capture heating as compared to results of more complete multi-component reaction networks \cite{haensel2008}.

$e^{-}$-capture cooling via neutrino-emission primarily occurs from scenario (c) in figure~\ref{figure:ECschematic}, where the odd-even mass staggering eliminates the possibility of an immediate subsequent $e^{-}$-capture following $e^{-}$-capture on an odd-$A$ nucleus. Instead, a $\beta^{-}$-decay may proceed (as opposed to case (a), where we assumed a much slower rate for the weak transition), 
resulting in an $e^{-}$-capture/$\beta^{-}$-decay cycle known as an Urca process. Nominally such a scenario is possible for an even-even nucleus, however a specific set of circumstances is required. 
Namely, as depicted in scenario (d) of figure~\ref{figure:ECschematic}, a low $J$ ($\sim0$$-$1) state with $E_\mathrm{xs}\sim k_\mathrm{B}T$ must be present for an odd-odd nucleus with a high $J$ ($\gtrsim2$) ground state. The only promising candidate so far for scenario (d) has been ruled-out~\cite{Meis15}.

The strength of Urca cooling is quantified by the luminosity of neutrinos produced by the pair of nuclides involved in the $e^{-}$-capture/$\beta^{-}$-decay Urca cycle. In its essence, the neutrino luminosity is determined by the quantity of nuclides in an Urca pair, which is 
set by the isobaric abundance and the Urca shell thickness, multiplied by the Urca cycle rate,
which in turn is determined by the weak transition rate. The shell thickness is proportional to the temperature $T$, since the Urca cycle can operate within the window $\mu_e\approx|Q_{\rm{EC}}|\pm k_\mathrm{B}T$, and inversely proportional to the local gravity $g$,
as follows from differentiating equation~(\ref{eq:z-mu})~\cite{Scha14}. The weak transition rates for the Urca cycle are  
mostly determined by the integral over the momentum phase-space~\cite{Beth47}. 
Due to the electron-degeneracy, the phase-space is limited to the small 
thermal window around $E_{e^{-}}\approx \mu_e$, which is several $k_\mathrm{B}T$ wide~\cite{Tsur70}.

The above considerations result~\cite{Tsur70,Deib16} in a neutrino cooling luminosity
\begin{equation}
\label{equation:Lnu}
L_{\nu}(Z,A) \approx L_{34}(Z,A)\times10^{34}\,{\rm{erg~s}}{}^{-1}X(A)T_{9}^{5}\left(\frac{g_{14}}{2}\right)^{-1}\left(\frac{R}{10~\mathrm{km}}\right)^{2}. \\
\end{equation}
Here $X(A)$ is the mass-fraction of the $e^{-}$-capture parent nucleus in the composition, $T_{9}$ is the temperature of the Urca shell in units of $10^{9} \, \mathrm{K}$, 
and $g_{14}\equiv g/(10^{14}~\rm{cm}~\rm{s}^{-2})$.
The intrinsic cooling strength of the Urca pair, $L_{34}(Z,A)$, is given by
\begin{equation}
\label{equation:L34}
L_{34}(Z,A)\approx 0.87\left(\frac{10^{6}~{\rm{s}}}{ft}\right)\left(\frac{56}{A}\right)\left(\frac{Q_{\rm{EC}}(Z,A)}{4~{\rm{MeV}}}\right)^{5}\left(\frac{\langle F\rangle^{*}}{0.5}\right), \\
\end{equation}
where $\langle F\rangle^{*}\equiv\langle F\rangle^{+}\langle
F\rangle^{-}/(\langle F\rangle^{+}+\langle F\rangle^{-})$, the Coulomb factor $\langle F\rangle^{\pm}\approx2\pi\alpha_f
Z/|1-\exp(\mp2\pi\alpha_f Z)|$, and $\alpha_f\approx1/137$ is the
fine-structure constant. $ft$ is the comparative half-life, which should be implemented as $ft=({ft_{e^{-}\rm{-capture}}+ft_{\beta^{-}}})/{2}$, since the degeneracy of the parent state impacts the transition rate. However, $ft$ from the $e^{-}$-capture and $\beta^{-}$-decay generally agree within a factor of a few~\cite{Paxt15e}, so using an estimate for one or the other transitions results in a negligible uncertainty compared to other contributions.

The cyclic nature of the  $e^{-}$-capture/$\beta^{-}$-decay enables the energy release from cooling to exceed energy deposition from the inherently one-way $e^{-}$-capture-only cases by more than an order of magnitude~\cite{Meis15}. We stress that this is 
plausible when the crust temperature is of the order of $1~$GK, given the $T^{5}$ dependence of equation (\ref{equation:Lnu}). For a typical outer crust $Q_{\rm{EC}}\approx12$~MeV and a not unreasonable estimate of ${\rm log}_{10}(ft)\approx5$, 
equation (\ref{equation:L34}) results in $L_{34}\approx10^{3}$. Similar $L_{34}$ values 
are obtained for $Q_{\rm{EC}}\approx8$~MeV and ${\rm log}_{10}(ft)\approx4$ or $Q_{\rm{EC}}\approx15$~MeV and ${\rm log}_{10}(ft)\approx5.7$.
For such nuclei, $L_{\nu}$ will be significant for $X(A)\gtrsim1$\%~\cite{Meis17} (compare with the $e^-$-capture heating estimate above). 
Therefore, it is critical for model calculations of near-surface nuclear burning to follow abundance evolutions even for relatively low-abundance nuclides. This requires multi-species and multizone nuclear reaction networks with precise nuclear physics input~\cite{Cybu16}. 
Furthermore, it is worth noting that cooling by  $e^{-}$-capture/$\beta^{-}$-decay cycling does not require continued accretion, whereas $e^{-}$-capture heating only occurs while accretion is active.

From equations~(\ref{equation:Lnu}) and (\ref{equation:L34}), it is evident that nuclear masses and low-lying nuclear structure of neutron-rich isobars of abundant surface-burning ashes are required to determine the strength of Urca cooling in accreted neutron star crusts. Ash abundances (discussed in detail in section~\ref{section:production}) are also critical, as $L_{\nu}(Z,A)$ scales linearly with the abundance of the Urca pair $X(A)$. Nuclear masses impact the location and strength of Urca cooling since $Q_{\rm{EC}}$ is 
related to atomic mass excesses $\rm{ME}$ via $Q_{\rm{EC}}=\rm{ME}(Z,A)-\rm{ME}(Z-1,A)$. Low-lying structure of Urca nuclides is key as $ft$ is directly related to the change in spin $\Delta J$ and parity $\Delta\pi$ required for a weak transition to occur~\cite{Sing98}.

In the absence of direct measurements, nuclear masses are estimated using global mass models. Popular choices for crust model calculations include the Finite Range Droplet Model (FRDM)~\cite{Moll12}, one of the many Hartree-Fock-Bogoliubov variations (e.g. HFB-21~\cite{Gori10}), and one of the Weisz\"{a}cker-Skyrme variations (e.g. WS3~\cite{Wang10}). Globally, the trend in the odd-even mass staggering for increasing neutron-richness is of particular interest. Large odd-even mass-differences along an isobar will result in greater $e^{-}$-capture heating, but will eliminate the possibility of Urca cooling for even-$A$ nuclides~\cite{Meis15}. Due to this effect, the largest crustal heating is predicted for FRDM, whereas HFB-21 
results in the most significant cooling~\cite{Scha14}. Nonetheless, the sensitivity of $L_{34}(Z,A)$ to $Q_{\rm{EC}}(Z,A)$ highlights the need for experimental constraints. Recently, such constraints have been the province of time-of-flight mass measurements, as this technique is able to access the most exotic nuclides as compared to alternatives~\cite{Meis13}. Already such measurements have constrained the heating for some of the strongest heating sources~\cite{Estr11,Meis16} and ruled-out Urca cooling for $^{56}\rm{Ti}\leftrightarrow^{56}\rm{Sc}$~\cite{Meis15}, which was once thought to be the strongest cooling source~\cite{Scha14}.

Presently, experimental data for $ft$ of weak transitions between neutron-rich nuclides are limited.
For several cases near to stability, $ft$ is available for ground-state to ground-state transitions and ground-state to excited-state (which would be $e^{-}$-capture parent excited state to $e^{-}$-capture daughter ground-state) from $\beta^{-}$-decay measurements. Developments are ongoing to enable $(d,^{2}\rm{He})$ charge-exchange measurements in inverse kinematics with radioactive ion beams, which would provide $ft$ for $e^{-}$-capture parent ground-state to $e^{-}$-capture daughter ground and excited states. However, the latter technique requires a radioactive beam rate of $\sim10^{6}$~Hz (e.g. Ref.~\cite{Noji15}), limiting the applicable distance from stability. For the large number of remaining cases lacking direct $ft$ measurements, one must resort to theoretical calculations or estimates derived from data-based systematics~\cite{Scha14,Deib16,Meis17}.

Among theoretical estimates, $ft$ from shell-model calculations tend to compare most favorably to experiment~\cite{Cole12}; however, the absence of large-scale calculations results in limited coverage of the nuclear chart~\cite{Sull16}. As such, predictions from quasi-random phase-approximation (QRPA) calculations have frequently been adopted instead~\cite{Gupt07,Scha14}. A more recent, alternative procedure has been to employ $ft$-values consistent with systematics based on the change in spin and parity $\Delta J^{\Delta\pi}$ for a weak transition~\cite{Deib16,Meis17}. Such a compilation is available in Reference~\cite{Sing98}. This 
means that improved predictions can be provided by spectroscopic measurements, which determine $E_{xs}$ and $J^{\pi}$ for low-lying states of interest. This is advantageous since spectroscopic measurements require far fewer statistics than measurements of $ft$ via $\beta^{-}$-decay. 
Therefore constraints can be provided for more exotic nuclides.

Given the dearth of available data, it is clear that many additional direct and indirect experimental constraints on $ft$ are necessary. In particular, measurements are critical for neutron-rich isobars of odd-$A$ nuclides predicted to have surface-burning ash abundances on the percent-level. Current estimates implicate $^{33}\rm{Al}\leftrightarrow^{33}\rm{Mg}$ and $^{55}\rm{Sc}\leftrightarrow^{55}\rm{Ca}$ as the most significant Urca pairs
for X-ray burst and superburst ashes, respectively~\cite{Meis17}.

 \subsection{Neutron Emission}\label{sec:nuc_emission}
 
With increasing $\mu_e$, $e^{-}$-captures shift the nuclear composition along isobaric chains towards the neutron drip line where the neutron separation energy $S_\mathrm{n}$ becomes negative. In that case, $(e^{-},x\mathrm{n})$ reactions 
(here $x$ means the number of neutrons emitted)
lead to the buildup of a free neutron abundance. In fact, neutron emission in this process can occur before the neutron drip line if excited states with $E_\mathrm{xs}> S_\mathrm{n}$ 
are populated during electron captures \cite{Gupt08}, as shown in case (b) in figure~\ref{figure:ECschematic} (at the second $e^{-}$-capture in double capture, or if the first allowed transition at the first step is to the excited state). After the neutron drip line, some neutron emission can proceed in rapid sequences called superthreshold electron-capture cascades (SEC) \cite{Gupt08}. In these sequences the products of $(e^{-}, x\mathrm{n})$ reactions can be highly unstable to more $e^{-}$-captures which can occur in a cascade faster than neutron captures.
This situation can repeat many times resulting in a diverse reaction sequence with large neutron exchange between isotopic chains. SEC are set by a competition between the $e^{-}$-capture and $(\mathrm{n},\gamma)-(\gamma, \mathrm{n})$ rates. Note that neutron capture reactions strongly depend on the 
neutron degeneracy and, since radiative processes are involved, can be enhanced by  plasma physics effects  \cite{Shternin2009PhRvD,Shternin2012PhRvC}. Some free neutrons can appear (to be immediately recaptured by the most abundant nuclei) from $(e^{-},x\mathrm{n})$ reactions as early as densities of $\mu_e\gtrsim 14$~MeV, but do not significantly alter the reaction sequences~\cite{Lau18}. The SEC processes are only found in multicomponent reaction networks \cite{Gupt08,Lau18} and do not appear in simplified one-species treatments \cite{Haen90,Haen03,haensel2008}. Neutron emission processes are an important source of heat since emitted neutrons are rapidly thermalized by mutual collisions. The presence of SEC shifts this type of heating to shallower depths \cite{Gupt08} than found in one-component models.

\subsection{Pycnonuclear Reactions}\label{sec:pycno}

\begin{figure}[ht]
\begin{center}
\begin{minipage}[t]{0.45\textwidth}
\includegraphics[width=\textwidth]{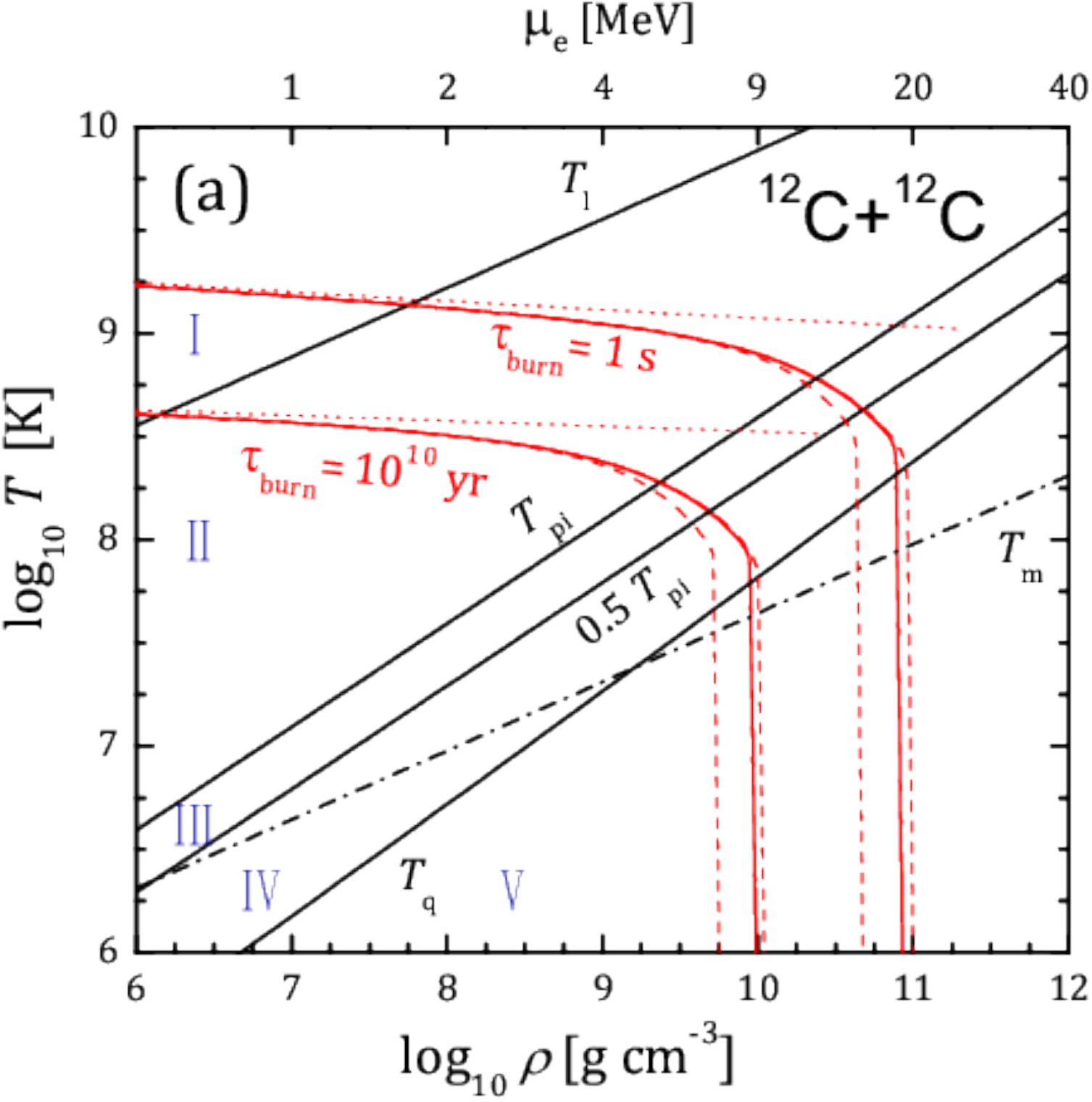}
\end{minipage}
\hspace{0.05\textwidth}
\begin{minipage}[t]{0.45\textwidth}
\includegraphics[width=\textwidth]{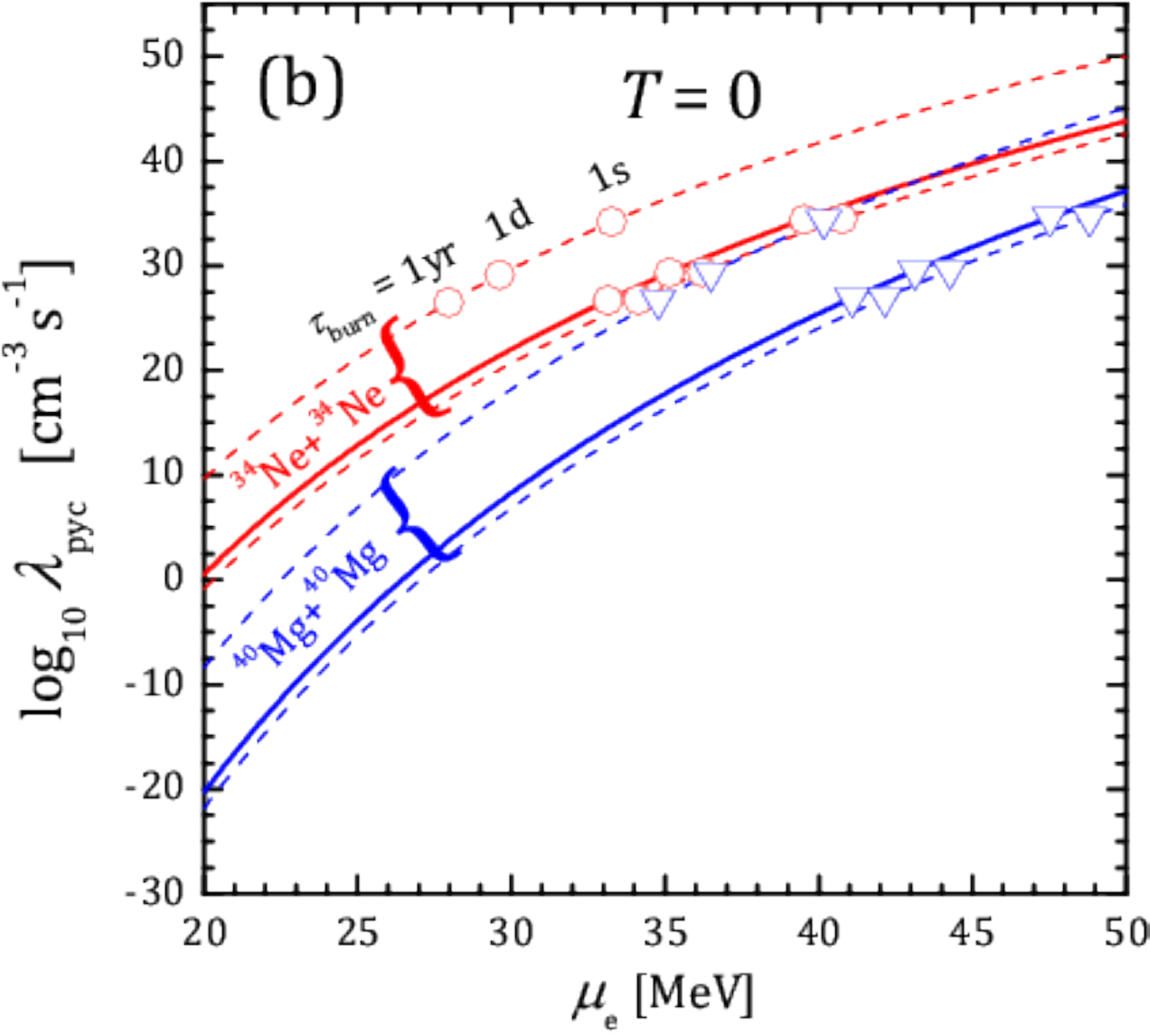}
\end{minipage}
\end{center}
\caption{Fusion burning in dense stellar matter. {\it (a)}: Phase diagram for carbon burning adapted from \cite{Chugunov2007PhRvD}. Black solid lines show relevant temperatures (marked in the plot) separating different burning regimes (I--V). The dash-dotted line shows solidification temperature in a classical liquid ($\Gamma=175$). 
Upper and lower red solid curves show the loci of the burning time $\tau_\mathrm{burn}=1\,\mathrm{s}$ and $10^{10}\,\mathrm{yr}$, respectively. Thin dashed lines bracket the uncertainty in calculations as discussed in the text. Dotted lines  represent the pure thermonuclear regime neglecting all medium effects. Values of $\mu_e$ are shown on the upper horizontal scale. {\it (b)}: Pycnonuclear reaction rates for $^{34}$Ne fusion (upper, red curves) and $^{40}$Mg fusion (lower, blue curves) in the neutron star crust assuming respective one-species composition and body-centered cubic lattice. Solid lines give the optimal reaction rates, while thin dashed lines bracket the theoretical uncertainties in Coulomb barrier penetration calculations. Symbols (open circles for $^{34}$Ne and open triangles for $^{40}$Mg) indicate in each case the depths where $\tau_\mathrm{burn}=1$~yr, 1~d, and 1~s. }\label{fig:pycno}
\end{figure}

Fusion reactions become significant when the reaction rate is faster than the rate at which matter is buried to a depth where another $e^{-}$-capture and/or neutron emission is/are energetically favorable~\cite{Sato79,Haen90}. As nuclei are charged, 
fusion reactions require tunneling through the Coulomb barrier. In dense stellar matter the probability of the barrier penetration, and so the rates of fusion reactions, are strongly affected by the environmental conditions. Only when ions are in the weak coupling regime, $T\gtrsim T_\mathrm{l}$, where $T_\mathrm{l}$ corresponds to $\Gamma=1$ (see section~\ref{section:structure}), the standard thermonuclear burning (regime I in figure~\ref{fig:pycno}a) operates. In this regime, the reaction rate is determined solely by the temperature and reacting nuclide abundances. In neutron star crusts most fusion reactions operate in the opposite, pycnonuclear limit, where the possibility to overcome the Coulomb barrier is driven by zero-point oscillations of the ions around their equilibrium positions. The rates of the pycnonuclear reactions are subject to huge uncertainties. In order to discuss those we first briefly address fusion in the familiar thermonuclear regime.

The key nuclear physics quantity for estimating a fusion reaction rate is the astrophysical $S$-factor~\cite{Mart09} 
related to the reaction cross section $\sigma(E)$ for charged-particle reactions by  $\sigma(E)= S(E) E^{-1}\,\rm{exp}\left(-2\pi\eta\right)$, where $E$ is the center of mass energy for the reaction,
$\eta=\alpha_f Z_{1}Z_{2}\sqrt{{\mu_{\rm{red}}c^{2}}/{(2E)}}$, is the Sommerfeld parameter, 
$Z_{i}$ and $\mu_{\rm{red}}c^{2}$ are the nuclear charges and reduced mass for the fusing nuclei, respectively. The advantage of introducing the $S$-factor is that it is a relatively smooth function of energy, where the main part of the strong energy-dependence resulting from the Coulomb tunneling probability is factored out by $\rm{exp}\left(-2\pi\eta\right)$. For a structureless (i.e. non-resonant) reaction, which most of the relevant reactions here are thought to be (though see the remark in section~\ref{sssec:superburstnuc} $^{12}$C$+$$^{12}$C), 
the counterbalance between the exponential increase in tunneling probability towards higher energies and the exponential decrease of the number of energetic nuclei results in that only nuclei in the vicinity of the 
Gamow peak energy $E_\mathrm{pk}$ in the tail of the thermal distribution actually fuse. The Gamow peak energy is given by
\begin{equation}\label{eq:Gamow_th}
E_\mathrm{pk}=\left(\frac{\pi^2 \mu_\mathrm{red} Z_1^2 Z_2^2 e^4 k_\mathrm{B}^2 T^2}{2\hbar^2}\right)^{1/3}\approx 0.5\,\mathrm{MeV}\ \left(\frac{A}{12}\right)^{1/3} \left(\frac{Z}{6}\right)^{4/3} \left(\frac{T}{10^8\,\mathrm{K}}\right)^{2/3},
\end{equation}
where the second equality is for the fusion of like nuclei. Under the assumption of weak $S(E)$ dependence, the thermonuclear rate becomes 
\begin{equation}\label{eq:fus_rate_therm}
\lambda_\mathrm{th}= 4\frac{n_i^2}{2} \sqrt{\frac{2 E_\mathrm{pk}}{3\mu_\mathrm{red}}}\frac{S(E_\mathrm{pk})}{k_\mathrm{B} T} \mathrm{exp}\left(-3E_\mathrm{pk}/(k_\mathrm{B} T)\right).
\end{equation}

Direct measurements of fusion between two neutron-rich nuclides is not possible, as this would require high-intensity low-energy beam-beam collisions between radioactive nuclides and no such facility exists. As such, experimental studies focus on fusion involving stable nuclides, which are useful in terms of benchmarking theoretical models~\cite{Bass77,Yako06a,Bear10}. Nonetheless, measurements of $S(E)$ for more systems are welcome, particularly in light of the fact that, while theoretical predictions and experimental results agree for fusion reactions involving some neutron-rich nuclides~\cite{Carn14}, intriguing discrepancies remain for other cases~\cite{Rudo12,Stei14,Sing17}.

The energies relevant for neutron stars are smaller than the current experimental measurements reach. 
Hence, determinations of the $S$-factor are either performed by extrapolations from fits to data measured at much higher energies or by theoretical calculations based on tunneling through a barrier corresponding to a theoretical nuclear potential~\cite{Gasq05,Yako06a,Yako06b,Bear10,Horo08}. Theoretical 
studies amount to calculating the tunneling probability for some angular momentum transfer $\ell$, as represented by the so called transmission coefficient $T_{\ell}$ ($\leq1$). The cross section results from the semi-classical relation $\sigma(E)=\pi\lambdabar^{2}\Sigma_{\ell=0}^{\ell,\rm{max}}(2\ell+1)T_{\ell}$,
where $\lambdabar$ is the reduced de Broglie wavelength. $\ell_{\rm{max}}={R_{\rm{max}}}/{\lambdabar}$ corresponds to the angular momentum at which the impact parameter is equal to the sum of the participating nuclear
radii $R_{\rm{max}}=r_{0}(A_{\rm{1}}^{1/3}+A_{\rm{2}}^{1/3}$) ($r_{0}\sim1.2$~fm), though generally the few lowest $\ell$ provide the dominant contributions~\cite{Gasq05}. 
A large set 
of astrophysical $S$-factors relevant to neutron star studies was collected recently in Reference \cite{Afanasjev2012PhRvC}, where an economical analytical parameterization was proposed.

Using $S$-factors from Reference~\cite{Afanasjev2012PhRvC}, in figure~\ref{fig:pycno}a we plot with dotted lines the locations in $T-\rho$ plane where the thermonuclear burning time $\tau_\mathrm{burn}\equiv n_\mathrm{i}/\lambda_\mathrm{th}$ is equal to $10^{10}$~yr and $1$~s for the $^{12}$C+$^{12}$C  reaction in a one-species carbon plasma. However, at large densities and/or low temperatures (when $\Gamma\gtrsim 1$), these rates become irrelevant and the effects of ion correlations start to play the dominant role. The respective lines of $\tau_\mathrm{burn}=\mathrm{constant}$ are shown with solid red curves and are clearly very different from the dotted lines. One finds five regimes of nuclear burning (figure~\ref{fig:pycno}a) \cite{Salp69}. In the regime of strong ion coupling $\Gamma\gtrsim 1$, screening of the Coulomb interaction eases barrier penetration enhancing the reaction rates as seen in figure~\ref{fig:pycno}a. This burning regime II, called thermonuclear with strong plasma screening, operates until quantum effects in ion motion become important at $T\lesssim T_\mathrm{pi}$. Here $T_\mathrm{pi}$ is the 
ion plasma temperature  (for a one-species plasma)
\begin{equation}\label{eq:Tpi}
 T_\mathrm{pi}\equiv \frac{\hbar \omega_\mathrm{pi}}{k_\mathrm{B}} = \frac{\hbar}{k_\mathrm{B}} \sqrt{\frac{4 \pi Z^2 e^2 n_i}{A m_u}}\approx1.9\times 10^9\,\mathrm{K}\ \sqrt{\frac{Z}{A}}\left(\frac{\mu_e}{20\,\mathrm{MeV}}\right)^{3/2},
\end{equation}
where $\omega_\mathrm{pi}$ is the ion plasma frequency.
The pycnonuclear regime V, already mentioned above, operates for the lowest temperatures $T\lesssim T_\mathrm{q}=0.5\, T_\mathrm{pi}/\ln\left(T_\mathrm{l}/T_\mathrm{pi}\right)$. The remaining regimes are the thermally enhanced pycnonuclear regime IV ($0.5T_\mathrm{pi}\gtrsim T\gtrsim T_\mathrm{q}$) and the thermopycnonuclear regime III ($T_\mathrm{pi}\gtrsim T\gtrsim 0.5T_\mathrm{pi}$), which is the most uncertain. We do not discuss intermediate regimes II--IV here, see detailed discussions in References \cite{Gasques2005PhRvC,Yako06a,Chugunov2007PhRvD,Pote12,Chugunov2009PhRvC}.

We now briefly describe the pycnonuclear regime mainly following the discussion in References \cite{Gasques2005PhRvC,Yako06a}. More details can be found there and in References \cite{Salp69,Yako06b,Afanasjev2012PhRvC,Chugunov2007PhRvD}. For simplicity we first address the single-species composition and then discuss the more relevant case of mixtures. 
The rate in the pycnonuclear regime $T\lesssim T_\mathrm{q}$ is temperature-independent. In analogy to equation (\ref{eq:fus_rate_therm}) it can be written
\begin{equation}\label{eq:fus_rate_pycno}
\lambda_\mathrm{pyc} =C_\mathrm{pyc} \frac{n_i^2}{2} S(E_\mathrm{pk}) \frac{\hbar}{m_i Z^2 e^2}\, \xi^\beta \exp\left(-\alpha \xi \right),
\end{equation}
where the typical interaction energy is $E_\mathrm{pk}\sim \hbar \omega_\mathrm{pi}$ [cf. equation~(\ref{eq:Gamow_th})]. Constants $C_\mathrm{pyc}$, $\alpha\sim 0.7$, and $\beta$ depend on the model for the Coulomb barrier penetration via zero-point vibrations, and the parameter 
$\xi=d^2/r_\mathrm{rms}^2$ characterizes the relative amplitude of these vibrations. Here $r_\mathrm{rms}^2=\hbar/(2\mu_\mathrm{red}\omega_\mathrm{pi})$ is the mean-square displacement of the oscillating ion and $d$ is the equilibrium distance between the closest reacting neighbors. It is assumed that ions in the one-species plasma form a body-centered cubic lattice. Then $d=(3\pi^2)^{1/6} a$, where $a$ is the ion-sphere radius defined in section~\ref{section:structure}. Putting numbers in,
\begin{equation}
\xi\approx 0.2 A^{1/2} Z^{7/6} \left(\frac{\mu_e}{20\,\mathrm{MeV}}\right)^{-1/2}.
\end{equation}
Under conditions relevant for the neutron star crust, $\xi$ is large and gradually decreases with depth. For pure $^{12}$C at $\mu_e=20$~MeV, $\xi=5.7$, and for $^{34}$Ne, the first species to fuse in the one-component model of Reference \cite{Haen90},  $\xi=17$ at the same depth. The pre-exponential factor in equation (\ref{eq:fus_rate_pycno}) is usually large so the pycnonuclear reactions start to be important at $\xi\gg1 $. Such reaction rate is very sensitive to variations in $\xi$ and in the exponential prefactor $\alpha$. Different models of tunneling resulting in small variations in $\alpha$ transform to huge (several orders of magnitude) variations in $\lambda_\mathrm{pyc}$ \cite{Gasq05}. If the type of the crystalline lattice is other than body-centered cubic, the value of $d$ can change producing huge variations again (see the MCP case below). To illustrate these uncertainties,  in figure~\ref{fig:pycno}b we plot the zero-temperature pycnonuclear reaction rates as a function of $\mu_e$ for two potentially important pycnonuclear reactions for the neutron star inner crust, $^{34}$Ne+$^{34}$Ne \cite{Haen90} and $^{40}$Mg+$^{40}$Mg, using $S$-factors from Reference~\cite{Afanasjev2012PhRvC}. Solid lines are the `optimal' rates, while thinner dashed lines correspond to maximal and minimal rates, see Reference~\cite{Gasq05} for an extensive discussion. Triplets of open symbols on each rate curve give depths where $\tau_\mathrm{burn}=1\,\mathrm{yr},\,1\,\mathrm{d},\,\mathrm{and}\, 1\,\mathrm{s}$. Note, that we use $\mu_e$ instead of $\rho$ here, since in the inner crust the latter depends on the amount of free neutrons and hence on the equation of state. As seen, the uncertainty in the rate can be as large as ten orders of magnitude. Therefore, depending on a model, a given pycnonuclear reaction can start at various points, can be delayed and start after the accretion ceased \cite{Yako06b}, or does not operate at all. 
The same uncertainties are shown with dashed lines in the left panel in figure~\ref{fig:pycno}a for carbon burning where all regimes are considered. It is clear that the theoretical uncertainty increases with regime number.

The situation is even more complicated in a multi-species plasma \cite{Yako06a}. 
The most important point is the availability of close neighbors. In the most popular uniform mixing model, the mean equilibrium distance is taken to be the same as in the one-component model. Then the rates are calculated in a similar way as in the one-component plasma (with proper renormalization of the model parameters, and this procedure is also not very certain \cite{Yako06a}). 
It is unclear, however, if the formalism  applicable for a one-component body-centered cubic lattice is equally good for potentially amorphous structures, or for less abundant chaotic impurities (if they have smaller charges they may start to fuse first). One can expect more frequent close encounters for this case than in a regular lattice, increasing the reaction rate. On the other hand, consider a regular lattice of two intermittent species. The closest distance $d$ for two nuclei of the same species will be a factor of 1.155 larger than in the one-component or uniformly mixed case, greatly suppressing the pycnonuclear rate due to exponential behavior in $d^2$ \cite{Yako06a}. Additional complications arise when the structural changes caused during the course of pycnonuclear burning are considered \cite{Salp69}. When two nuclei in a lattice fuse, this clearly creates a defect. What will happen next, how the lattice will react to a number of such defects and in what respect this can affect the burning rates, is basically unexplored.

Nuclear physics enters the pycnonuclear reaction rate (equation (\ref{eq:fus_rate_pycno})) through the astrophysical factor $S(E_\mathrm{pk})\approx S(0)$. As the pycnonuclear reactions are thought to occur in the inner crust of the neutron star,  
the presence of free neutrons provides another potential source of uncertainty. It is not unreasonable to assume that the neutron gas can alter the properties of the Coulomb barrier, making it lower and/or thicker. It is impossible to study this experimentally. In Reference~\cite{Afanasjev2012PhRvC}, the authors addressed this problem by introducing phenomenological modifications of the barrier shape. They found, taking $^{34}$Ne as an example, that even slight perturbations of the barrier shape lead to another ten orders of magnitude variations in the $S$-factor at small energies and, as a consequence, in the reaction rate. 

The onset of pycnonuclear burning is thus strongly model-dependent. The lighter nuclei can burn even in the ocean \cite{Horo08}. However the main pycnonuclear fusion is expected to occur in the inner crust \cite{Haen90,haensel2008,steiner2012}. The importance of the pycnonuclear burning is that it is the main deep crustal heating source. For instance, in a one-species-per-depth model of \cite{Haen90,Haen03,haensel2008}, the first pycnonuclear reaction ($^{34}$Ne+$^{34}$Ne) releases about 0.5~MeV/u, or $25-35$ percent of the total heating. In some cases, similar to superthreshold electron capture cascades discussed in the previous section, the product of the pycnonuclear fusion may become immediately unstable to a cascade of electron capture reactions which drives it back to the initial nucleus by a chain of neutron emission reactions. In other words, one of the nuclei acts as a catalyst for the conversion of the other one into free neutrons \cite{steiner2012,Lau18}. Note, that like superthreshold electron capture cascades this scenario is only possible when multi-component reaction networks are employed. 

At first glance, the strong variation in the properties of pycnonuclear burning make impossible any conclusions about the nuclear transformations and heat release in the inner crust. However, this is not the case. As it has been shown \cite{haensel2008}, the total heat released in the crust is remarkably independent of the particular sequence of nuclear reactions an accreted element encounters. Indeed, close to the crust-core boundary the accreted crust composition is expected to merge with the cold-catalyzed one. This means that the energy released in the accreted crust is determined by the heat reservoir the initial ashes have with respect to the ground (cold-catalyzed) state. During compression, this extra energy is released in one way or another, but the net result is the same \cite{haensel2008}. The total heat release in the accreted crust is found to be about $2$~MeV/u. Looking from the other side, the total heating is governed by the properties of the cold-catalyzed state. Reference~\cite{steiner2012} analyzed several equations of state in the inner crust and found that some of them have lower ground states predicting twice the heating that is assumed in traditional models. Unfortunately, the pycnonuclear uncertainties strongly impact the heating sources' distribution, which may considerably affect the crustal thermal state during and after an accretion outburst.

\section{Observable Impact of Nuclei in the neutron star Ocean and Crust} \label{section:impact}

 \subsection{Thermal Timescale of the Accreted Crust }\label{subsection:impact_inner}

As shown above, nuclear transformations during accretion modify the composition of the crust and are the source of non-equilibrium heating/cooling processes. Below we discuss the importance of these processes for observable systems  -- cooling X-ray transients and for ignition conditions and lightcurves of X-ray superbursts. To begin, we consider the heat propagation in the crust.

The thermal diffusion timescale  
between a column depth $y$  
and the surface is %
\cite{Henyey1969ApJ}
\begin{equation}\label{eq:tau_diff}
\tau = \frac{1+z}{4}\left[\int_0^y \left(\frac{c_P}{\rho K}\right)^{1/2} \mathrm{d } y' \right]^2,
\end{equation}
where $c_P$ is the specific heat capacity per unit mass\footnote{Under neutron star crust conditions, the heat capacity at constant pressure $c_P$ and heat capacity at constant volume $c_V$ are the same, $c_P\approx c_V$ \cite{HPY2007}.} and $K$ is the thermal conductivity. The heat capacity in the ocean and outer crust is set by ions, except for the outer ocean at high temperatures, where the electrons dominate. In the inner crust, free neutrons dominate the heat capacity unless they are superfluid (see section~\ref{section:structure}). In the superfluid regions, the neutron contribution is suppressed and $c_P$ is determined by ions and electrons. 

Degenerate electrons provide the dominant contribution to the thermal conductivity and it is mainly set by electron-ion scattering:
\begin{equation}\label{eq:kappa_e}
K=K_e=\frac{\pi^2 c^2 k_\mathrm{B}^2 T n_e}{3 \mu_e \nu_{\rm ei}},
\end{equation}
where $\nu_{\rm ei}$ is the effective electron-ion collision frequency. The character of the electron-ion collisions depends strongly on the phase state of the ionic system. In the multicomponent liquid ocean, under the linear mixing rule approximation (e.g., \cite{pote99})
\begin{equation}\label{eq:nuei_liq}
\nu_{\rm ei} = \frac{4\alpha_f^2}{3\pi} \frac{\mu_{\rm e}}{\hbar} \frac{\langle Z^2 \Lambda_{\rm ei}\rangle }{\langle Z\rangle}, 
\end{equation}
where $\Lambda_{\rm ei}\sim 1$ 
is the Coulomb logarithm of the species in mixture. For estimates, the mean ion model with $\langle Z^2 \Lambda_{\rm ei}\rangle\approx \langle Z \rangle^2$ can be used \cite{pote99}. Consider, for example, thermal wave propagation in the hot ocean, where the heat capacity is set by electrons, $c_P=\pi^2 \langle Z\rangle k_\mathrm{B}^2 T/(\langle A \rangle m_u \mu_e)$. The integrand in equation~(\ref{eq:tau_diff}) is then temperature-independent and behaves like $y^{-5/8}$ (neglecting the composition dependence on $y$). Thus the resulting estimate is
\begin{equation}\label{eq:time_ocean}
\tau \approx 16.5\ {\rm hr} \left(\frac{y}{10^{12}~\mathrm{g}~\mathrm{cm}^{-2}}\right)^{3/4}
\left(\frac{g_{14}}{2.44}\right)^{-5/4} \frac{\langle Z\rangle \langle Z^2 \Lambda_{\rm ei}\rangle}{26^3} \left(\frac{56}{\langle A\rangle}\right)^2
\end{equation}
for $1+z=1.31$.
This gives 
an order of magnitude estimate of the timescale for superburst decay \cite{cumming2001,2004CummingMacBeth,Stro02}.

The situation is different in the crust. If a perfect crystal is formed (as expected for the pristine crust, section~\ref{subsection:pristine}), the thermal conductivity is set by electron-phonon scattering. The corresponding collision frequency in the classical limit ($T\gtrsim 0.2 T_\mathrm{pi}$) in the single-phonon approximation is \cite{Yakovlev1980,Schmitt17}
\begin{equation}\label{eq:nuei_phon}
\nu_{\rm ei}=\nu_{\rm e-ph}\approx 13\;\alpha_f \frac{k_\mathrm{B}T}{\hbar}.
\end{equation}
Note that this expression neglects various corrections (e.g., \cite{pote99}) but is sufficient for the qualitative discussion here. Clearly, since $\nu_{\rm e-ph}\propto T$, it is much smaller than $\nu_{\rm e i}$ in the liquid phase [equation~(\ref{eq:nuei_liq})]; accordingly, the thermal conductivity is high. The reason is that the elastic (Bragg) part of the scattering off the ordered crystalline lattice leads to renormalization of the electron states to Bloch waves, and as such does not lead to dissipation.

Consider for example the diffusion timescale of the crystalline outer crust. This is relevant for studies of crustal cooling in young ($\lesssim 100$~yr) neutron stars 
(e.g., \cite{Gnedin2001MNRAS,Shternin2008AstL}) and 
in the so-called quasi-persistent transients (section \ref{sec:quasipers}). 
For a classical crystal,
the ion heat capacity is $c_P=3 k_\mathrm{B}/(A m_u)$ and from equation~(\ref{eq:tau_diff}) one obtains \cite{Brow09}
\begin{equation}\label{eq:time_crust}
\tau\approx 40\ \mathrm{days}\times y_{14}^{3/4} \left(\frac{g_{14}}{2.44}\right)^{-5/4} \left(\frac{56}{\langle A \rangle}\right)^2 \frac{\langle Z \rangle}{26},
\end{equation}
where now $y_{14}$ is the column density measured in units of $10^{14}~\mathrm{g}~\mathrm{cm}^{-2}$ and again $1+z=1.31$. The pure crust cools relatively fast. However, if weakly correlated impurities are present, they contribute to the charge fluctuations that electrons scatter off. It is customary to write the total conductivity using  Mathiessen's rule $\nu_{\rm ei}=\nu_{\rm e-ph}+\nu_{\rm e-imp}$, where $\nu_{\rm e-imp}$ is given by equation~(\ref{eq:nuei_liq}) with the substitution $\langle Z^2 \Lambda_{\rm ei}\rangle \to Q_\mathrm{imp} \Lambda_\mathrm{imp}$, where $\Lambda_\mathrm{imp}\sim 1$ is the Coulomb logarithm for uncorrelated impurities scattering. 
Comparing $\nu_{\rm e-imp}$ with  $\nu_{\rm e-ph}$ we find that the former dominates if
\begin{equation}
Q_\mathrm{imp} \Lambda_\mathrm{imp}\gtrsim 31\ \frac{\langle Z \rangle}{26} \frac{30\ \mathrm{MeV}}{\mu_e}\, \frac{T}{10^8~\mathrm{K}}.
\end{equation}
Thus, the impurity scattering becomes more important in the inner crust (large $\mu_e$) at low temperatures (since impurity scattering is elastic and thus temperature-independent). Moreover, at $T\lesssim 0.2 T_\mathrm{pi}$, the electron-phonon scattering is further suppressed by quantum effects roughly as $\propto T/(0.2 T_\mathrm{pi})$ (e.g., \cite{Schmitt17}), making the impurity contribution dominant even for a ``small'' impurity parameter $Q_\mathrm{imp}\sim 1$. These estimates show that large impurity content would increase the thermal diffusion time. 

The large impurity parameter of the ashes from most surface burning processes (see figure~\ref{figure:ECcrustcomposition} and table~\ref{table:AvgZAQTable}) raises doubts about the applicability of the simple prescription of the pure crystal with uncorrelated impurities to the multicomponent mixture of the accreted crust. Most of the recent molecular dynamics studies suggest that crystallization occurs even for large $Q_{\rm imp}$ in a way that ions of large charge occupy lattice sites (of a body-centered cubic lattice), while smaller charge nuclei are found in interstitial regions \cite{Horo09,Horowitz2007PhRvE,Hughto2011PhRvE,Horowitz2009PhRvE}. The appearance of the Bragg structure in scattering is clearly seen in simulations. In addition, it was found that the impurities are actually correlated. Calculations of the thermal conductivity \cite{Horowitz2007PhRvE,Horo09,Rogg16} have shown that the uncorrelated impurities limit underestimates the impurity contribution. Recently, the authors of Reference~\cite{Rogg16} used the Path Integral Monte Carlo approach to find that the electron-impurity scattering can be adequately approximated by $\nu_\mathrm{e-imp}$, where the {\it effective} impurity parameter $\tilde{Q}_\mathrm{imp}=L(\Gamma) Q_\mathrm{imp}$ is used in place of $Q_\mathrm{imp}$. 
According to~\cite{Rogg16},  
the correction factor $L(\Gamma)$ is about $1.5-2$ at $\Gamma\sim 300-500$ and increases to $3-4$ at lower temperatures (higher densities) where $\Gamma\sim 10^{4}$. 

Finally, 
if the mixture is so diverse that no crystal is formed and the solid phase is in an amorphous state, then no regular lattice exists and equation~(\ref{eq:nuei_liq}) is applicable as well (e.g., \cite{Daligault2009}). This is thought to be a lower limit for the thermal conductivity and results in much longer diffusion timescales than given by equation~(\ref{eq:time_crust}) (e.g., \cite{Brow00}). Nevertheless, it has been shown that diffusion in a multicomponent Coulomb plasma is sufficiently fast to relax an initially amorphous structure  to a crystalline state \cite{Hughto2011PhRvE}.

The presence of a pasta layer at the base of the inner crust (section~\ref{section:structure}) with anisotropic nuclear clusters can drastically change the thermal behavior of the inner crust. Transport properties of the pasta were studied by several authors during the last decade (e.g., \cite{horowitz2008,Schn16,NandiSchramm2018ApJ,Yakovlev2015MNRAS}), but a consistent picture has not been developed yet. It is expected that transport in the pasta phase is anisotropic, with conductivity along one of the symmetry directions being much more effective than along another \cite{Yakovlev2015MNRAS,Schn16}. The resulting thermal conductivity will depend on the orientation of the pasta domains in the star. If these domains are oriented chaotically, or the lower-conduction axis is aligned with the radius,  the conductivity can be significantly reduced. The pasta phase is thought to have a regular structure within one domain, but can contain topological defects that can act like impurities \cite{Schn16}. This can make pasta a highly resistive phase. Taking into account these complications, one usually describes thermal conductivity in the pasta phase by introducing the phenomenological (usually large) impurity parameter  $Q_\mathrm{imp,\mathrm{pasta}}$. Note that some studies do not find reduced conductivity in pasta (e.g., \cite{NandiSchramm2018ApJ}).

An additional uncertainty that affects thermal relaxation in the inner crust is contained in the heat capacity $c_P$. If neutrons are superfluid (section \ref{section:structure}), they do not contribute to the heat capacity. However, in regions where neutrons are normal, they 
comprise the dominant contribution to $c_P$. As follows from equation~(\ref{eq:tau_diff}), the presence of normal neutrons will delay thermal relaxation; additionally, they can store more heat. Thus, the results are sensitive to the profile of the neutron singlet $^1$S$_0$ pairing gap. This is especially important if the pairing gap does not penetrate the core and closes at lower mass densities than the crust-core transition (e.g., \cite{gandolfi2008}). In this case a layer with a high abundance of normal neutrons would be present at the base of the inner crust. In the regions where normal neutrons exist,  the neutron thermal conductivity $K_\mathrm{n}$ may play a role. It was found, however, that generally $K_{\rm n} < K_e$ throughout 
the inner crust~\cite{Bisnovaty1982,Deib17}. Still, the neutron thermal conductivity can be important near the crust-core interface and is worthy of future study  \cite{Deib17}.

\subsection{Cooling Transients}
\label{sec:cooling_transients}

Accreting neutron star 
transients in low-mass X-ray binaries (section~\ref{section:accretion}) are
bright ($L_\mathrm{X}\sim 10^{36}-10^{39}$~erg~s$^{-1}$) 
X-ray sources during accretion outbursts (see table 1 of Reference~\cite{Dege15} for a summary). 

As described in section~\ref{section:interaction}, non-equilibrium 
reactions 
deposit $\approx 1 \textrm{--} 2 \, \mathrm{MeV}$ per accreted nucleon in the neutron star 
crust during an accretion episode. When an accretion outburst ends, the X-ray luminosity drops by several orders of magnitude, the neutron star 
enters quiescence, and the fainter ($L_\mathrm{X}\sim 10^{31}-10^{34}$~erg~s$^{-1}$) 
thermal emission from the now cooling surface can be measured by sensitive X-ray observatories. The subsequent evolution depends on the duration of the accretion episode. One distinguishes normal transients, where accretion lasts for several weeks, and 
quasi-persistent transients, where accretion can proceed for years or even decades. A recent detailed discussion of the properties of cooling transients can be found in Reference~\cite{Wijnands2017JApA}. 
Here we briefly describe the properties of transients necessary to understand the nuclear physics impact on these objects.

\subsubsection{Normal transients}
Before  the start of an accretion episode, the neutron star is isothermal inside (except a thin outer 
heat blanketing envelope). Energy release from nuclear reactions during accretion breaks this equilibrium and heats the crust creating an inward and outward energy flux from the heating regions. During a relatively short accretion outburst, the crust is not heated strongly and quickly relaxes to thermal equilibrium with the core after accretion ceases. In this process, a fraction $f\lesssim 1$ of the total heat release enters the neutron star core, typically $\approx 90\,\%$~\cite{Brow00}, and a smaller fraction $1-f$ of heat diffuses toward the surface and is radiated away during crustal cooling and by increased neutrino emission from the heating regions. Many cycles of accretion/quiescence result in secular heating of the whole star with the secular heating rate (as measured by a distant observer) of $L_\mathrm{heat}^\infty=fQ \langle \dot{M} \Delta t\rangle/(m_u t_\mathrm{rec} (1+z))$,
where $Q$ is the deep crustal heating power, $\langle\dot{M} \Delta t\rangle$ is the mass accreted during an outburst, on average, and $t_\mathrm{rec}$ is the average recurrence time. 
This heating is enough to balance the energy loss due to surface emission $L_\gamma^\infty$ and neutrino cooling $L_\nu^\infty$ from the bulk. If the transient accretion behavior persists for a long time, the neutron star is found in a steady state set by the condition $L_\mathrm{heat}^\infty=L_\nu^\infty+L_\gamma^\infty$, with the temperature much higher than expected for a passively cooling neutron star of comparable age \cite{BBR1998}. This steady state point is thus determined by the (unknown) neutrino emission rate from the core. For higher rates, the steady state will be reached at lower temperatures (and thus surface luminosities) for a given $L_\mathrm{heat}^\infty$. Measurements of the quiescent luminosity with sensitive X-ray observatories
thus in principle  allow one to constrain the neutrino emission mechanisms operating in the neutron star core \cite{Yakovlev2004ARA&A,Wijnands2013,Wijnands2017JApA}, as well as the core heat capacity~\cite{Cumm17}.
At the moment, the strongest constraint comes from the coldest (in quiescence) 
transient in the observed sample, SAX~J1808.4$-$3658 \cite{Heinke2009ApJ}, whose conditions suggest that the powerful direct Urca neutrino emission process is operating in the core of the neutron star in this binary. In addition, the need for a direct Urca process in about 1 percent of the core was recently proposed to explain the low inferred core temperature in the quasi-persistent transient (section \ref{sec:quasipers}) MXB~1659$-$29 \cite{Brown2018arXiv}. However, this case is more uncertain as the accretion duty cycle is not known.

The main nuclear input here is the 
power of the deep crustal heating $Q$; however, a robust inference is plagued by other uncertainties such as distance measurements,  stability of accretion duty cycles, and composition of the 
heat blanketing envelope of the neutron star. The latter strongly affects the relation between the measured surface temperature and the internal temperature 
which determines
$L_\nu^\infty$. 
This uncertainty does not apply for hot and weakly accreting systems where $L_\gamma^\infty\gg L_\nu^\infty$ so that the steady state is determined by the directly measurable neutron star surface emission (e.g., IGR~J00291+5934, \cite{Wijnands2017JApA}). Such systems in principle have a potential to constrain $Q$, but the observational uncertainties are still large.

\subsubsection{Quasi-persistent transients}\label{sec:quasipers}
During the long accretion outbursts in quasi-persistent transients,
the  heat deposited in the crust is so large that it brings the crust out of thermal equilibrium with the core~\cite{Page13}. When an accretion outburst ends, the neutron star %
enters quiescence, 
the crust cools toward thermal equilibrium with the core, and the neutron star's 
surface thermal emission powers an X-ray light curve~\cite{ushomirsky2001,rutledge2002}. The shapes of quiescent cooling curves depend on 
the thermal structure of the neutron star's 
outer layers at the 
beginning of quiescence. Confronting  thermal evolution models of the neutron star's 
outer layers 
with observations leads to constraints on crustal properties such as thermal conductivity and specific heat of dense matter, as discussed in section~\ref{subsection:impact_inner}. 

Active studies of crust cooling in quasi-persistent transients started about two decades ago when the first source KS~1731$-$260 ended a $\gtrsim 12.5$~yr outburst and was observed in quiescence with Chandra \cite{wijnands2001c,rutledge2002}. The continuous monitoring of this source in quiescence revealed the crust cooling towards thermal equilibrium with the core on a $\sim$yr timescale \cite{cackett2006}. This relatively short cooling timescale 
suggests a high thermal conductivity in the crust consistent with an ordered lattice  
contaminated by a small impurity content \cite{cackett2006,Shte07,Brow09}. An amorphous crust with low thermal conductivity, which would have a much longer thermal time (section~\ref{subsection:impact_inner}), contradicts observations.

\begin{figure}
\centering
\includegraphics[scale=1.2]{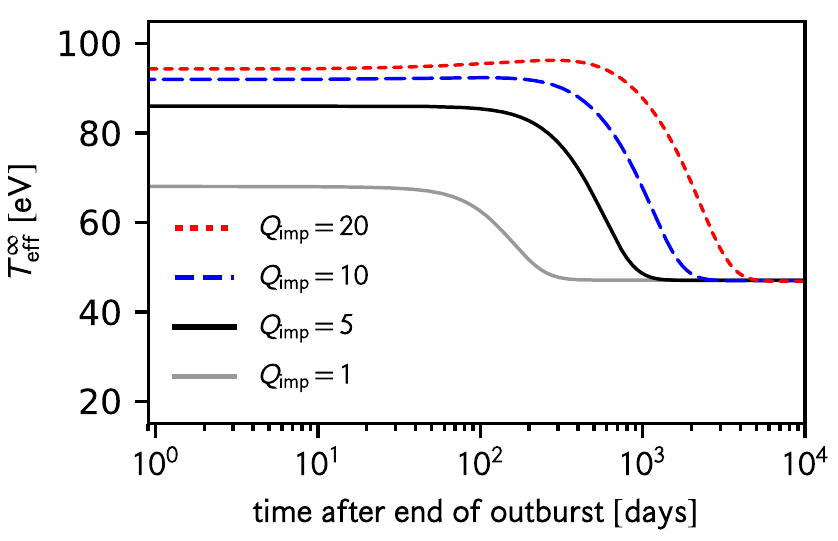}
\caption{Crust cooling models, showing the temperature decline of a neutron star's surface in quiescence,
employing various choices of $Q_{\mathrm{imp}}$ for the entire crust. This crust cooling model uses a neutron star mass of $M= 1.4 \, \mathrm{M}_{\odot}$, neutron star radius of $R = 12 \, \mathrm{km}$, and a core temperature of $T_{\rm core} = 3 \times 10^{7} \, \mathrm{K}$.
\label{figure:NScool_Qimp}}
\end{figure}

Since 2001, quiescent cooling curves have been observed in several low-mass X-ray binaries and provide important data on the thermal evolution of the neutron star outer layers across accretion regimes and  
gravities \cite{Wijnands2017JApA}. Interestingly, crustal cooling is observed now not only in the quasi-persistent sources with long $>1$~yr outbursts, but also for several `ordinary' transients with shorter accretion outbursts. The results of these observations and crustal light curve modeling are extensively reviewed elsewhere (e.g., \cite{Wijnands2017JApA}). Here we summarize only the main points:

\paragraph{The crust is pure:} The analyses of all cooling sources generally suggest that a high thermal conductivity and hence a low effective impurity parameter $\tilde{Q}_{\rm imp}\lesssim 4-7$ throughout the crust is favored (see, however, Reference~\cite{Degenaar2014}). Taking into account the results of Reference~\cite{Rogg16}, this indicates even smaller actual impurity content $Q_{\rm imp}\lesssim 1-3$. By contrast, the distribution of ashes produced from X-ray bursts and stable surface burning can have $Q_{\mathrm{imp}} \sim 70-100$ (see figure~\ref{figure:ECcrustcomposition} and table~\ref{table:AvgZAQTable}) suggesting that a purification mechanism exists in the crust. 
We illustrate the impact of the impurity parameter on cooling curves in figure~\ref{figure:NScool_Qimp}. This crust cooling model uses the accreted crust composition from Reference~\cite{Haen90} and contains nuclear heating of $Q = 0.3 \, \mathrm{MeV \ u^{-1}}$ in the outer crust, and deep crustal heating of $Q = 1.5 \, \mathrm{MeV \ u^{-1}}$ in the inner crust. As can be seen in figure~~\ref{figure:NScool_Qimp}, large values of $Q_{\rm imp}$ lower the thermal conductivity of the inner crust and increase the thermal time. Therefore, it takes longer for heat to diffuse out of the inner crust during quiescence and for the crust to reestablish thermal equilibrium with the core.

The estimates obtained in this way are more relevant for the inner crust where the impact of impurities is more pronounced (section~\ref{subsection:impact_inner}). In addition, the constraints on the outer crust properties are contaminated by the mysterious shallow heating source (discussed momentarily). 
The impurity parameter in most models and in our illustrative calculations is set constant. Reference~\cite{Page13} allowed for a variable $Q_{\rm imp}$ with a higher value in the outer crust, lower value in the inner crust, and a smooth transition between them. This allowed the authors  
to explain the observations of the source XTE~J1701$-$462; the adopted model is also consistent with observations of other transients. 

\paragraph{Neutron superfluidity is favored:} The intermediate part of the transient cooling curve is usually fit better when neutron superfluidity is taken into account (e.g., \cite{Shte07,Brow09}), which lowers the heat capacity in the inner crust causing it to store less heat and cool more rapidly (section~\ref{subsection:impact_inner}). 

\paragraph{Extra shallow heating is required:} It has been found that the standard heat release from deep crustal heating models is usually insufficient to heat the crust to the high effective temperatures observed at the early phase of quiescence \cite{Brow09,Turl15}. Although the uncertainties in the nuclear physics may vary the heat deposited in the inner crust~\cite{steiner2012}, analyses of quiescent light curve shapes demonstrate that the additional heating source must be located at densities $\rho_\mathrm{shallow}\lesssim 10^{11}$~g~cm$^{-3}$. For a standard model of deep crustal heating, the strength of this additional source is found to be on the $Q_\mathrm{shallow}\sim 1 \, \mathrm{MeV}$ level. However, this estimate is model-dependent, varies from source to source, and may vary from one outburst to another in the same source~\cite{Turl15,Pari17}. 

\begin{figure}
\centering
\includegraphics[scale=1.2]{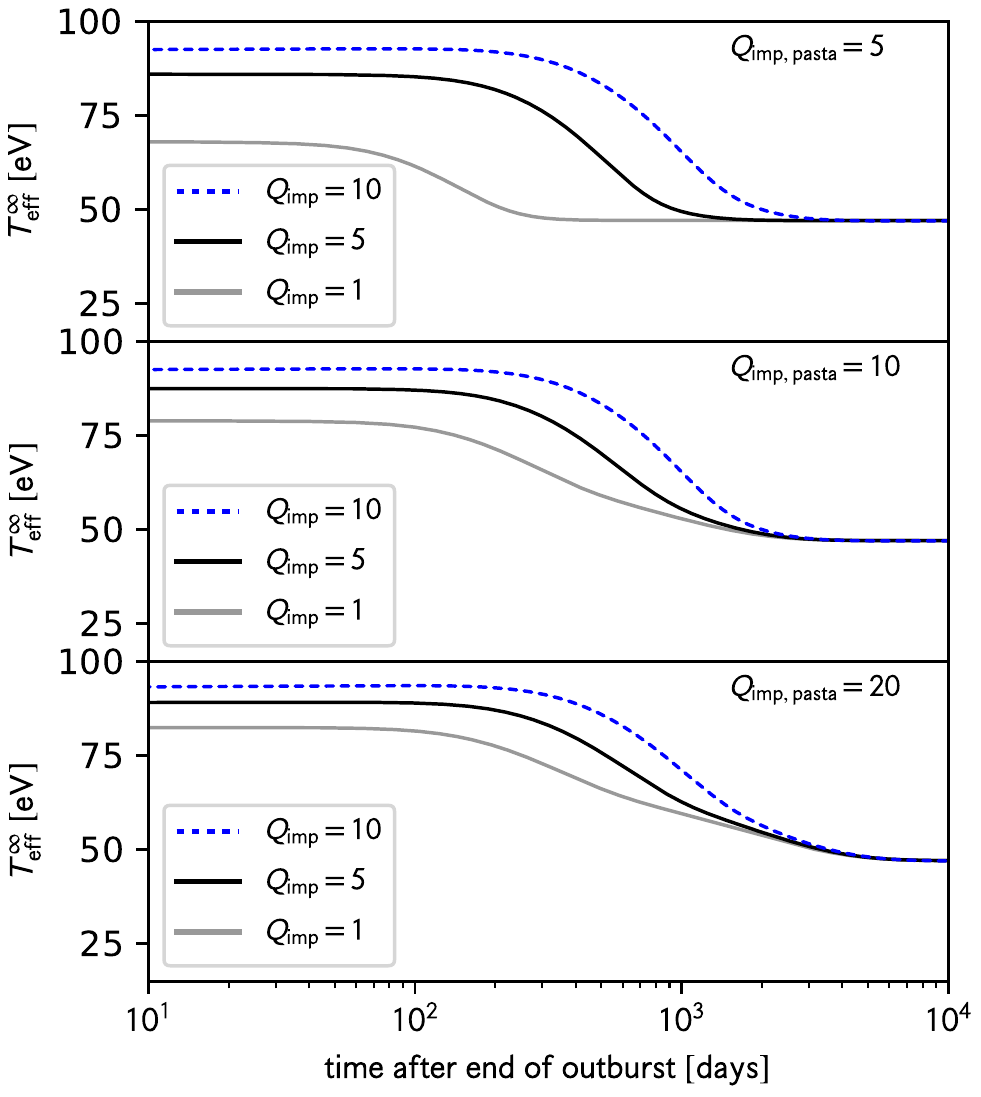}
\caption{Crust cooling models, showing the temperature decline of a neutron star's surface in quiescence, employing various choices of $Q_{\mathrm{imp}}$ for the crust and the pasta layer at $\rho \gtrsim 8 \times 10^{13} \, \mathrm{g \ cm^{-3}}$. This crust cooling model uses a neutron star mass of $M= 1.4 \, \mathrm{M}_{\odot}$, neutron star radius of $R = 12 \, \mathrm{km}$, and a core temperature of $T_{\rm core} = 3 \times 10^{7} \, \mathrm{K}$. \label{figure:NScool_pasta}}
\end{figure}

\paragraph{Nuclear pasta is favored:} It was found that the cooling curve of MXB~1659$-$29 is better explained if a low-conducting region is included in the inner crust, that can be attributed to the pasta layer \cite{Horo15,Deib17}. We illustrate the impact of the presence of  pasta in figure~\ref{figure:NScool_pasta}, where the thermal conductivity in the pasta is modeled by a large impurity parameter $Q_{\mathrm{imp},\mathrm{pasta}}$ for $\rho>8\times10^{13}$~g~cm$^{-3}$ and the model uses the $^1$S$_0$ neutron singlet pairing gap from Reference~\cite{gandolfi2008}. 
The presence of the heat insulator in the pasta layer may explain the continuous cooling observed in MXB~1659$-$29 after  11 yr in quiescence \cite{Cackett2013ApJ}. If the nuclear pasta layer has a low thermal conductivity, some regions of the inner crust may remain above the neutron superfluid critical temperature depending on the choice for the pairing gap \cite{Deib17}. As discussed in section~\ref{subsection:impact_inner},  there may be normal neutrons present in the pasta layer and the heat stored by normal neutrons will impact the late-time cooling behavior of neutron star 
transients which remain in quiescence for more than $ > 1000 \, \mathrm{days}$~\cite{Deib17}. Note, however, that the temperature drop observed in  MXB~1659$-$29 may not be caused by continuous crustal cooling, but could be a result of an increase in the absorption column density prior to the last observation \cite{Cackett2013ApJ}.

\paragraph{}
It is important to note that strong constraints on the composition of the crust are currently difficult to obtain because the shape of the cooling light curve is degenerate with other neutron star 
parameters, such as the  
gravity and the core temperature, which are unknown a priori. Also, accurate accounting for a variable accretion rate during a long outburst can be important \cite{Ootes2016MNRAS}.
Finally, the observations, especially of the early stages of crustal cooling, may be contaminated by residual accretion.

The hottest neutron star 
transient MAXI~J0556$-$332 \cite{matsumura2011,Homa11,sugi13,homan14} is an excellent test bed for observational signatures of Urca cooling nuclei pairs \cite{Deib15} because of the strong ($T^5$) temperature dependence of Urca neutrino cooling (see equation~(\ref{equation:Lnu})). The inferred surface temperature at the onset of quiescence is nearly twice as large as the next hottest transient observed~\cite{homan14}, and it is the only transient thought to have a hot enough crust the Urca cooling could produce an appreciable $L_{\nu}$. It was found that MAXI~J0556$-$332 likely does not have Urca cooling operating in its crust because the quiescent cooling trend would follow a different behavior than what is observed \cite{Deib15}.

\begin{figure} [h]
\centering
\includegraphics[scale=0.5]{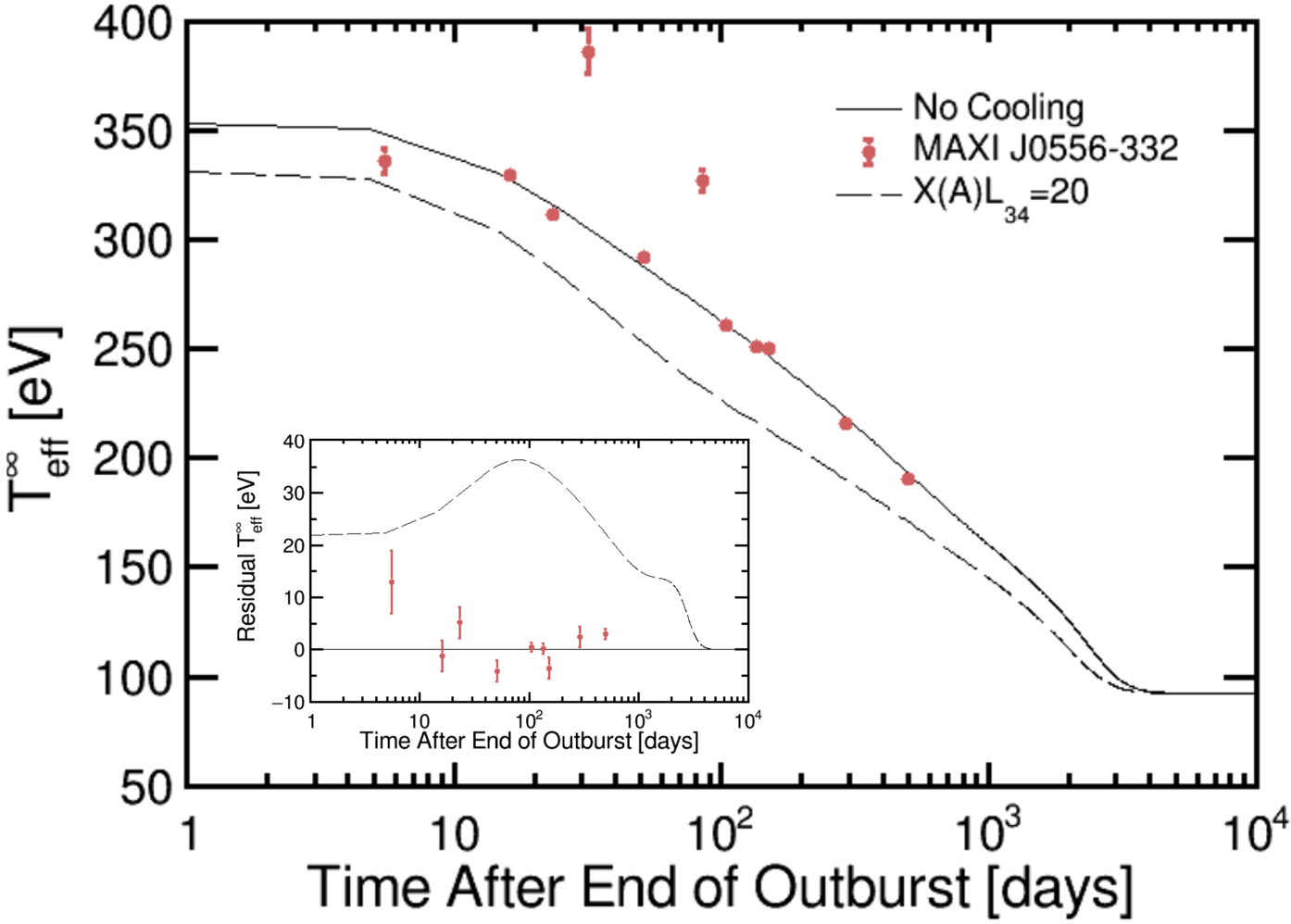}
\caption{Light curve of MAXI J0556$-$332 after the end of an accretion outburst~\cite{homan14} compared to {\tt dStar}~\cite{Brow15} model calculations with (dashed line) and without (solid line) Urca cooling expected for $^{33}\rm{Al}\leftrightarrow^{33}\rm{Mg}$~\cite{Meis17}. Models employ $Q_{\rm{shallow}}=8$~MeV per accreted nucleon,  $Q_{\rm{imp}}=1$ throughout the crust, and an outburst duration and accretion rate that match observations~\cite{homan14}. The inset shows the residuals to the no-cooling calculation.\label{figure:urcaLC}}
\end{figure}

The authors of Reference~\cite{Meis17} reexamined the quiescent cooling of MAXI~J0556$-$332 and self-consistently added Urca cooling nuclei to the ocean and crust composition of a thermal relaxation model. 
They found that Urca cooling would be
apparent for X-ray burst or superburst ashes, but not for the ashes of stable burning. Interestingly, stable burning ashes produce percent-level $X(A)$ for odd-$A$ nuclides in the $70\lesssim$$A$$\lesssim80$ range, but these isobars appear to lack significant Urca pairs due to their relatively large $ft$~\cite{Scha14,Meis17}, see equations~(\ref{equation:Lnu})-- (\ref{equation:L34}). Because of the large ocean and crust temperature in MAXI~J0556$-$332, it is indeed likely that stable burning has occurred 
in the outer layers. As such, the absence of an Urca cooling signature in the MAXI~J0556$-$332 light curve, as shown in figure~\ref{figure:urcaLC}, is consistent with model calculations. Further consistency checks will require the remaining nuclear physics uncertainties, particularly $ft$-values (section~\ref{ssec:ECrxns}) of abundant nuclides, to be determined experimentally. Spectroscopy studies at the limits of experimental accessibility are underway for this purpose.

Well-constrained estimates of Urca cooling strengths, coupled with models of cooling transient light curves, enable surface nuclear burning to be constrained over vast timescales 
for neutron stars 
with high-temperature crusts. By analyzing cooling transient light curves for signatures from Urca cooling, the composition of the accreted neutron star crust
can be constrained by determining the presence or absence of Urca pairs~\cite{Meis17}. As the composition is determined by
past nucleosynthetic activity on the neutron star  
surface, this method provides a theoretical
approach to constrain surface nuclear burning on accreting neutron stars 
over the past millenia.

\subsection{X-ray Superbursts}\label{subsection:obs_superbursts}
\begin{figure}
\centering
\includegraphics[scale=0.6]{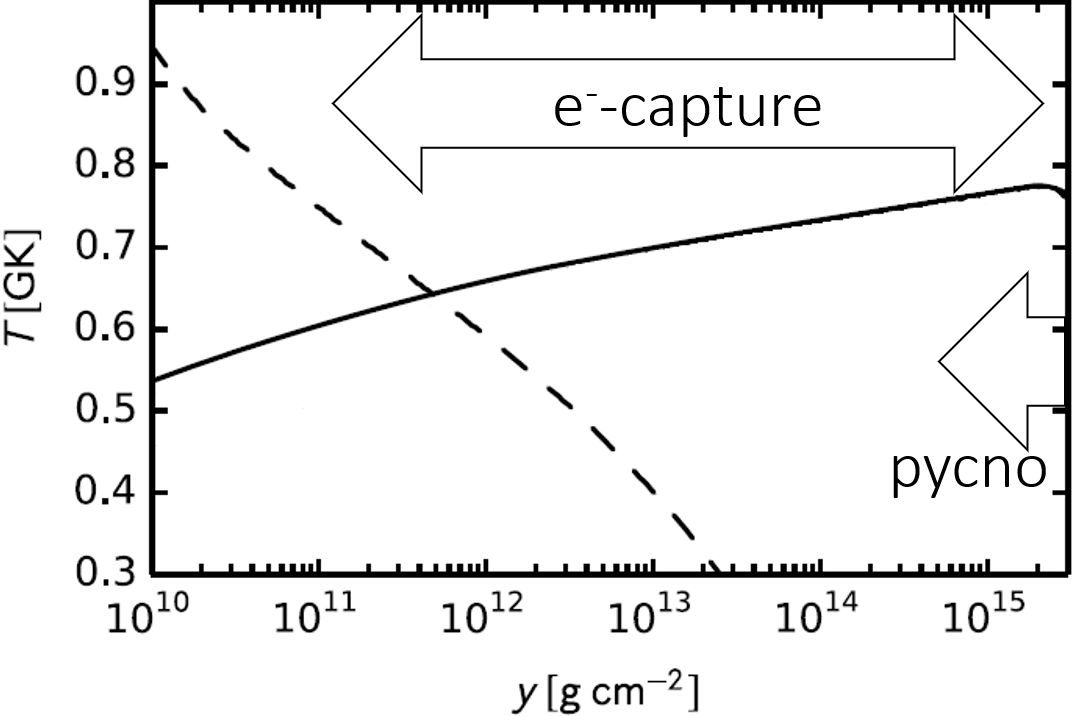}
\caption{Steady-state ocean temperature as a function of column depth (solid line) for a neutron star with $\dot{M}=0.3~\dot{M}_{\rm{Edd}}$, $M=1.4~M_{\odot}$, $R=10$ km, and
$T_{\rm{core}}=3\times10^{7}$ K. A shallow heating of 5~MeV per nucleon is assumed, resulting in $\sim$0.25~MeV per accreted nucleon in the ocean. The dashed black line indicates the temperature and density at which ignition of unstable
carbon burning in a mixed iron-carbon ocean with $X_{\rm{Fe}}=0.8$ and $X_{\rm{C}}=0.2$ will ensue. 
The open arrows indicate the regions where $e^{-}$-capture heating/cooling and pycnonuclear heating are active, where the region of pycnonuclear heating extends to depths beyond the figure extent.
(Adapted from Reference~\cite{Deib16}).
\label{figure:NStemp}}
\end{figure}

Recall that  X-ray superbursts are thermonuclear runaways triggered by carbon ignition in the accreted neutron star 
ocean (section~\ref{sec:superbursts}). 
Their long duration is set by the thermal time scale of deep neutron star ocean layers, indicating an ignition column depth of $10^{11}-10^{12}\ \mathrm{g\ cm^{-2}}$. This depth is relatively close to the outer crust. Therefore, the occurrence of superbursts provides a measure of the thermal properties at the crust-ocean interface as a function of depth. This is illustrated in figure~\ref{figure:NStemp} where the dashed line shows temperature required to ignite carbon at a given column depth. Its position is sensitive to the $^{12}$C$+$$^{12}$C reaction rate, and enhancements in this rate are of particular interest, as they may help to explain the discrepancy between carbon ignition depths inferred from observations and required in simulations (section~\ref{sssec:SuperburstObs}).
Furthermore, superbursts quantify the carbon content of the hydrogen/helium burning ashes, which cannot be inferred directly from observations of short bursts (section \ref{sec:burst_observation}).

The thermal profile around the ignition layer is set by the heating sources in the ocean and the crust.
The impact of  the nuclear reactions discussed in section~\ref{section:interaction} 
is highlighted in figure~\ref{figure:NStemp}, where the locations for $e^{-}$-capture and pycnonuclear reactions are shown. 
Heat from these reactions, which is typically produced at depths below carbon ignition,
diffuses upwards, raising the ocean temperature and enabling carbon to ignite at some depth~\cite{Brow04,Keek11}.
This heat is insufficient, though, to raise the ocean temperature to the level required to explain observations. The deep crustal heating uncertainty of $\sim 2$~MeV/u, discussed above \cite{steiner2012}, modifies the heating in the `pycno' region just beyond the right bound of figure~\ref{figure:NStemp}. 
This heat is mainly conducted inwards to the core due to high conductivity and provides insufficient impact on the ignition layer temperature. The temperature profile shown by a solid line in figure~\ref{figure:NStemp}, which crosses the ignition curve (dashed line) at reasonable densities, is obtained by adding a strong ($\sim 5$~MeV/u) shallow heat source at $y_\mathrm{shallow}\sim 10^{15}$~g~cm$^{-2}$ \cite{Deib16}.
The need for such a shallow heat source is in line with the findings for the quasi-persistent transients discussed in section~\ref{sec:quasipers}. Some heating can also come from compositionally-driven convection proposed in the ocean~\cite{Medin2011} (see section~\ref{sec:pure}). 

Additional complications come from the 
Urca pairs that can be present  
in the neutron star's ocean and 
crust.
They form a thermal barrier between heat released deeper in the crust and the carbon ignition layer. The strong temperature sensitivity of the neutrino luminosity from Urca cooling (see equation~(\ref{equation:Lnu})) means that even 
strong deep crustal heating will not raise the ocean temperature if such a barrier exists. 

A recent study investigated the interplay between heating and cooling reactions in the accreted ocean and crust and the carbon ignition depth~\cite{Deib16}.
The authors self-consistently calculated the Urca cooling nuclei pairs \cite{Scha14} one would expect 
from the compressed 
ashes of Type I X-ray bursts and superbursts \cite{Scha99,Scha03,Keek08}. Urca cooling layers were then implemented into a superburst ignition model \cite{Pote12} and it was found that Urca cooling in the neutron star's 
crust lowers the ocean's steady state temperature during an accretion outburst, while the ocean Urca pairs produce little effect. 
As a consequence, superburst ignition occurs deeper than it would otherwise (see dashed curve in figure~\ref{figure:NStemp}).

The highest mass-fraction odd-$A$ Urca nuclides produced in superbursts are of interest for this scenario, i.e. $A=53,55$, and $57$ (see figure~\ref{figure:ashes}), assuming recurring superbursts would erase the signature of X-ray burst ashes. Of these, $A=55$ is by far the most significant due to the predicted percent-level abundance. Theoretical estimates vary largely 
for $L_{\nu}$ for $A=55$ nuclides, where the primary uncertainty comes from the assumptions used to determine the values of ${\rm log}_{10}(ft)$ for the $^{55}$Ti$\rightarrow^{55}$Sc$\rightarrow^{55}$Ca $e^{-}$-capture sequence~\cite{Scha14,Deib16}. As such, measurements are required to resolve this issue. Should ${\rm log}_{10}(ft)\lesssim5$ be found for either of these transitions, this would imply that the corresponding 
thermal barrier 
prevents a significant heat flow from deep crustal heating 
to the ocean. 
This would deepen the mystery of the superburst ignition problem. Furthermore, the presence of such an Urca layer would 
limit any shallow heating source, discussed above, to locations above existing Urca layers.

\subsection{Accreting neutron stars as Sources of Gravitational Waves}\label{sec:gw}
The first detection of the gravitational wave transient GW~150914 from a double black hole coalescence with LIGO in 2015 \cite{Abbott2016PhRvLGW150914} has started the era of gravitational wave astronomy. Only two years after the first detection, a double neutron star 
merger GW~170817 was reported \cite{Abbott2017PhRvLGW170817} opening a new window for neutron star 
astrophysics and providing exciting new insights into the physics of these objects.

Not only binary neutron stars 
can emit
gravitational waves. In fact, fast spinning solitary neutron stars 
have long been considered as possible {\it persistent} 
gravitational wave sources if they possess some degree of asymmetry; for a recent review see Reference~\cite{Glampedakis2017arXiv}. This can result from the certain types of oscillation modes possibly excited in the star, or because of the presence of static density inhomogeneities (``mountains'') that the neutron star crust can hold \cite{Glampedakis2017arXiv}. Let us focus on the latter case. If the mountains result in neutron star ellipticity $\varepsilon$, a star rotating with a frequency $\nu$ would emit 
gravitational waves mainly with a frequency $f=2\nu$ and the energy loss is
\begin{equation}\label{eq:E_gw}
\dot{E}_\mathrm{GW}=2\pi\nu N_\mathrm{GW} = - \frac{32}{5} \frac{G(2\pi \nu)^6 (I\varepsilon)^2}{c^5},
\end{equation}
where $I$ is the neutron star moment of inertia and $N_\mathrm{GW}$ is the associated braking torque. Thus, the emission of 
gravitational waves results in effective spin down which strongly scales with frequency. The maximal ellipticity that crustal mountains can produce is quite large $\varepsilon_\mathrm{max}\sim (0.1-1)\times10^{-4}\, \sigma_\mathrm{max}$ \cite{Johnson2013,Ushomirsky2000MNRAS,Haskell2006MNRAS} where 
$\sigma_\mathrm{max}\lesssim 0.1$ is the maximal strain \cite{Horowitz2009PhRvL,Chugunov2010MNRAS} the crust can sustain.
The largest mountains could produce a detectable signal. However no 
gravitational wave emission has been detected yet from known pulsars and current upper limits on ellipticity with advanced LIGO are of the order of $10^{-7}$ with a minimum value of few of $10^{-8}$ \cite{Abbott2017ApJaLIGO}. 

Mountains can be built up in accreting neutron stars 
once some asymmetry in the accretion process is assumed. One of the proposed mechanisms is directly related to nuclear reactions in the crust (for other possibilities such as ``magnetic'' mountains see, e.g., Reference~\cite{Glampedakis2017arXiv}). The idea is that if lateral thermal or compositional gradients are present in the crust, this radially shifts the positions of the electron capture layers and hence the associated density jumps \cite{Bildsten1998ApJL,Ushomirsky2000MNRAS}. The physics of the temperature sensitivity of the $e^{-}$-capture layers is based on the observation that at sufficiently high temperatures $T\gtrsim 2\times 10^{8}$~K the $e^{-}$-capture rates become faster than the accretion timescale considerably before the corresponding threshold. Calculations show that in fact, most $e^{-}$-capture transitions occur subthreshold at $|Q_\mathrm{EC}|-\mu_e=\Upsilon k_\mathrm{B} T$, where $\Upsilon\approx 10-20$ and depends logarithmically on $T$, comparative half-life $ft$, 
threshold $Q_\mathrm{EC}$, and local accretion rate 
$\dot{m}\equiv \dot{M}/(4\pi R^2)$ \cite{Bildsten1998ApJL,Ushomirsky2000MNRAS,Bild98}. 
The value of $\Upsilon$ gives a measure of the thermal sensitivity of the position of the $e^{-}$-capture layer. The radial amplitude of the layer variation is then $\Delta \zeta \approx \Upsilon Y_e k_\mathrm{B} \delta T/(m_u g) $, where $\delta T$ is the amplitude of lateral temperature variations \cite{Ushomirsky2000MNRAS}. If these variations are quadrupole, the resulting ellipticity is proportional to $\Delta \zeta \Delta\rho$, where in the outer crust the density jump at a given $e^{-}$-capture layer is $\Delta \rho/\rho= 2/Z$. Based on this, the authors of Reference~\cite{Ushomirsky2000MNRAS} obtain the following fiducial estimate
\begin{equation}\label{eq:ellipt_fid}
\varepsilon \approx 1.7\times 10^{-9}\, \frac{\Upsilon}{Z} \frac{\delta T}{10^7\,\mathrm{K}} \frac{2.44}{g_{14}}\, \left(\frac{R}{10~\mathrm{km}}\right)^4 \,I_{45}^{-1}\left(\frac{|Q_\mathrm{EC}|}{30\,\mathrm{MeV}}\right)^3,
\end{equation}
where $I_{45}$ is the neutron star moment of inertia in units of $10^{45}$~g~cm$^{2}$.
This estimate is strictly speaking applicable for the outer crust, but numerical calculations give similar orders of magnitude also when the reactions in the inner crust are considered (there the capture layers are thicker and their structure is modified since degenerate neutrons dominate the pressure). In addition, equation~(\ref{eq:ellipt_fid}) does not account for the elastic response of the crust which in fact can result in an order of magnitude smaller values \cite{Ushomirsky2000MNRAS}.
Similar $\Delta \zeta$ can be provided by compositional gradients instead of thermal ones, if, for instance, the ashes that are compressing are not symmetrically distributed. The source of the temperature asymmetry can stem from the asymmetry of deep crustal heating. Therefore the largest gradients can be expected from the regions 
of the largest heat release  -- i.e. in the inner crust where pycnonuclear reaction or superthreshold electron-capture cascades operate. Notice that too large of gradients would result in flux variations that are currently constrained at the level of $\delta T/T\lesssim 0.1$ \cite{Haskell2015MNRAS}. 

Could nuclear mountains be detected with gravitational observatories? When accretion ceases, the thermal gradients and thus mountains are thought to be erased on the thermal timescale (section~\ref{subsection:impact_inner}) \cite{Bildsten1998ApJL,Haskell2015MNRAS}. Thus, although the mountains in the inner crust can be sustained longer, the chance of the gravitational wave detection from transient sources is small. However, for the persistent accretion sources the situation is more optimistic with the future instrumentation \cite{Haskell2015MNRAS}. There is also a possibility that the mountains are frozen in the crust and build up incrementally. However, this can lead to too large of a spin-down rate due to gravitational wave emission which contradicts observations for at least some sources \cite{PatrunoWatts2012}. See Reference~\cite{Haskell2015MNRAS} for a detailed discussion.

Even if they are undetected by gravitational wave observatories, crustal mountains have important astrophysical implications. Initially the gravitational emission from accreting neutron stars 
was
invoked as a mechanism for limiting the neutron star spin-up due to accretion \cite{Bildsten1998ApJL}. The fastest rotator among the low-mass X-ray binaries is 4U~1608$-$52 with a frequency measured from bursts oscillations of 619~Hz \cite{Hartman2003HEAD} and the fastest  radio millisecond pulsar spins at  716~Hz \cite{Hessels2006Sci}  while basically one expects that the accretion torque can spin a star up until the stability limiting frequency $\sim 1$~kHz \cite{HPY2007}. The strong frequency dependence of gravitational wave spin-down torque in equation~(\ref{eq:E_gw}) thus provides a natural explanation of the cutoff existence. Estimating the accretion torque as $N_a=\dot{M}\sqrt{ G M R}$, we get the ellipticity which allows gravitational wave torque to balance $N_a$: 
\begin{equation}\label{eq:ellip_a}
\varepsilon_a\approx 8\times 10^{-9} I_{45}^{-1} \left(\frac{600\,\mathrm{Hz}}{\nu}\right)^{5/2} \frac{\dot{M}}{10^{-9}\, M_\odot\, \mathrm{yr}^{-1}}
\end{equation}
for $M=1.4 M_\odot$ and $R=10$~km.
In fact, the accretion torque at high frequencies may be smaller, so the estimate from equation~(\ref{eq:ellip_a}) is the upper limit (see Reference~\cite{Patruno2017ApJ} for further discussion). This estimate is generally consistent with equation~(\ref{eq:ellipt_fid}) if the mountains are built in the inner crust (note also that equation~(\ref{eq:ellipt_fid}) is for a single capture layer so many asymmetric capture layers can combine in larger mountains). 
Although it is not clear now that 
gravitational wave emission is even needed for the explanation of the spin-up limit and, instead, 
the physics of accretion possibly plays the major role \cite{Patruno2017ApJ,D'Angelo2017MNRAS}, the bimodal distribution of the low-mass X-ray binaries over spin frequencies reveals a sharp spike clustering around $600$~Hz which is hard to explain without assuming that gravitational wave 
emission becomes important around this frequency \cite{Patruno2017ApJ}.

Another piece of evidence came recently from observations of the spin down of the transient millisecond pulsar PSR~J1023$+$0038 which showed a transition between radiopulsar and low-mass X-ray binary stages. According to Reference~\cite{Haskell2017PhRvL},
the PSR~J1023$+$0038 spin-down rate is faster in the low-mass X-ray binary state by $\dot{\nu}_\mathrm{diff}=-6.428\times 10^{-16}$~Hz~s$^{-1}$. This could be the first evidence that a mountain is built up during accretion and the 
gravitational wave torque provides additional spin down. The required ellipticity (estimated from equation~(\ref{eq:E_gw})) is about $5\times10^{-10}$, in line with the distortions nuclear mountains can provide according to equation~(\ref{eq:ellipt_fid}). If this is indeed the case, then the gradual decrease of the spin-down rate on thermal timescales is expected in the radiopulsar stage. On the other hand, if continuous monitoring of X-ray oscillations during the low-mass X-ray binary stage reveals a spin-down rate increase, this could indicate mountain(s) build-up and the nuclear mountain mechanism can be tested.

\section{Summary and Outlook} \label{section:summary}

Remarkable progress has been made in the 50 years since the neutron star 
crust was first proposed. We now know the outer layers of accreting neutron stars 
host a variety of nuclear phenomena, many of which can be studied in terrestrial laboratory experiments. Decades of efforts with X-ray telescopes have yielded a number of complementary observations which, with insight provided by astrophysics model calculations, have enabled 
the construction of the tomographic picture of the accreted neutron star 
outer layers that we have today. These advances and current efforts in observation, experiment, and theory have been 
discussed
in the previous sections. Here we summarize major remaining challenges and planned efforts to address them.

\subsection{What is Made In Surface Burning Processes?}
Surface burning processes provide the seeds for crust processes, determining major features of the crust composition. Therefore they indicate which nuclear reaction heating/cooling processes are relevant for a particular neutron star. 
However, major uncertainties remain as to the ashes produced in various burning regimes. 
 Models struggle to reproduce observational signatures of surface burning processes when employing the inferred environment conditions.
For example, among H-rich H/He-burning X-ray bursters, only GS~1826$-$24 has had its light curve successfully modeled and for that case models need a higher accretion rate than is inferred from observations (possibly due to a high inclination angle of this source)~\cite{Hege07,Meis18}.

Type-I X-ray burst ash compositions from model calculations are sensitive to a large number of poorly constrained reaction rates~(section \ref{sec:rpprocess}). 
Rates of particular interest are those that lead to global changes in $\langle Z\rangle$, which would alter the depth at which pycnonuclear fusion could proceed~\cite{Wijn13}, and rates affecting odd-$A$ abundances, which may result in significant Urca cooling in the crust~\cite{Meis17}.
While some studies have 
investigated which rates these may be~\cite{Pari08,Pari09,Cybu16,Scha17}, a relatively small range of astrophysical conditions has been explored in self-consistent calculations. In principle the rates which impact superburst ash production, namely rates operating after freeze-out from nuclear statistical equilibrium, would be of interest, but computational challenges have made sensitivity studies prohibitively 
expensive.

The range in which different burning regimes operate is also not clear. One challenge facing the study of Type I X-ray bursts is a discrepancy between theory and observations for the boundary between stable and unstable nuclear burning (section~\ref{sec:burst_theory}). Observational surveys of thousands of Type I X-ray bursts find a peak burst rate near a mass accretion rate of  $\dot{M} \approx 0.3 \, \dot{M}_{\rm Edd}$~\cite{Paradijs1988,Cornelisse2003,galloway08} after which the burst rate decreases with increasing $\dot{M}$. Numerical models, however, predict an increase in burst rate with increasing $\dot{M}$. Furthermore, studies of stable and explosive burning fail to produce adequate amounts of carbon needed to ignite superbursts 
seen in some objects exhibiting Type-I bursts~\cite{Stev14}.

\subsection{What are the Major Heating/Cooling Sources in the Accreted Crust?}
The thermal structure of the crust influences the transitions between burning regimes, the depth at which fuel is ignited for unstable burning, and the rate of cooling after accretion turns off. Therefore identifying and quantifying the most significant heat sources and heat sinks is paramount to constructing accurate models of near-surface phenomena on accreting neutron stars.
While the general features of nuclei that make them susceptible to generating substantial heating or cooling are known, identifying specific nuclei is presently 
problematic (see section~\ref{section:interaction}).

The masses, low-lying structure, and weak transition rates for neutron-rich nuclides with $A\lesssim100$ lack sufficient constraints to definitively quantify the presence and strengths of $e^{-}$-capture heating and cooling. For pycnonuclear fusion, the properties of the nuclear potential for low-$Z$ nuclides far from stability needs further study to establish whether observed enhancement in fusion rates, e.g. Reference~\cite{Sing17}, are the exception or the rule. The myriad of theoretical uncertainties associated with estimating pycnonuclear rates (section~\ref{sec:pycno}) need to be reconciled if the region of deep crustal heating is to be determined. More modeling efforts for crust composition evolution need to be coupled to surface-burning models for the same objects in order to remove degrees of freedom and more rigorously test our understanding of heating and cooling reactions in the crust.

A related and even 
less certain issue is the identity of the shallow heating mechanism required in model-observation comparisons for nearly all accreting neutron star 
observables. Heating beyond the scale of nuclear physics uncertainties (as we understand them) appears to be necessary to reproduce observed properties of some Type-I X-ray bursts~\cite{Keek17},
cooling transients~\cite{Turl15,Deib15}, and possibly superbursts~\cite{Keek11,Reic17}. Better constraints on reaction-based heating and establishing the presence or absence of Urca cooling layers will be key to restricting the strength and location, and therefore mechanism, of shallow heating in accreting neutron stars. Furthermore, a proper accounting of all possible heat sources needs to be done~\cite{Fattoyev2017arXiv}.

\subsection{How does the Crust Become So Pure?}\label{sec:pure}
Inferences of the accreted crust thermal conductivity from observed cooling indicate that 
minimal crust impurities are allowed (section ~\ref{sec:cooling_transients}).
However, Type-I X-ray burst
and stable burning ashes have more than 10 times larger 
$Q_{\rm{imp}}$ (table~\ref{table:AvgZAQTable} and figure~\ref{figure:ashes}). 
Models suggest that phase separation in the ocean and crust can be crucial for 
formation of a purified crust. 
Molecular dynamics simulations
\cite{Horowitz2007PhRvE,Horowitz2007PhRvE,Schneider2012PhRvE,Hughto2012PhRvE,Capl17} and semi-analytical calculations \cite{Medi10,Mcki16} show 
that at the crystallization interface in most cases the solid phase is enriched in heavier (higher $Z$) elements, while lighter elements remain in the liquid phase. In some cases, however, (when the ashes average charge is low) the inverse situation was found \cite{Mcki16,Capl17}. 

The consequences of phase separation are not entirely clear. It was proposed that enrichment of the ocean with light elements will 
result in compositionally-driven convection~\cite{Medin2011,Medin2014ApJ,Medin2015ApJ}.
However once a steady state is achieved, the composition of the freezing solid  should match on average the composition of matter entering
the top of the ocean \cite{Medin2011}. 
Alternatively, the accumulation of light nuclei in the ocean eventually results in their solidification so that alternating crystalline layers or crystalline domains of different composition can form, reducing considerably $Q_{\mathrm{imp}}$ in each layer or domain \cite{Capl17}

Another possibility is that nuclear reactions can purify the ashes as they sink deep in the crust (section~\ref{section:interaction}). 
Recall that the observational constraints on impurities are more relevant for the inner crustal regions (section~\ref{sec:cooling_transients}). Pycnonuclear reactions burn lighter nuclei before the neutron drip point and neutron emission reactions around neutron drip simplify the composition towards a few closed-shell nuclei. Interestingly, this results in lowering the impurity for rp-process ashes, but on the contrary increases the diversity of species if one starts from a pure one-element composition \cite{Gupt08,Lau18,steiner2012}.  Furthermore, very recent network calculations that follow the 
evolution of ashes beyond neutron drip indicate that these multicomponent configurations are quickly destroyed and all matter at $\rho\gtrsim 1.5\times 10^{12}$~g~cm$^{-3}$ is converted into a single nucleus and a neutron gas, regardless of the initial surface abundances~\cite{Lau18}. The only exception is the case when heavy nuclei ($A\ge 106$) are present in the initial ashes. In this case the $N=82$ nuclei are locked and still present in the inner crust; the impurity parameter remains high in this case \cite{Lau18}.

We also note, that binary (or multinary) crystalline structures can form, for instance, if two species dominate the composition in a certain proportion \cite{Chamel2017JPhCS}. 
Then the impurity parameter can formally  be high (depending on the charges in the mixture) but the pure crystalline structure will lead to a high conductivity. Indeed, the compositions dominated by two or few species in similar fractions are found at some point in the reaction network evolved to the inner crust \cite{Lau18}.

Clearly, a solid understanding of the accreted crust structure and composition does not exist yet. This would require time-dependent calculations which couple models 
of the structural changes in the crust (e.g. molecular dynamics) 
to a nuclear reaction network, following the reaction sequence all the way to the crust-core transition, including all relevant nuclear transitions. 

\begin{figure} [h]
\centering
\includegraphics[scale=0.6]{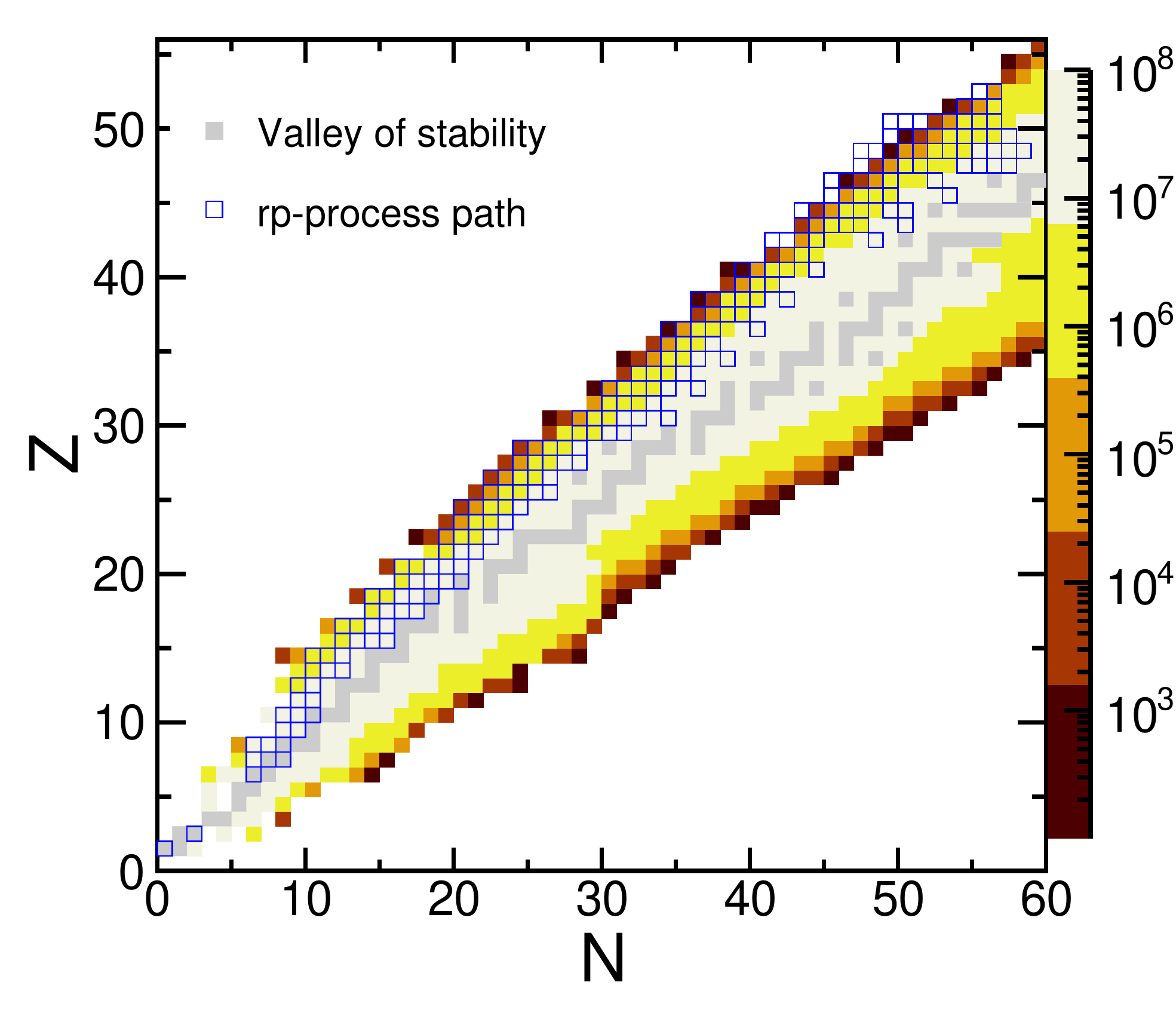}
\caption{Predicted production rates in particles per second, 
as indicated by the color,
for ReA3 beams at the Facility for Rare Isotope Beams~\cite{Boll11} compared to the rp-process path~\cite{Scha06}. See Reference~\cite{Meis16b} for a similar figure with anticipated fast beam rates.
\label{figure:reaproduction}}
\end{figure}

\subsection{What is Being Done to Solve These Problems?}
Ongoing efforts in experiment, observation, and theory are constantly advancing the frontier of what is known about the nuclear processes occurring in accreted neutron star 
crusts and their observable impacts. Recent and near-future advances in instrumentation and the open-source software movement promise to accelerate the pace of progress.

X-ray observations, on which neutron star 
studies rely, are 
performed with
a number of relatively new and advanced telescopes. These include stalwarts like the {Neil Gehrels Swift Observatory}~\cite{swift}, INTEGRAL~\cite{integral}, the Chandra X-ray observatory~\cite{chandra}, XMM-Newton~\cite{newton}, 
and { MAXI}~\cite{maxi} and newcomers {ASTROSAT}~\cite{astrosat}, {NUSTAR}~\cite{nustar}, and {NICER}~\cite{nicer}. The problems discussed here would benefit from additional observational data of Type-I X-ray bursts, superbursts, and cooling transients. For instance, more regular bursters like GS~1826$-$24 would provide extra tests for models that successfully reproduce observations~\cite{Hege07,Meis18,Gall17}. Superbursts and cooling transients have small populations, making our conclusions about these phenomena susceptible to small sample biases. Another cooling transient with a hot crust like MAXI~J0556$-$332 would provide 
a new 
test for the presence of Urca cooling in the accreted crust. The holy grail from this perspective would be a source with a variable accretion rate that exhibits Type-I X-ray bursts and/or superbursts prior to an extended accretion outburst at high (near Eddington) accretion rate. Further progress will also require next-generation telescopes, e.g. the planned missions eXTP~\cite{extp} and Strobe-X~\cite{strobex}, in order to enable high-precision observations that will obviate the need for light-curve averaging and possibly allow time-resolved spectroscopy.

Nuclear physics experimental efforts are beginning to benefit from newly-developed techniques and will soon benefit from next-generation facilities. Improved constraints on masses, structure, and reaction rates are required from nearly dripline to dripline on the nuclear chart for $A\lesssim100$. A consequence of the large number of nuclei involved is that many useful indirect measurements are left to be done at presently existing stable and radioactive ion beam facilities. The reach of these facilities is being expanded by studies leveraging recently developed capabilities such as in-ring reactions~\cite{Mei15}, the $\beta$-Oslo technique~\cite{Spyr14}, phase-imaging penning trap mass measurements~\cite{Elis15}, and ion beams produced in projectile fragmentation that have been stopped and re-accelerated to astrophysical energies coupled to windowless gaseous targets~\cite{Bard16}. Nonetheless, the next-generation facilities FRIB~\cite{frib} and FAIR~\cite{fair} will provide unprecedented access to nuclei participating in surface burning and crust reactions (see figure~\ref{figure:reaproduction}).

Most aspects of the accreted neutron star 
atmosphere, ocean, and crust have been modeled, including various regimes of nuclear burning, phase separation, and exotic reaction processes. Some of the most valuable ongoing and near-future efforts are and will be those which seek to combine the major aspects of several models, e.g. time-dependent accretion, phase separation, and a nuclear reaction network. More rigorous tests of crust models could be achieved by engaging in consistent multi-observable modeling for sources such as KS~1731$-$260 which exhibits Type-I X-ray burst, superburst, and cooling transient phases. Aside from reproducing observables, more parameter studies assessing the sensitivity of models to nuclear processes are desired to move beyond the small set of conditions that have been explored thus far~\cite{Pari08,Pari09,Cybu16,Scha17}. Open-source astrophysics model codes, such as {\tt MESA}~\cite{Paxt11} and {\tt dStar}~\cite{Brow15} are invaluable resources for these studies by enabling more members of the nuclear astrophysics community to contribute.

It 
goes without saying that efforts in observation, experiment, and theory will benefit tremendously from cross-discipline collaboration. In this regard the nuclear astrophysics community
is performing exceedingly well. Continuing and strengthening these bonds (through focal points such as the Joint Institute for Nuclear Astrophysics) will be key to solving the many remaining issues of nuclear physics in the outer layers of accreting neutron stars and the affected astronomical observables.

\section*{Acknowledgements}
The authors thank M. Prakash, H. Schatz, and D.~G. Yakovlev for useful discussions. Z.~M. was supported by the U.S. Department of Energy under grant No. DE-FG02-88ER40387. P.~S. was supported by the BASIS Foundation.
This work was supported in part by the U.S. National Science Foundation under grant No. PHY-1430152 (Joint Institute for Nuclear Astrophysics--Center for the Evolution of the Elements).

\section*{References}
\bibliographystyle{iopart-num}
\bibliography{References}

\end{document}